\documentclass{aa}
\usepackage{graphicx}
\usepackage[varg]{txfonts}
\usepackage{comment}
\usepackage{placeins}
\makeatletter
\renewcommand*\aa@pageof{, page \thepage{} of \pageref*{LastPage}}
\makeatother

\def \lcdm {{\hbox{$\Lambda$CDM}}}
\def \omegam {{\hbox{$\Omega_{\rm m}$}}}
\def \omegal {{\hbox{$\Omega_\Lambda$}}}
\def \hzero {{\hbox{$H_0$}}}
\newcommand{\kmsmpc }{\mbox{km s$^{-1}$ Mpc$^{-1}$}}

\def \kmsmpc      {km\,s$^{-1}$\,Mpc$^{-1}$}

\def \deg         {\text{$^{\circ}$}}

\def \mjybeam     {mJy\,beam$^{-1}$}

\def \mach	  {\mathcal{M}}

\newcommand{\Msun}{\text{$\rm M_\odot$}}

\newcommand{\planck }{{\em Planck}}
\newcommand{\lofar }{LOFAR}
\newcommand{\lotss }{LoTSS}
\newcommand{\xmm }{{\em XMM-Newton}}
\newcommand{\chandra }{{\em Chandra}}

\begin{document}

\title{The \planck{} clusters in the \lofar{} sky}
\subtitle{VI. \lotss-DR2: Properties of radio relics\thanks{FITS images in Fig.~\ref{fig:radio_collage} and Fig.~\ref{fig:xray_collage} are only available in electronic form at the CDS via anonymous ftp to cdsarc.cds.unistra.fr (130.79.128.5) or via \url{https://cdsarc.cds.unistra.fr/cgi-bin/qcat?J/A+A/}}}
\author{A. Jones\inst{\ref{hamburg}}, F. de Gasperin\inst{\ref{ira},\ref{hamburg}}, V. Cuciti\inst{\ref{hamburg}}, A. Botteon\inst{\ref{bologna},\ref{ira},\ref{leiden}}, X. Zhang\inst{\ref{leiden},\ref{sron}}, F. Gastaldello\inst{\ref{inaf}}, T. Shimwell\inst{\ref{astron},\ref{leiden}}, A. Simionescu\inst{\ref{sron},\ref{leiden},\ref{kavli}}, M. Rossetti\inst{\ref{inaf}}, R. Cassano\inst{\ref{ira}}, H. Akamatsu\inst{\ref{sron}}, A. Bonafede\inst{\ref{bologna},\ref{ira}}, M. Br\"uggen\inst{\ref{hamburg}}, G. Brunetti\inst{\ref{ira}}, L. Camillini\inst{\ref{inaf},\ref{milano}}, G. Di Gennaro\inst{\ref{hamburg}}, A. Drabent\inst{\ref{tautenburg}}, D. N. Hoang\inst{\ref{hamburg}}, K. Rajpurohit\inst{\ref{bologna},\ref{ira},\ref{tautenburg}}, R. Natale\inst{\ref{inaf},\ref{milano}}, C. Tasse\inst{\ref{paris},\ref{rhodes}}, R. J. van Weeren\inst{\ref{leiden}}}

\institute{
Hamburger Sternwarte, Universit\"{a}t Hamburg, Gojenbergsweg 112, D-21029 Hamburg, Germany \label{hamburg} \\
\email{ajones@hs.uni-hamburg.de}
\and INAF - Istituto di Radioastronomia, via P. Gobetti 101, 40129, Bologna, Italy \label{ira}
\and DIFA - Università di Bologna, Via Gobetti 93/2, I-40129 Bologna, Italy \label{bologna}
\and Leiden Observatory, Leiden University, PO Box 9513, 2300 RA Leiden, The Netherlands \label{leiden}
\and SRON Netherlands Institute for Space Research, Niels Bohrweg 4, 2333 CA Leiden, The Netherlands \label{sron}
\and INAF, IASF-Milano, via A. Corti 12, I-20133 Milano, Italy \label{inaf}
\and ASTRON, the Netherlands Institute for Radio Astronomy, Postbus 2, 7990 AA Dwingeloo, The Netherlands \label{astron}
\and Kavli Institute for the Physics and Mathematics of the Universe (WPI), The University of Tokyo, Kashiwa, Chiba 277-8583, Japan \label{kavli}
\and Dipartimento di Fisica, Università degli Studi di Milano, Via Celoria 16, 20133 Milano, Italy \label{milano}
\and Th\"uringer Landessternwarte, Sternwarte 5, D-07778 Tautenburg, Germany \label{tautenburg}
\and GEPI \& USN, Observatoire de Paris, Université PSL, CNRS, 5 Place Jules Janssen, 92190 Meudon, France \label{paris}
\and Department of Physics \& Electronics, Rhodes University, PO Box 94, Grahamstown, 6140, South Africa \label{rhodes}
}

\authorrunning{A. Jones et al.} 
\titlerunning{The \planck{} clusters in the \lofar{} sky}

\date{Received XXX; accepted YYY}

\abstract
% context heading
{It is well-established that shock waves in the intracluster medium launched by galaxy cluster mergers can produce synchrotron emission, which is visible to us at radio frequencies as radio relics. However, the particle acceleration mechanism producing these relics is still not fully understood. It is also unclear how relics relate to radio halos, which trace merger-induced turbulence in the intracluster medium.}
% aims heading
{We aim to perform the first statistical analysis of radio relics in a mass-selected sample of galaxy clusters, using homogeneous observations.}
% methods heading
{We analysed all relics observed by the Low Frequency Array Two Metre Sky Survey Data Release 2 (\lotss{} DR2) at 144 MHz, hosted by galaxy clusters in the second \planck{} catalogue of SZ sources (PSZ2). We measured and compared the relic properties in a uniform, unbiased way. In particular, we developed a method to describe the characteristic downstream width in a statistical manner. Additionally, we searched for differences between radio relic-hosting clusters with and without radio halos.}
% results heading
{We find that, in our sample, $\sim$ 10\% of galaxy clusters host at least one radio relic. We confirm previous findings, at higher frequencies, of a correlation between the relic-cluster centre distance and the longest linear size, as well as the radio relic power and cluster mass. However, our findings suggest that we are still missing a population of low-power relics. We also find that relics are wider than theoretically expected, even with optimistic downstream conditions. Finally, we do not find evidence of a single property that separates relic-hosting clusters with and without radio halos.}

\keywords{galaxies: clusters: general -- galaxies: clusters: intracluster medium -- radiation mechanisms: nonthermal -- radiation mechanisms: thermal -- catalogs}

\maketitle

\section{Introduction}
\label{sec:intro}

Mergers of galaxy clusters generate shock waves that propagate through the intracluster medium (ICM). Since the characterisation of a cluster shock by \citet{Markevitch2002} with \chandra{}, many more have been found using measurements of the X-ray surface brightness, entropy, and temperature profiles of cluster outskirts \citep[e.g.][]{Ogrean2013Coma,Shimwell2015,Eckert2016,Akamatsu2017,Urdampilleta2018}.
The connection between radio relics (RRs) and merger shocks has been well-established by a number of shocks detected in X-rays, as ICM density and temperature discontinuities, at the location of a RR \citep[e.g.][]{Finoguenov2010,Bourdin2013,Akamatsu2013,Botteon2016_A115,Botteon2016}. There is also clear evidence of the relation between RRs and galaxy cluster merger events from both weak lensing studies \citep[e.g.][]{Jee2016,Finner2017} and optical spectroscopy \citep[e.g.][]{Golovich2019Spec}.
Not all RRs have a known associated shock, though this is likely the result of difficulties in shock detection from the low X-ray counts in cluster outskirts, where relics are typically located \citep[][]{Vazza2012,Ogrean2013XMM}.
Fermi-I, diffusive shock acceleration (DSA) is typically adopted to explain the generation of RRs from cluster shocks \citep[][]{Ensslin1998}. In DSA, charged particles are accelerated to relativistic energies by scattering upstream and downstream off magnetic inhomogeneities \citep[][]{Fermi1949,Blandford1987}. Due to the presence of cluster-scale magnetic fields, they emit synchrotron emission, which is observable as diffuse, roughly arc-like RRs \citep[also known as cluster radio shocks, see][for reviews]{Brunetti2014,VanWeeren2019}.
The power-law energy spectrum of cosmic-ray electrons (CRes) produced by DSA generates a radio brightness profile in line with observations \citep[e.g.][]{Hoeft2007,Kang2012}. Additionally, relics are typically observed to have high polarisation fractions ($\gtrsim 20\% - 60\%$), matching expectations of magnetic field alignment along the shock surface \citep[][]{Ensslin1998}.
However, DSA from the thermal pool cannot entirely explain the properties of RRs. A study by \citet{Botteon2020Shock} found that acceleration of CRes from the thermal pool via DSA, in such weak shocks ($\mach{} \lesssim 3$), is in most cases insufficient to explain the acceleration efficiencies required to produce the luminosity of relics. Re-acceleration of a pre-existing population of mildly relativistic CRes could relieve some of this tension \citep[e.g.][]{Markevitch2005,Kang2011,Pinzke2013}. There is morphological and spectral evidence that the tails of radio galaxies can provide seed electrons that are re-accelerated by shocks \citep[e.g.][]{Bonafede2014,VanWeeren2017,DiGennaro2018}. For example, \citet{VanWeeren2017} found that the RR in Abell 3411-3412 is connected to the tail of a cluster-member radio galaxy. Moreover, the energy spectrum steepens along the tail, consistent with radiative losses, and subsequently flattens again at the inner boundary of the RR, implying re-acceleration. 
There are, however, still relatively few relics for which there is evidence of a connection/re-acceleration in general.

In addition to giant shock waves, galaxy cluster mergers generate turbulence in the ICM. This turbulence cascades down to smaller scales and can (re-)accelerate CRes and produce radio synchrotron emission, in the form of a radio halo \citep[RH, see][for a review]{Brunetti2014}. These RHs are typically located in the cluster centre and follow the morphology of the X-ray-emitting gas. Numerous statistical studies of RHs have been performed and have shown a correlation between the RH power and its host cluster mass \citep[e.g.][]{Basu2012,Cassano2013,VanWeeren2021,Cuciti2021}, as well as with the X-ray luminosity \citep[e.g.][]{Liang2000,Brunetti2009}.
In contrast, there have been relatively few statistical studies of RRs. Such investigations of relic properties are more challenging than for RHs due to a number of observational constraints.
For example, the lower abundance of RRs \citep[in $ \sim 5\%$ of clusters,][]{Kale2015}, compared to that of RHs \citep[in $\sim 40\%$ of clusters, e.g.][]{Cuciti2021}, and the typical location in the cluster periphery, where X-ray counts are low, make such studies challenging. Additionally, an unbiased measurement of the properties of relics is difficult, due to projection effects and their irregular morphologies.
In the first statistical study of RRs, \citet{VanWeeren2009} compiled all of those discovered (26 individual RRs) at that time from the literature. They discovered a correlation between the longest linear size (LLS) of a relic and its distance from the cluster centre. This finding was corroborated by \citet{DeGasperin2014} (hereafter FdG14) and was found at low significance by \citet{Bonafede2012}, who both restricted their analysis to only double radio relics (dRRs), that is to say pairs of diametrically opposed relics in the same cluster. The advantage of using dRRs is that the merger axis is relatively well-known and approximately on the plane of the sky \citep[][]{VanWeeren2011,Golovich2019Spec}, minimising projection effects. 
FdG14 also reported a correlation between host-cluster mass and RR power at 1.4 GHz, that is to say that more powerful relics are typically located in higher-mass clusters. However, simulations by \citet{Nuza2017} and \citet{Bruggen2020} suggest that we are missing a significant number of low-power relics, likely only detectable at low frequencies, and that the cluster mass provides a maximum radio power a relic can reach, rather than directly determining its power. The advent of sensitive, all-sky surveys at low radio frequencies will enable discovery of these low-power relics, if such a population exists.

In this paper we present the first statistical study of RRs at 150 MHz and their connection to RHs, using galaxy clusters covered by both the \planck{} PSZ2 catalogue \citep[][]{Planck2016} and the Low Frequency Array (\lofar{}) Two Metre Sky Survey Data Release 2 \citep[\lotss{} DR2, ][]{Shimwell2022}. Wherever possible, this was supplemented with archival \chandra{} and \xmm{} data, to determine the X-ray properties of the sample clusters, which are described fully in \citet{Zhang22}.
The biggest advantage of using such a sample is that it allows us to study RRs observed by the same telescope, and therefore approximately the same $uv$-coverage, observing frequency and sensitivity to compact and diffuse emission. Additionally, the observations were calibrated and the images produced in a uniform manner \citep[see][]{Tasse2021,Shimwell2022}.
The use of the PSZ2 catalogue allows us to produce a mass-selected sample, ensuring that our results are representative of those RRs observable at the sensitivity of \lofar{}.
This paper is the sixth in a series of papers\footnote{\url{https://lofar-surveys.org/planck_dr2.html}} utilising the \lotss{} DR2 - PSZ2 cluster sample to explore the properties of diffuse radio emission in the ICM. \citep[][]{Botteon2022} describes the sample in detail, the methods and data used and the source classification. The occurrence and scaling relations of all RHs in this sample are presented in \citet{Cassano22} and Cuciti et al. (in prep.), respectively, whilst upper limits on RH power in clusters with no detected diffuse emission are in \citet{Bruno22}.
An analysis of the X-ray properties of the sample is presented in \citet{Zhang22}.
See also \citet{Hoang2022} for analysis of diffuse radio emission within \lotss{} DR2 in non-PSZ2 clusters.

The paper is structured as follows. Sec~\ref{sec:sample} describes the sample of relics and how we measured their properties. In Sec.~\ref{sec:results} we present our results. In Sec.~\ref{sec:discussion} and Sec.~\ref{sec:conclusions} we discuss our results and conclude. 
We adopt a fiducial \lcdm\ cosmology with $\omegal = 0.7$, $\omegam = 0.3$, and $\hzero = 70$ \kmsmpc. All errors are at $1\sigma$, unless otherwise stated.

\section{The sample}
\label{sec:sample}

We provide here a summary of the sample of relics used in this paper, its composition and the measurement of relic properties. For a full description of the cluster sample, including the data calibration, imaging procedure, and radio source classification, we refer the reader to \citet{Botteon2022}.

\subsection{Relics In \lotss{} DR2}
\label{sec:sample:classification}

Of the 1653 galaxy clusters contained in the \planck{} PSZ2 catalogue \citep[][]{Planck2016}, 309 lie within the \lotss{} DR2 footprint \citep[][]{Shimwell2022}. The \lotss{} DR2 data were reprocessed to produce 144 MHz radio images, at various resolutions, for each cluster \citep[][]{Botteon2022}.
The radio images were then visually inspected for evidence of diffuse emission not associated with an AGN.
Elongated ($\gtrsim$ 300 kpc) diffuse emission with a sharp radio edge, lying outside the bulk of X-ray emission, in the cluster outskirts, was classified as a RR. Of those, relics diametrically opposed to another relic on the opposite side of its cluster were defined as dRRs. Some clusters host more than one RR, but do not fit this criterion and were therefore classified as multiple radio relics (mRRs). We use archival \chandra{} and \xmm{} data to determine the cluster X-ray properties. The data are processed and used to produce images, smoothed to 30~kpc at the cluster redshift \citep[see][for further details]{Botteon2022}. A full analysis of the X-ray properties of the \lotss{} DR2 - PSZ2 sample will follow in \citet{Zhang22}. For clusters with no \chandra{} or \xmm{} observation, the emission was classified as a candidate radio relic (cRR), since it is not possible to define the location of the diffuse emission with respect to the ICM. In this case, the position of the radio emission with respect to the cluster optical overdensity was used.
The resulting sample consists of 26 relic-hosting clusters, of which 20 have accompanying X-ray observations. Of the 35 individual relics residing in these clusters, 12 were defined as dRRs, 5 as mRRs and 6 as cRRs. The rest (12) were defined as RRs. There are no double, or multiple, candidate radio relics.
For an image gallery of all RRs in our sample, and, where possible, their location with respect to the ICM X-ray emission, see Appendix~\ref{app:collages}. Images and tables are taken from \citet{Botteon2022} and can be found at full resolution on the project website \footnote{\url{https://lofar-surveys.org/planck_dr2.html}}.

Radio halos in the \lotss{} DR2 - PSZ2 sample were classified by \citet{Botteon2022}. They were defined as extended radio sources occupying the same region as either the bulk ICM X-ray emission (RHs) or an overdensity of optical galaxies (cRHs). Of the 26 relic-hosting clusters, 12 also host a radio halo (11 RHs and 1 cRH). For some radio objects, originally classified as RHs, the low signal-to-noise ratio did not allow fitting of the model used to estimate RH flux \citep[denoted RH*s/cRH*s, see][for more detail]{Botteon2022}. We treated clusters which host such RH*s/cRH*s as hosting a RH, though there is only 1 relic-hosting cluster for which this is the case (PSZ2\,G116.50-44.47, classified as RH*).

\begin{figure}
    \resizebox{\hsize}{!}{\includegraphics{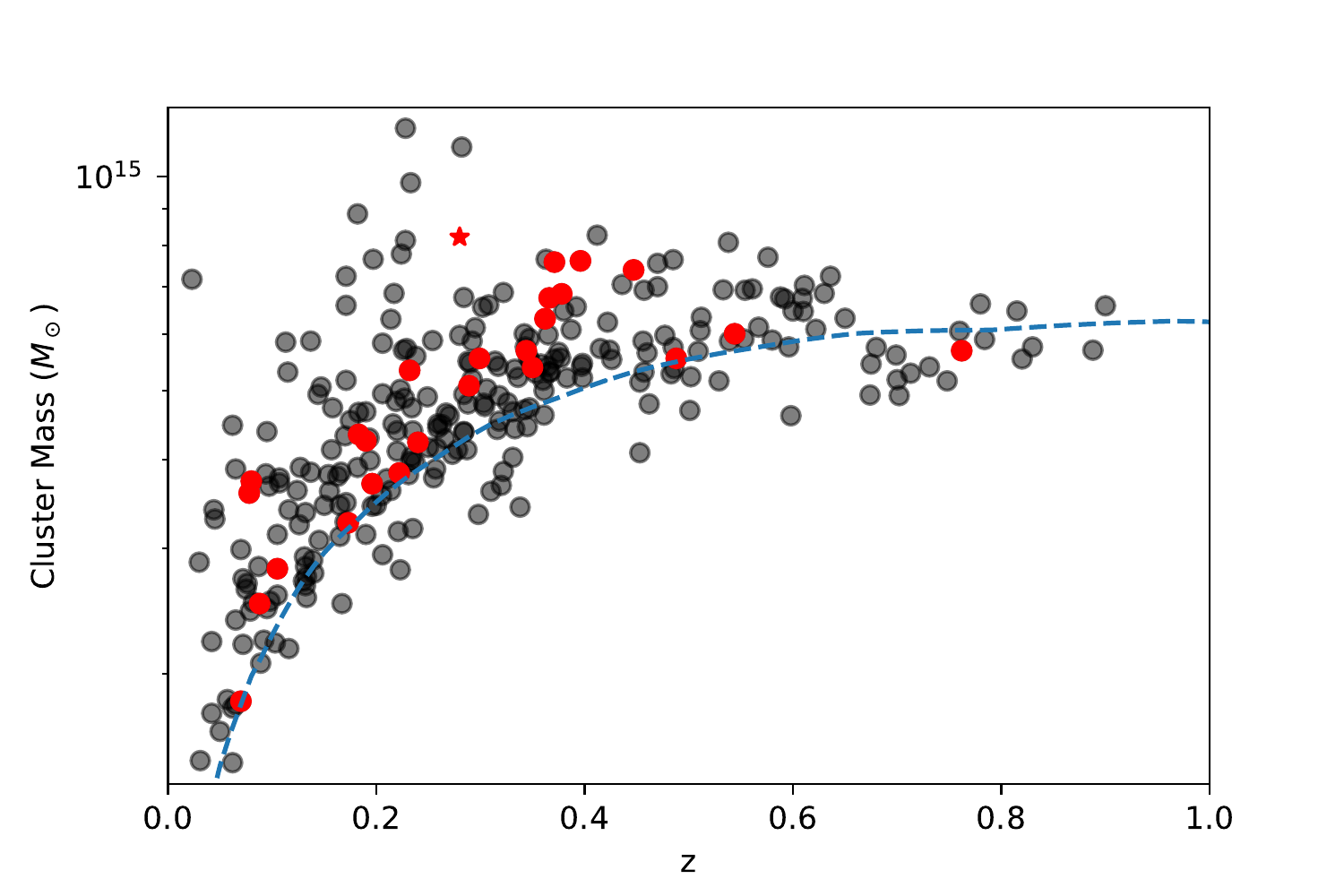}}
    \caption{Cluster $M_{500}$ mass vs. redshift for all clusters in the \lotss{} DR2 - PSZ2 sample. Red circles denote clusters which host at least one RR. Black circles denote all other clusters in the \lotss{} DR2 - PSZ2 sample \citep[][]{Botteon2022}. PSZ2\,G107.10+65.32 S is plotted as a red star, since the mass reported in the PSZ2 catalogue likely comes from both sub-clusters, PSZ2\,G107.10+65.32 N and PSZ2\,G107.10+65.32 S. The 50\% PSZ2 completeness line is shown in blue.}
    \label{fig:M_z}
\end{figure}

Fig.~\ref{fig:M_z} shows the mass distribution of all clusters in the \lotss{} DR2 - PSZ2 sample, as a function of redshift, where the mass is $M_{500}$ from the PSZ2 catalogue \citep[][]{Planck2016}. Clusters which host at least one RR are shown as red points. PSZ2\,G107.10+65.32 is comprised of two sub-clusters, each undergoing its own merger \citep[Abell 1758, ][]{Botteon2020A1758}. We have plotted this cluster separately, as a red star, since only the S sub-cluster hosts a RR, but, since the resolution of \planck{} is not sufficient to separate the two sources, the mass given in the PSZ2 catalogue is likely from a combination of the two sub-clusters. The blue dashed line shows the mass at which Planck is 50\% complete, as a function of redshift. This line comes from converting the selection function from the \planck{} archive from SZ signal - cluster size to mass - redshift \citep[][]{Planck2016} and taking the boundary at which the probability of detecting a cluster is 50\%. The only relic-hosting cluster which lies below this line is PSZ2\,G069.39+68.05.
Throughout this paper, we consider only clusters lying above this line as our representative cluster sample and restrict our analysis to only these clusters. However, since there is only one relic-hosting cluster below the line (PSZ2\,G069.39+68.05) and one for which we cannot be certain (PSZ2\,G107.10+65.32), we plot these two relics whenever possible and label them accordingly, despite their absence from our analysis.

\subsection{Relic measurements}
\label{sec:sample:measurements}

All measurements used in this paper are presented in \citet{Botteon2022} (Table A.4.). For completeness, we describe the procedures used below.

Radio relics, owing to their often complex morphologies, do not lend themselves well to parametric model-fitting procedures, such as those used for RHs \citep[][]{Boxelaar2021}. Connections of relics to radio galaxies, or even RHs, makes automatically separating relic from non-relic emission very challenging. Therefore, to best enable fair comparison of relic properties, we adopted a hybrid approach to their measurement. For each relic we manually defined a region which best covers it, whilst avoiding non-relic emission, by visually inspecting the radio and X-ray images. In general, we uniformly computed the properties of each relic from the 50 kpc-taper, compact-source-subtracted image of each cluster, where we defined relic emission as emission above $2\sigma_{rms}$ within the pre-defined region. $\sigma_{rms}$ is the rms noise of the image.
A few relics required slightly different treatment. The relics PSZ2\,G089.52+62.34 N2 (Abell 1904), PSZ2\,G091.79-27.00, PSZ2\,G113.91-37.01 S, PSZ2\,G166.62+42.13 E (Abell 746), and PSZ2\,G205.90+73.76 N/S are not fully visible in the 50 kpc-taper images. We therefore chose to use the 100 kpc-taper images instead. Additionally, visual inspection of the model used to subtract compact sources in images of PSZ2\,G190.61+66.46 revealed that it included some relic emission. Since there are no compact sources within the relic, we chose to use the 50 kpc-taper image without compact-source subtraction for this relic.

\begin{figure*}
    \includegraphics[width=17cm]{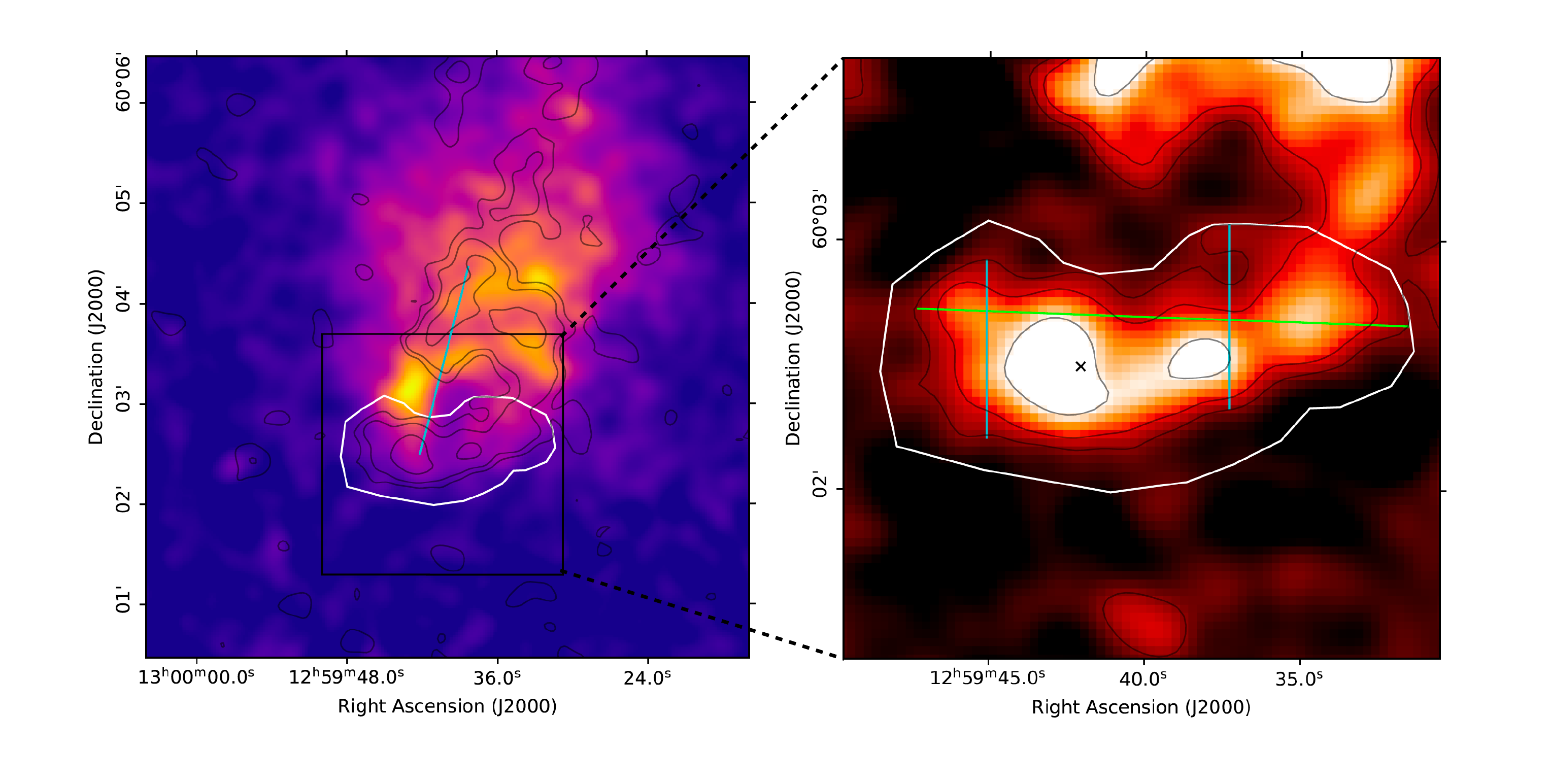}
    \caption{Reference images of the relic in PSZ2\,G121.03+57.02. The white region shows the region used to compute the properties of the relic. Black lines show the 2$\sigma_{rms} \times [1,2,4,..]$ contours from the 50 kpc-taper, compact-source-subtracted \lofar{} image from \citet{Botteon2022}. \textbf{Left}: \chandra{} X-ray image. The cyan line scales the distance between the relic position and the cluster centre, as defined in Sec.~\ref{sec:sample:measurements}. \textbf{Right}: Zoom-in of the 50-kpc-taper, compact-source-subtracted \lofar{}, centred on the relic. The green line shows the LLS of the relic. Two example lines used to measure the relic width are shown in \textbf{cyan}. The black cross is where we define the relic coordinate.}
    \label{fig:Ref}
\end{figure*}

Fig.~\ref{fig:Ref} shows a reference image of PSZ2\,G121.03+57.02, demonstrating the measurement of relic properties, as detailed below. The left panel shows the location of the relic, outside the bulk of ICM X-ray emission. The right panel shows a zoom-in of the relic from the 50 kpc-taper \lofar{} image. The region used for this relic and the $\geq 2\sigma_{rms}$ contours are shown as white and black lines, respectively.

Since the \lofar{} images are at a nominal frequency of 144 MHz, we computed the relic flux and power at 150 MHz, assuming $\alpha = -1$, where $\alpha$ is the spectral index ($S_{\nu} \propto \nu^{\alpha}$). The spectral index of RRs is typically in the range $-1 \lesssim~\alpha~\lesssim -1.5$ \citep[e.g.][FdG14]{Feretti2012}. Since the frequency conversion is small and the clusters in our sample are relatively nearby, the choice of $\alpha$ in this range is somewhat arbitrary. If we instead choose $\alpha = -1.5$, corresponding to a steep-spectrum RR, the k-corrected 150 MHz RR power is $\lesssim 10\%$ greater, in most cases.
Since the k-correction factor for a source with spectral index $\alpha = -1$ is zero, we did not need to k-correct our luminosities. The errors take into account a combination of the error from the rms noise and a 10\% calibration error \citep[][]{Shimwell2022}.
We also included an error to account for residuals from the compact-source subtraction process. We split the images into four groups, based on their total discrete-source flux density, $S_{discrete}$, and assigned an appropriate fractional error which increases with $S_{discrete}$ \citep[see][]{Botteon2022}.

The LLS was calculated as the distance between the two pixels with maximum separation which were defined as part of the relic emission. Since the synthesised beam is the smallest angular scale across which we can trust the flux, the error in the LLS corresponds to one beam width. We used the same approach for all other distance measurements presented in this paper. We note that this can be considered a lower limit of the LLS, since we are limited by how much of the RR is detected. The LLS of PSZ2\,G121.03+57.02 is shown in Fig.~\ref{fig:Ref} as a green line.

Due to the often complex relic morphologies, the measurement of the relic extent downstream of the shock front, or relic width, is strongly dependent on the location at which it is measured. This makes it extremely difficult to measure a single width value in a consistent way which is fair for all relics in our sample. We therefore took a statistical approach, by measuring the width at many positions along the relic. 
In the case of a shock propagating outwards, the LLS should be oriented approximately perpendicular to the direction of propagation, that is perpendicular to the upstream - downstream direction. This orientation was verified by eye, though we do not account for any curvature of the relic. We could therefore, at each pixel along the LLS, draw a line perpendicular to the LLS and calculate the maximum distance between relic pixels which lie on it. The blue lines in Fig.~\ref{fig:Ref} (right) show two example lines used to calculate the width of the relic in PSZ2\,G121.03+57.02. We then took the median of all values measured as the characteristic relic width and one standard deviation as its error. We chose to take the median as our characteristic width because it minimises the effect of small width measurements at the relic edges and areas with abnormally large widths.
As with the LLS, the width measurements we made are lower limits, since the entire downstream extent of the RRs may be too faint to detect.

Without a direct detection of a shock front, it is not necessarily clear where the shock front producing a RR is located. There is still debate over the nature of the bright filaments often observed in RRs. However, a detailed, high-resolution study of the relics in Abell 3667 by \citet{DeGasperin2022} suggests that the filaments trace regions of shock acceleration. Recent simulations support the scenario that the brightest RR regions correspond to the highest Mach numbers \citep[e.g.][]{Dominguez-Fernandez2021substructure,Wittor2021}. We therefore took the flux-weighted centre of the brightest 10\% of relic pixels as the location of the RR. The coordinate calculated for PSZ2\,G121.03+57.02 is shown as a black cross in Fig.~\ref{fig:Ref} (right). We subsequently used this point to calculate the distance to the cluster centre, $D_{RR-c}$, where we considered the X-ray centroid of a cluster, measured within $R_{500}$, its centre (shown as black crosses in Fig.~\ref{fig:xray_collage}) and to other relics, $D_{RR-RR}$, for dRRs. We note that, since the cRRs in our sample are those without accompanying X-ray images (see Sec.~\ref{sec:sample:classification}), we do not measure $D_{RR-c}$ for any cRRs.
We included an additional error in these distance measurements to account for possible projection effects. The merger axes of dRR-hosting clusters are expected to lie on, or close to, the plane of the sky \citep[][]{VanWeeren2011}. We therefore set this additional error as the distance corresponding to a 10\deg offset. For all other relics, we use an offset of 30\deg.

\section{Results}
\label{sec:results}

In this section, we report our results on the statistical properties of RRs and their host clusters.
In general, throughout the figures, cRRs are denoted by triangular data points, dRRs by plusses, and all other relics, including mRRs, by circles. This notation extends to graphs with one data point per cluster, that is to say that clusters which host dRRs are plotted as a plus, etc.. Additionally, though not included in all graphs, data points with surrounding red circles denote clusters which also host an RH/cRH.
Due to the ambiguity in PSZ2\,G069.39+68.05 (below 50\% \planck{} completeness, see Fig.~\ref{fig:M_z}) and PSZ2\,G107.10+65.32 (double cluster, see Sec.~\ref{sec:sample:classification} for more details), whenever possible, we label the points as '069' and '107' respectively, if they are included. We did not include these relics at all when plotting histograms, since they are not easily labelled. Since we did not include them when assessing the presence of a correlation, we also excluded them from the corresponding plots.

\subsection{X-ray morphological disturbance}
\label{sec:res:c-w}

The dynamical state of the clusters in our sample is assessed using the 30~kpc-smoothed \chandra{} and \xmm{} images. With these we are able to calculate the cluster concentration parameter \citep[][]{Santos2008},
\begin{equation}
\label{eq:c}
    c = \dfrac{F(r < R_{core})}{F(r < R_{ap})},
\end{equation}
where $F$ is the X-ray flux, $R_{core}$ the aperture of the core region and $R_{ap}$ the outer aperture,
and centroid shift \citep[][]{Mohr1993,Poole2006},
\begin{equation}
\label{eq:w}
    w = \left[\dfrac{1}{N_{ap}-1}\sum_{i}\left(\Delta_{i}-\bar\Delta\right)^{2}\right]^{1/2} \dfrac{1}{R_{ap}},
\end{equation}
where $N_{ap}$ is the number of apertures, $\Delta_{i}$ the centroid of the \textit{i}th aperture, and $\bar\Delta$ the average centroid.
$R_{core}$ and $R_{ap}$ were set following the convention of \citet{Cassano2010}, that is 100~kpc and 500~kpc, respectively.

\begin{figure}
    \resizebox{\hsize}{!}{\includegraphics{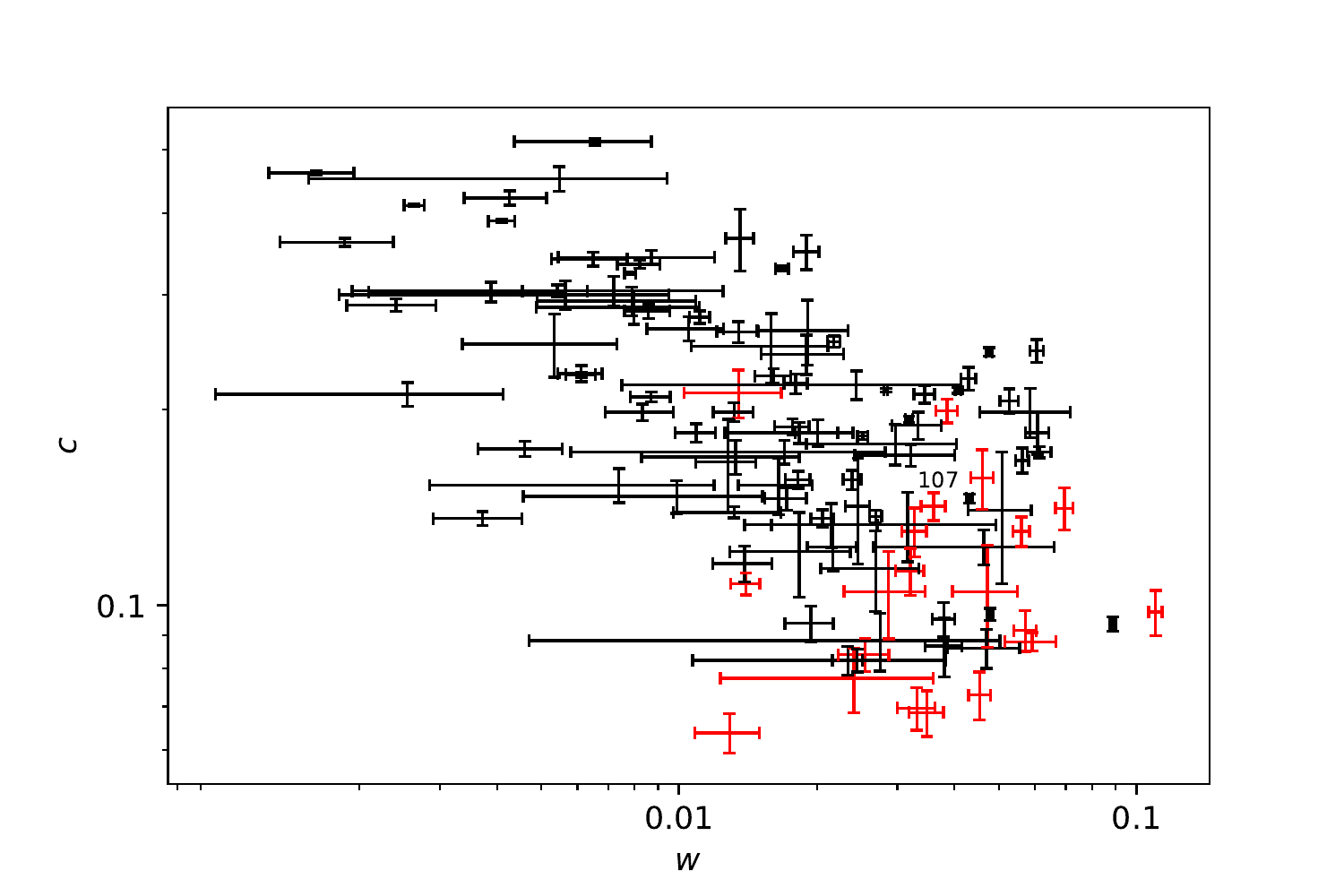}}
    \caption{Concentration parameter, $c$ vs. centroid shift, $w$ for all clusters with X-ray observations above the \planck{} 50\% completeness line in the DR2 sample. Relic-hosting clusters are denoted by red points. All other clusters are black. PSZ2\,G107.10+65.32 is labelled, with the label directly above the data point.}
    \label{fig:c_w}
\end{figure}

In Fig.~\ref{fig:c_w} we plot the cluster concentration parameter, $c$ against the centroid shift, $w$, for all clusters in the \lotss{} DR2 - PSZ2 sample above the \planck{} 50\% completeness line. The values, and corresponding errors, for all clusters are given in \citet{Botteon2022} and \citet{Zhang22}.

The plot shows the $c$ and $w$ parameters with their corresponding errors. We note that the combination of \chandra{} and \xmm{} measurements and corresponding uncertainties, where available, gives rise to large errors in clusters where the two instruments disagree significantly. This is likely caused by differences in PSF and X-ray count rate between the two instruments. In general, however, there is good agreement of the concentration parameter and centroid shift between the two instruments \citep[see][for a full discussion]{Zhang22}. However, there are no relic-hosting clusters which have large discrepancies between their \chandra{}-derived and \xmm{}-derived morphological parameters. There is no clear bi-modal distribution representing disturbed and relaxed clusters, but, in general, clusters with smaller $c$ and larger $w$ are more dynamically disturbed. Relic-hosting clusters are shown in red, with all other clusters in black.
We see that relic-hosting clusters primarily reside in the bottom right corner of the plot, corresponding to the most disturbed systems. Interestingly, the least disturbed cluster in the RR sample, PSZ2\,G205.90+73.76, hosts both a RH and a dRR pair.

\begin{figure}
    \resizebox{\hsize}{!}{\includegraphics{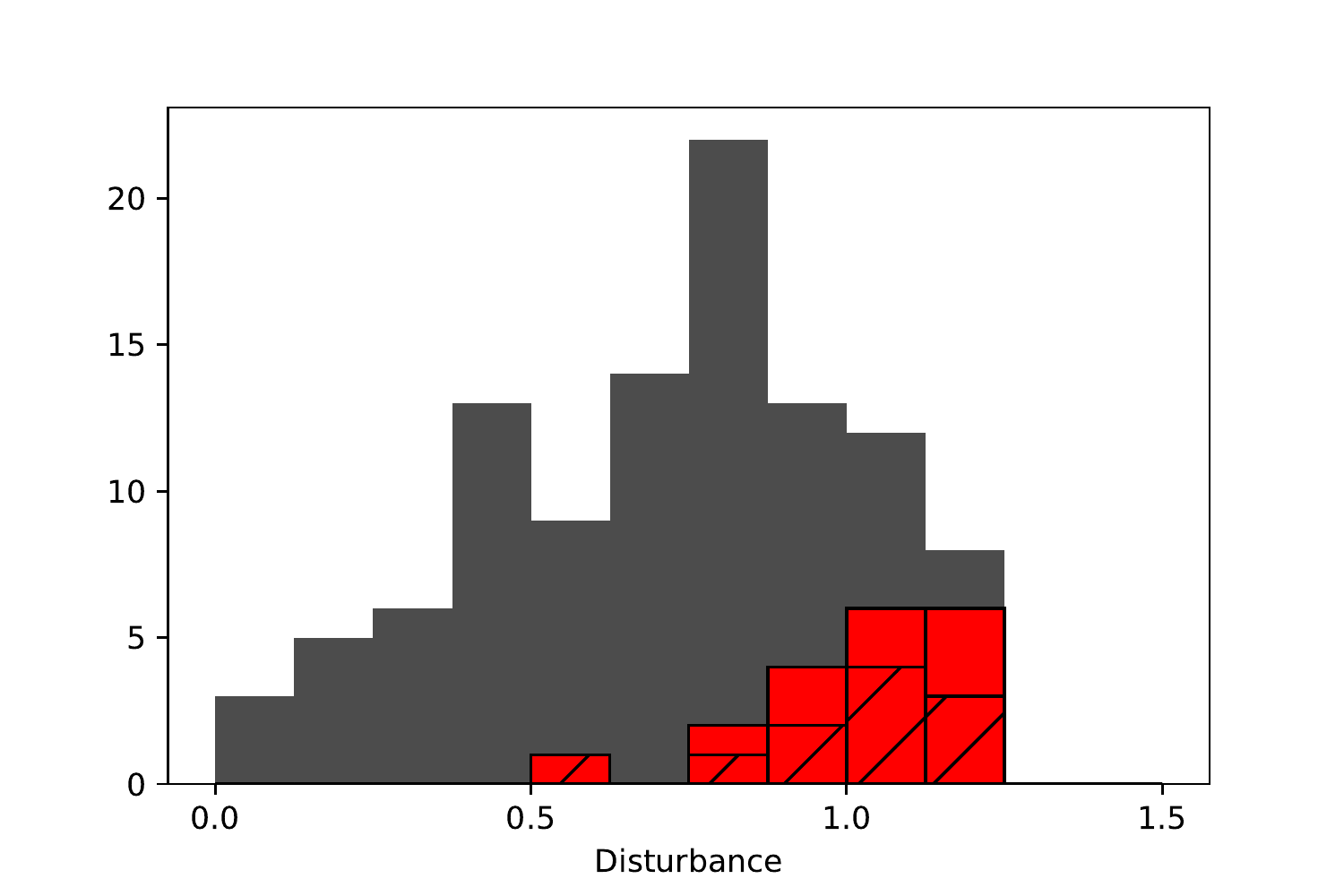}}
    \caption{Histogram of cluster disturbances. The grey bars show the distribution of all clusters above the \planck{} 50\% completeness line in the \lotss{} DR2 - PSZ2 sample with $c$ and $w$ measurements. The red bars show the distribution of all clusters which host a RR and the hatched bars show the clusters which host both a RR and RH. PSZ2\,G107.10+65.32 is excluded entirely.}
    \label{fig:Disturb_hist}
\end{figure}

Cuciti et al. (in prep.) define a quantity, 'disturbance', using the $c$ and $w$ values of the clusters in our sample. This quantity has no physical meaning, but is useful to compare 'disturbed' and 'relaxed' clusters, since larger values correspond to more dynamically disturbed clusters. For clarity, we summarise its calculation here.
We first normalised the values of $c$ and $w$, to account for the different ranges covered by each, with
\begin{equation}
    \mathcal{P}_{\rm norm} = \frac{\mathrm{log}(\mathcal{P}_i) - \mathrm{min}(\mathrm{log}(\mathcal{P}))}{\mathrm{max}(\mathrm{log}(\mathcal{P})) -\mathrm{min}(\mathrm{log}(\mathcal{P}))},
\end{equation}
where $\mathcal{P}$ represents either $c$ or $w$.
We then fit a line of the form $\log_{10}(c_{\rm norm}) = m\log_{10}(w_{\rm norm}) + q$ to the $c_{\rm norm}$ and $w_{\rm norm}$ data for all clusters with accompanying X-ray observations in the full \lotss{} DR2 - PSZ2 sample.
We derived the projected position of each cluster along this line and assumed that the cluster with X-ray disturbance\,$ = 0$ is the first along the line starting from the top left corner of the plot. The disturbance of the other clusters was calculated as the distance along the same line from the cluster with disturbance\,$ = 0$.
We note that this quantity is similar to the relaxation score, $\mathcal{R}$, calculated by \citet{Zhang22}. Both quantities combine the two morphological parameters we have, $c$ and $w$, into one, which describes the dynamical state of a cluster. The two quantities approximately anti-correlate, that is to say that lower $\mathcal{R}$ is associated with higher values of disturbance.
In our analysis, we chose to use the disturbance, since we remain consistent with the disturbance values for the RHs in the \lotss{} DR2 - PSZ2 sample from Cuciti et al. (in prep.) and \citet{Cuciti2021}.
This disturbance represents the same information as the disturbance calculated by \citet{Cuciti2021}, that is the logarithmic distance from the bisector of the median $c$ and $w$ values from \citet{Cassano2010} ($c$ = 0.2, $w$ = 0.012). The advantage of our method is that it does not rely on the somewhat arbitrary bisector slope and median $c$ and $w$ values.
In Fig.~\ref{fig:Disturb_hist} we plot the cluster disturbance distribution for all clusters above the \planck{} 50\% completeness line in the \lotss{} DR2 - PSZ2 sample with $c$ and $w$ measurements. The disturbance of all clusters is shown in grey, with all clusters hosting a RR in red. The clusters which host both a RR and a RH are shown by hatched black bars.
As seen in the $c$ - $w$ plot, relic-hosting clusters are among the most disturbed in our sample.
There is no obvious difference between the disturbances of relic-hosting clusters which also host a RH and those that don't.

\subsection{Radio relic scaling relations}
\label{sec:res:corrs}

For each of the relations reported in the following subsections, we calculated Spearman's rank correlation coefficient and report its associated p-value. We assessed the presence of a correlation both including and excluding cRRs. PSZ2\,G069.39+68.05 and PSZ2\,G107.10+65.32 were excluded in both cases. The p-values for the correlations in the following subsections are reported in Tab.~\ref{tab:corr_p_values}. We considered the null hypothesis to be rejected if $p < 0.05$, where the null hypothesis we are testing is that the variables $X$ and $Y$ are not correlated.

\begin{table*}
    \centering
    \begin{tabular}{ccc}
        \hline
        Correlation & \multicolumn{2}{c}{p-value} \\
        \\
        & cRRs Excluded & cRRs Included \\
        \hline
        $P_{150\rm{MHz}}$ - $M_{500}$ & 0.003 & 0.003 \\
        $P_{150\rm{MHz}}$ - LLS & 0.261 & 0.029\\
        LLS - $D_{RR-c}$ & 0.002 & -\\
        LLS - $D_{RR-c}/R_{500}$ & < 0.001 & - \\
     \hline
    \end{tabular}
    \caption{Spearman rank correlation coefficient p-values for RR power - cluster mass, RR power - LLS, LLS - distance from cluster centre and LLS - distance from cluster centre as a fraction of $R_{500}$ correlations. The p-value is calculated separately with and without candidate relics included. There is no p-value including cRRs in the LLS - $D_{RR-c}$, nor the LLS - $D_{RR-c}/R_{500}$ correlation, since the distance from the cluster centre is computed using the cluster X-ray centroid. We note that PSZ2\,G069.39+68.05 and PSZ2\,G107.10+65.32 are not included in either sub-sample.}
    \label{tab:corr_p_values}
\end{table*}

If the null hypothesis was rejected, we fit our data, using BCES linear regression methods \citep[][]{Akritas1996}, to the equation $\log_{10}(X) = B\log_{10}(Y) + A$. This was done separately, both including and excluding the cRRs in our sample. We then calculated the 95\% confidence interval of our line of best fit as
\begin{equation}
    \Delta Y = \pm \sqrt{\left[ \sum_{i=0}^N \frac{(Y_i-Y_m)^2}{N-2} \right] \left[ \frac{1}{N} + \frac{(X-X_m)^2}{\Sigma_{i=0}^N (X_i-X_m)^2)} \right]},
    \label{eq:confidence}
\end{equation}
where $Y_m = BX_i + A$ and $X_m = \Sigma_{i=0}^N X_i/N$ for each observed $X_i$.

\subsubsection{Radio power - cluster mass}
\label{sec:res:corrs:P-M}

\begin{figure*}
    \includegraphics[width=17cm]{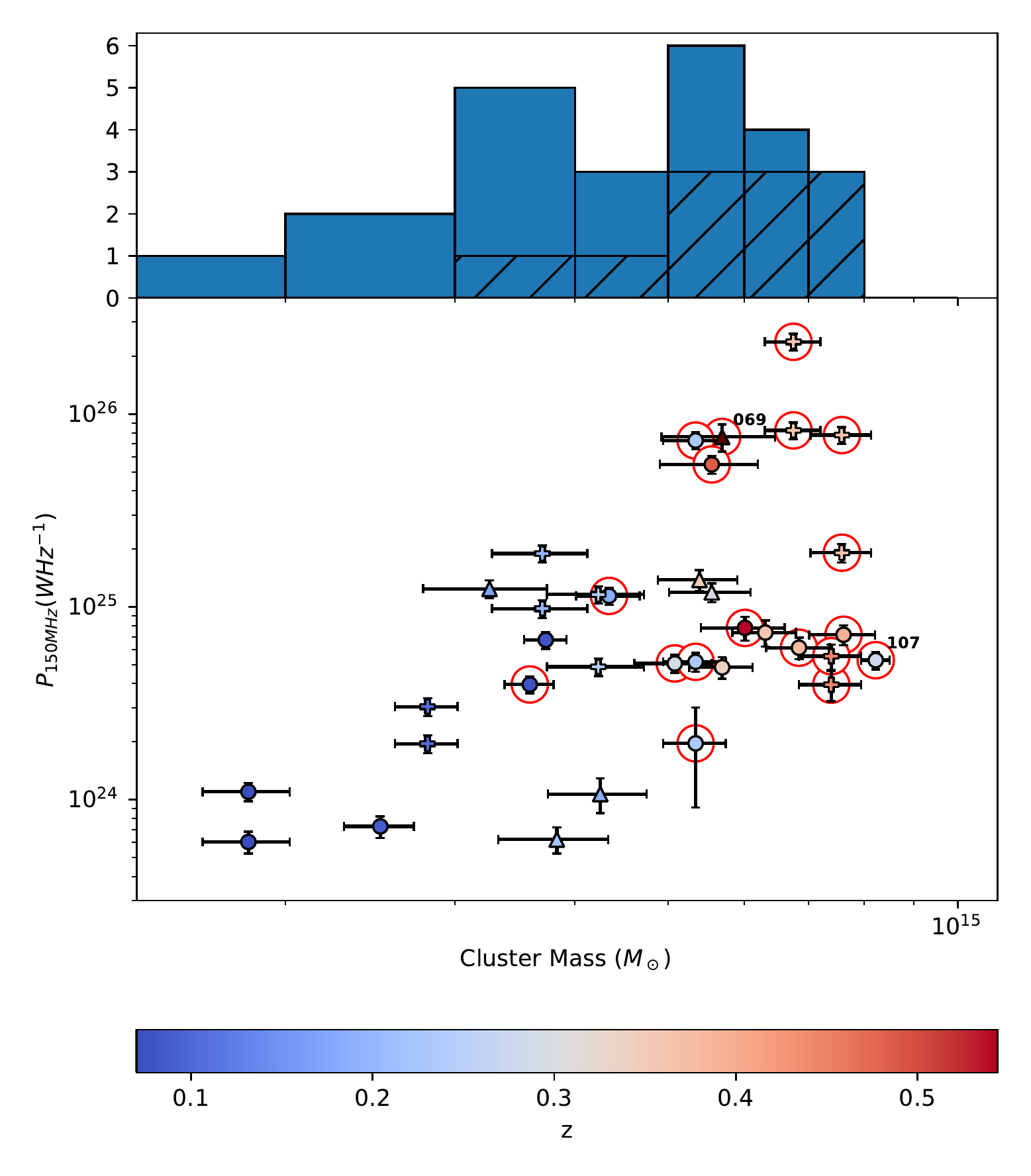}
    \caption{Mass distribution of our RR sample. \textbf{Top}: As a histogram. The hatched bars show the distribution for only clusters which also host a RH. PSZ2\,G069.39+68.05 and PSZ2\,G107.10+65.32 are both excluded. 
    \textbf{Bottom}: Relic power vs. cluster mass with redshift on the colour bar. Triangles denote candidate relics and plusses those relics which are part of a double relic pair. All other relics are plotted as circles. Red circles surround relics in clusters which also host a RH. PSZ2\,G069.39+68.05 and PSZ2\,G107.10+65.32 are labelled, with the labels above and right of the data points.}
    \label{fig:P_M_RH_hist}
\end{figure*}

Fig.~\ref{fig:P_M_RH_hist} shows the 150 MHz radio power of each relic in our sample against the host cluster mass ($M_{500}$, from \planck{}). The cluster redshift is shown on the colour bar. More massive relic-hosting clusters tend to be found at higher redshifts, due to the \planck{} cluster selection function (see Fig.~\ref{fig:M_z}). Low-mass clusters in our sample host only low-power RRs. Relics in more massive clusters span a larger range in radio power, but tend to host more powerful RRs than low-mass clusters.
To quantify the scatter in the power distribution of RRs residing in high-mass ($>5.2 \times 10^{14} M_{\odot}$, where $5.2 \times 10^{14} M_{\odot}$ is the median cluster mass in our sample) and low-mass clusters ($\leq5.2 \times 10^{14} M_{\odot}$), we calculate the coefficient of variance for each sub-sample. The coefficient of variance, defined as $c_{v} = \sigma / \mu $, where $\sigma$ is the standard deviation and $\mu$ the mean, allows us to compare the scatter in two sub-samples with very different mean values. We find that in the high-mass bin, $c_v = 1.5$ and in the low-mass bin $c_v = 0.9$.

We find a positive correlation between the relic power and cluster mass (p = 0.003, both with, and without, cRRs). Tab.~\ref{tab:P_M_fits} shows the best-fit gradient (B) and y-intercept (A) values for the different fitting methods used, for the sample with and without cRRs included.

\begin{table*}
    \centering
    \begin{tabular}{ccccc}
        \hline
        Fit Method & \multicolumn{2}{c}{cRRs Excluded} & \multicolumn{2}{c}{cRRs Included} \\
        \\
        & B & A & B & A \\
        \hline
        Y$\vert$X & $2.24\pm0.44$ & $-7.92\pm6.46$ & $2.30\pm0.45$ & $-8.94\pm6.59$\\
        X$\vert$Y & $5.45\pm1.25$ & $-55.00\pm18.32$ & $6.11\pm1.35$ & $-64.70\pm19.93$\\
        Bisector & $3.22\pm0.14$ & $-22.37\pm2.05$ & $3.40\pm0.15$ & $-25.04\pm2.13$\\
        Orthogonal & $5.19\pm1.20$ & $-51.27\pm17.60$ & $5.84\pm1.32$ & $-60.84\pm19.39$\\
     \hline
    \end{tabular}
    \caption{Radio relic power - cluster mass line of best-fit parameters for different fitting methods. The parameters are calculated separately with and without candidate relics included. We note that PSZ2\,G069.39+68.05 and PSZ2\,G107.10+65.32 are not included in either sub-sample.}
    \label{tab:P_M_fits}
\end{table*}

\begin{figure}
        \resizebox{\hsize}{!}{\includegraphics{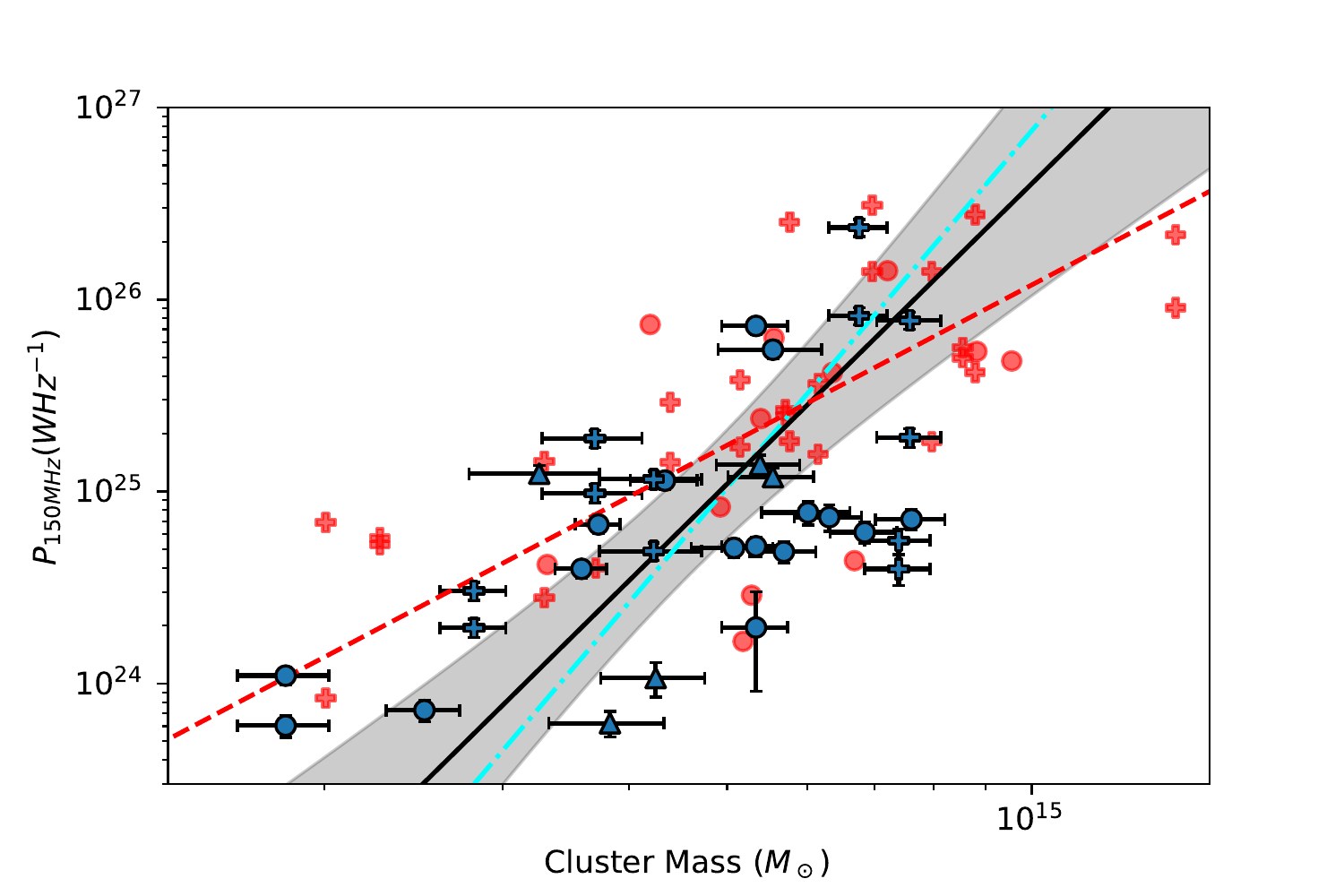}}
        \resizebox{\hsize}{!}{\includegraphics{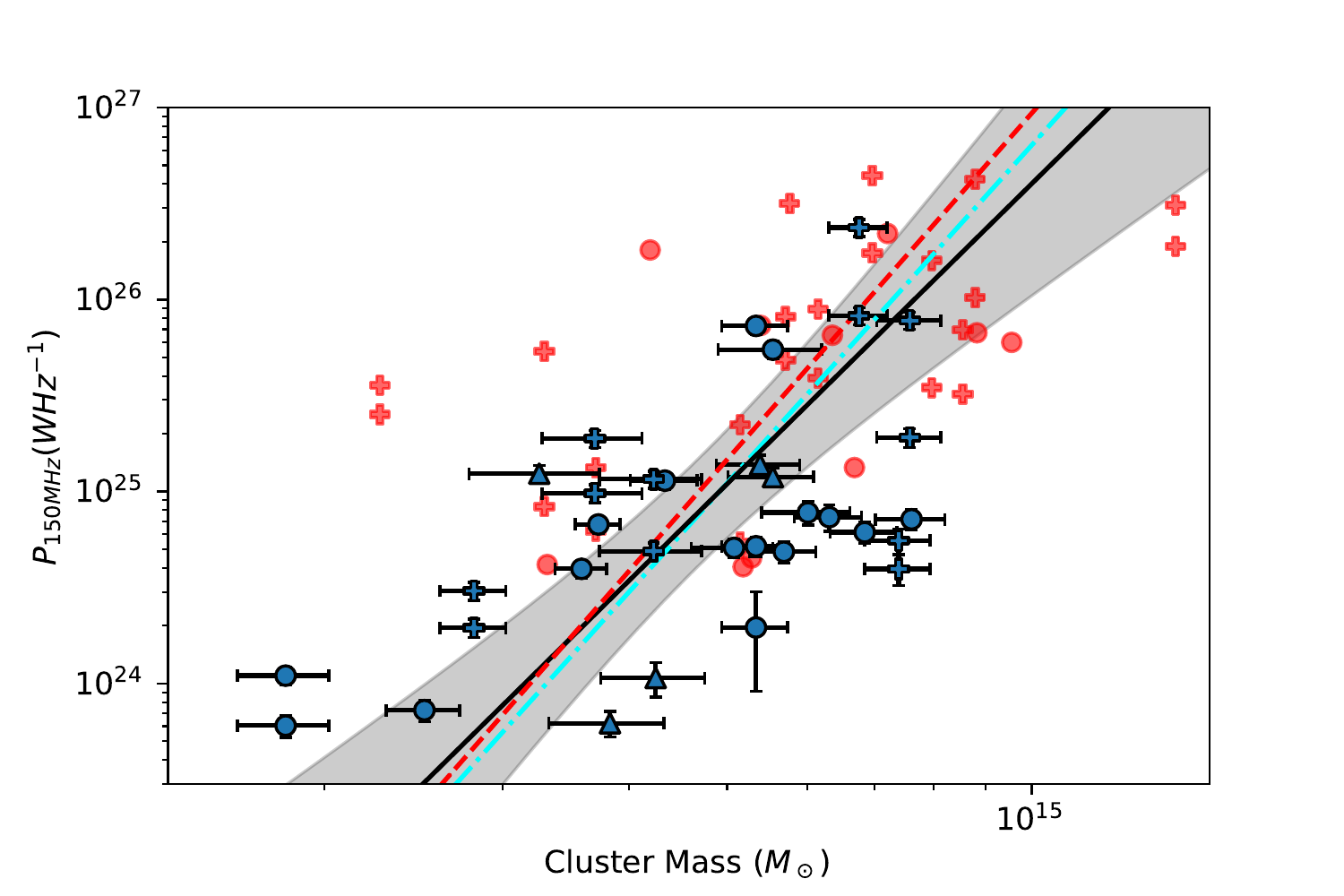}}
        \caption{Relic power vs. cluster mass with regression lines. Triangles denote candidate relics and plusses those relics which are part of a double relic pair. The black (solid) and cyan (dot-dash) lines are the orthogonal least squares regression lines for the relics in the DR2 sample. Black is if no candidate relics are included, with its corresponding confidence interval shaded, and cyan is if they are. Red points are the RRs from FdG14, with their corresponding regression line in red (orthogonal fit). Both PSZ2\,G069.39+68.05 and PSZ2\,G107.10+65.32 are excluded, since they are not used to calculate the regression lines. \textbf{Top}: FdG14 relic powers and regression line scaled to 150 MHz assuming $\alpha=-1$ for all relics. We note that the cosmologies used to calculate powers are slightly different between this sample data and the FdG14 data. See Fig.~\ref{fig:P_M_fdg_cosmo}.
        \textbf{Bottom}: FdG14 powers scaled to 150 MHz using the actual spectral indices of each relic. Relics with no spectral information in FdG14 are excluded. 
        The regression line (orthogonal fit) is recomputed on the scaled data.}
            \label{fig:P_M_fdg}
\end{figure}

In Fig.~\ref{fig:P_M_fdg} we plot the lines of best fit with (cyan dashed line) and without (black solid line) cRRs, from the orthogonal fit. The confidence interval, calculated using Eq.~\ref{eq:confidence}, is shown by the grey shaded region. We choose to plot the orthogonal-fit line, that is to say the line that minimises the orthogonal distances, as this is the same method used by FdG14 to compute their lines of best fit and enables fair comparison of the two relic studies. Their line of best fit is plotted as a dashed red line. For consistency, we plot the orthogonal-fit line for all other correlations in this paper.

The FdG14 sample selection is considerably different to ours. Their sample is comprised of all dRRs known at the time, in addition to the "elongated" relics of \citet{Feretti2012} (41 individual RRs, of which 30 are part of a dRR pair). The dRR-cluster sample of FdG14 contains only two clusters included in our sample, PSZ2 G071.21+28.86 (MACS J1752.0+4440) and PSZ2 G165.46+66.15 (Abell 1240). Additionally, PSZ2 G048.10+57.16 (Abell 2061) is also contained in the sample of \citet{Feretti2012}.
The cosmology used to calculate distances and luminosities (flat $\Lambda$CDM, $H_0 = 71$~\kmsmpc{}, $\Omega_m = 0.27$) and the frequency (1.4 GHz) are both different to our sample. The cluster masses are also the $M_{500}$ values given by \planck{}, though from the PSZ1 catalogue \citep[][]{Planck2014}.
The line, and corresponding red data points, were taken at a frequency of 1.4 GHz, so must be converted to 150 MHz, for comparison with our dataset. We do this using two slightly different approaches, shown in the top and bottom sections of Fig.\ref{fig:P_M_fdg}. The first method (top) is to assume a constant spectral index, $\alpha = -1$, for all relics and scale the full FdG14 dataset and the line of best fit to 150 MHz uniformly.
The advantage of using a constant spectral index to scale the FdG14 sample data is that we can also scale the line of best fit by the same factor, thereby allowing direct comparison of the slopes measured at both frequencies.
We keep the original cosmology of the FdG14 data, since it was used to derive the line of best fit. Though the cosmology used for our dataset is slightly different to that used for FdG14, the results are almost identical (see Appendix~\ref{app:fdg+14 P comparison}). 
Radio relics typically have spectral indices in the range $-1 \lesssim~\alpha~\lesssim -1.5$. Choosing the flattest spectrum in this range allows the closest comparison of the two samples, since even with this flatter spectral index, the RRs in our sample are, on average, less powerful.
In this case, the slope of the line of best fit we obtain does not overlap with that of FdG14 ($B_{orth}=5.19\pm1.20$ vs. $B_{fdg}=2.83\pm0.39$). The second rescaling method (bottom) is to use only the relics with measured spectral index in FdG14 and scale each relic power by its actual spectral index. Since, in this case, the line of best fit reported by FdG14 does not fit the rescaled data, we then recomputed the line of best fit (orthogonal method) ourselves. We obtained best-fit parameters of $B_{fdg}=5.99\pm1.58$ and $A_{fdg}=-62.85\pm23.38$ for the rescaled FdG14 data. We do not include any of our sample in the fitting procedure. In this case, the gradient of the line is within the errors of the line calculated for our sample.
Qualitatively, we see that our dataset contains more low-mass clusters and more low-power relics in relatively high-mass systems, though it does not contain clusters as massive as in FdG14.

\subsubsection{Radio power - LLS}
\label{sec:res:corrs:P-LLS}

\begin{figure}
    \resizebox{\hsize}{!}{\includegraphics{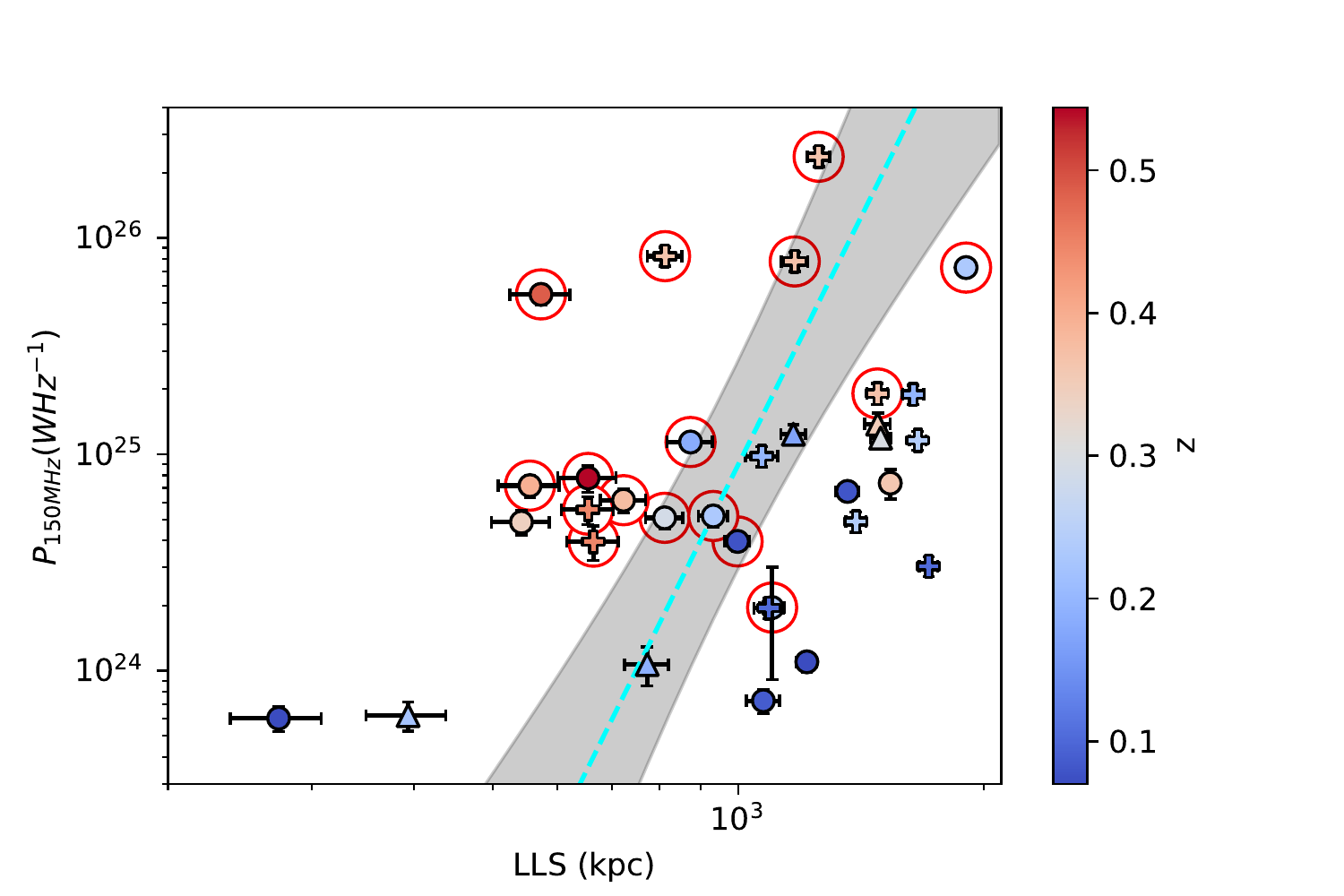}}
    \caption{Radio relic power vs. LLS. The host cluster redshift is on the colour bar. Triangles denote candidate relics and plusses those relics which are part of a double relic pair. All other relics are plotted as circles. Red circles surround relics in clusters which also host a RH. The dotted cyan line is the regression line (orthogonal fit) for our sample, including cRRs, with its corresponding confidence interval shaded. We note that there is no line of best fit excluding cRRs, since the null hypothesis could not be rejected in this case. Both PSZ2\,G069.39+68.05 and PSZ2\,G107.10+65.32 are excluded.}
    \label{fig:P-LLS}
\end{figure}

We did not find a correlation between the power of RRs and their LLS, with cRRs excluded, that is to say we found that the null hypothesis could not be rejected (p = 0.261). However, when we included cRRs, we found that the null hypothesis was rejected (p = 0.029). Tab.~\ref{tab:P_LLS_fits} shows the best-fit gradient (B) and y-intercept (A) values for the different fitting methods used. Since the null hypothesis was only rejected when cRRs were included, we only calculated the best-fit parameters, A and B, for the entire sample.
In Fig.~\ref{fig:P-LLS} we plot the power against the relic LLS. The line of best fit (orthogonal fit) is plotted as a dotted cyan line and its corresponding confidence interval the grey shaded region.

\begin{table*}
    \centering
    \begin{tabular}{ccccc}
        \hline
        Fit Method & B & A \\
        \hline
        Y$\vert$X & $1.23\pm0.46$ & $21.19\pm1.36$ \\
        X$\vert$Y & $8.28\pm4.10$ & $0.112\pm12.32$ \\
        Bisector & $2.36\pm0.19$ & $17.82\pm0.58$ \\
        Orthogonal & $7.60\pm3.80$ & $2.15\pm11.42$\\
     \hline
    \end{tabular}
    \caption{Radio relic power - LLS line of best fit parameters for different fitting methods. The values quoted are only for cRRs included, since the null hypothesis was only rejected with their inclusion. PSZ2\,G069.39+68.05 and PSZ2\,G107.10+65.32 are not included.}
    \label{tab:P_LLS_fits}
\end{table*}

\subsubsection{Longest linear size - cluster centre distance}
\label{sec:res:corrs:LLS-D}

We find that there is a positive correlation between the LLS of a relic and its distance from the cluster centre, $D_{RR-c}$, (p = 0.002) and its distance as a fraction of the cluster $R_{500}$ (p < 0.001), that is to say larger relics are preferentially found further from the cluster centre. Tab.~\ref{tab:LLS_D_fits} shows the best-fit gradient (B) and y-intercept (A) values for the different fitting methods used. The slopes of both correlations are within errors for all fitting methods. We did not compute the Spearman rank correlation coefficient and perform fitting including cRRs, since we do not have any $D_{RR-c}$ or $R_{500}$ measurements (see Sec.~\ref{sec:sample:measurements}). 

\begin{table*}
    \centering
    \begin{tabular}{ccccccc}
        \hline
        Fit Method & \multicolumn{2}{c}{LLS - $D_{RR-c}$} & \multicolumn{2}{c}{LLS - $D_{RR-c}/R_{500}$} \\
        \\
         & B & A & B & A \\
        \hline
        Y$\vert$X & $0.93\pm0.21$ & $0.14\pm0.64$ & $0.89\pm0.21$ & $2.97\pm0.03$ \\
        X$\vert$Y & $1.88\pm0.38$ & $-2.76\pm1.16$ & $1.96\pm0.37$ & $2.95\pm0.04$ \\
        Bisector & $1.30\pm0.13$ & $-1.00\pm0.38$ & $1.29\pm0.16$ & $2.96\pm0.03$ \\
        Orthogonal & $1.49\pm0.27$ & $-1.55\pm0.82$ & $1.50\pm0.35$ & $2.96\pm0.03$ \\
     \hline
    \end{tabular}
    \caption{Radio relic LLS - cluster-centre distance and LLS - cluster-centre distance as a fraction of $R_{500}$ line of best fit parameters for different fitting methods. PSZ2\,G069.39+68.05 and PSZ2\,G107.10+65.32 are not included.}
    \label{tab:LLS_D_fits}
\end{table*}

\begin{figure}
    \resizebox{\hsize}{!}{\includegraphics{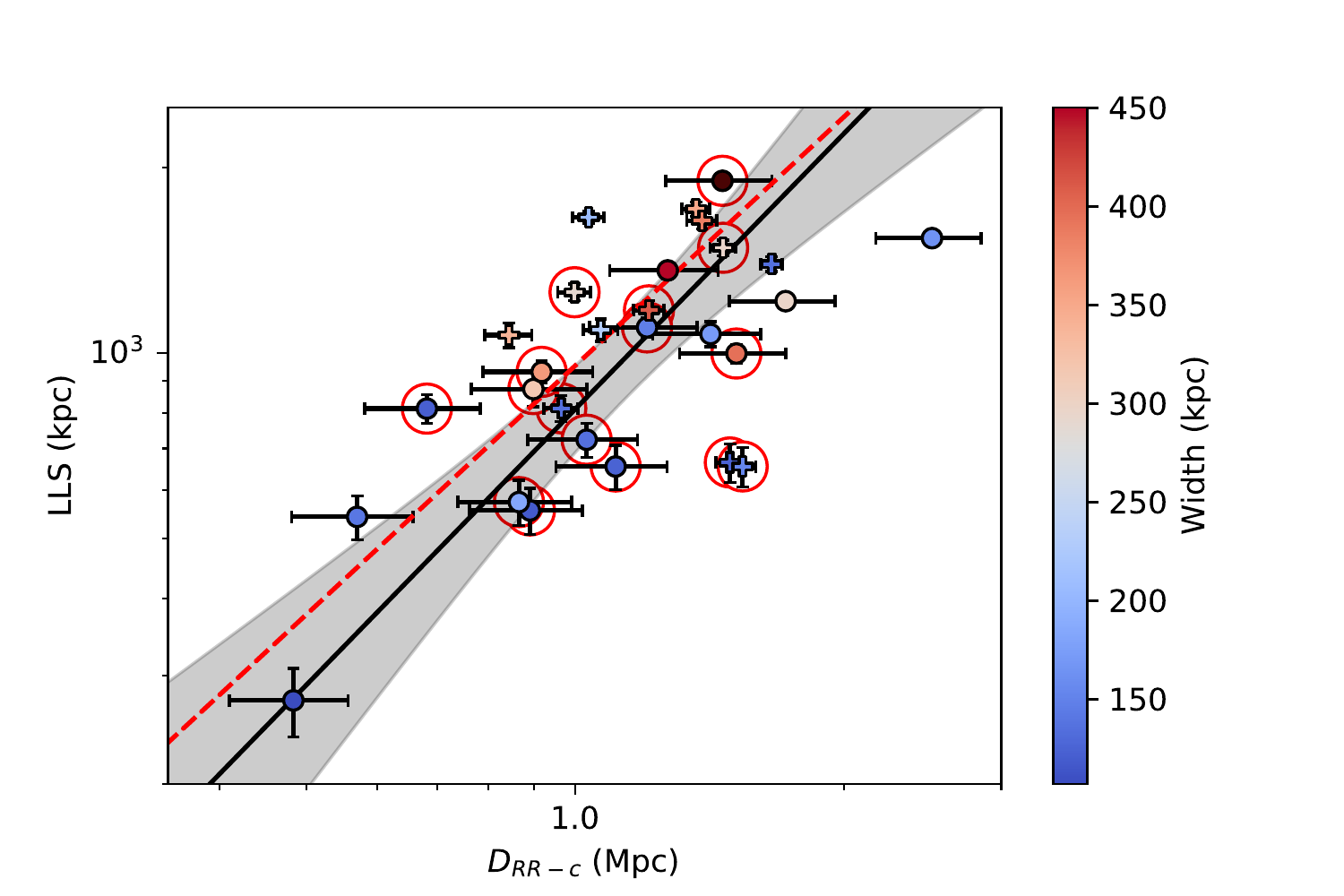}}
    \resizebox{\hsize}{!}{\includegraphics{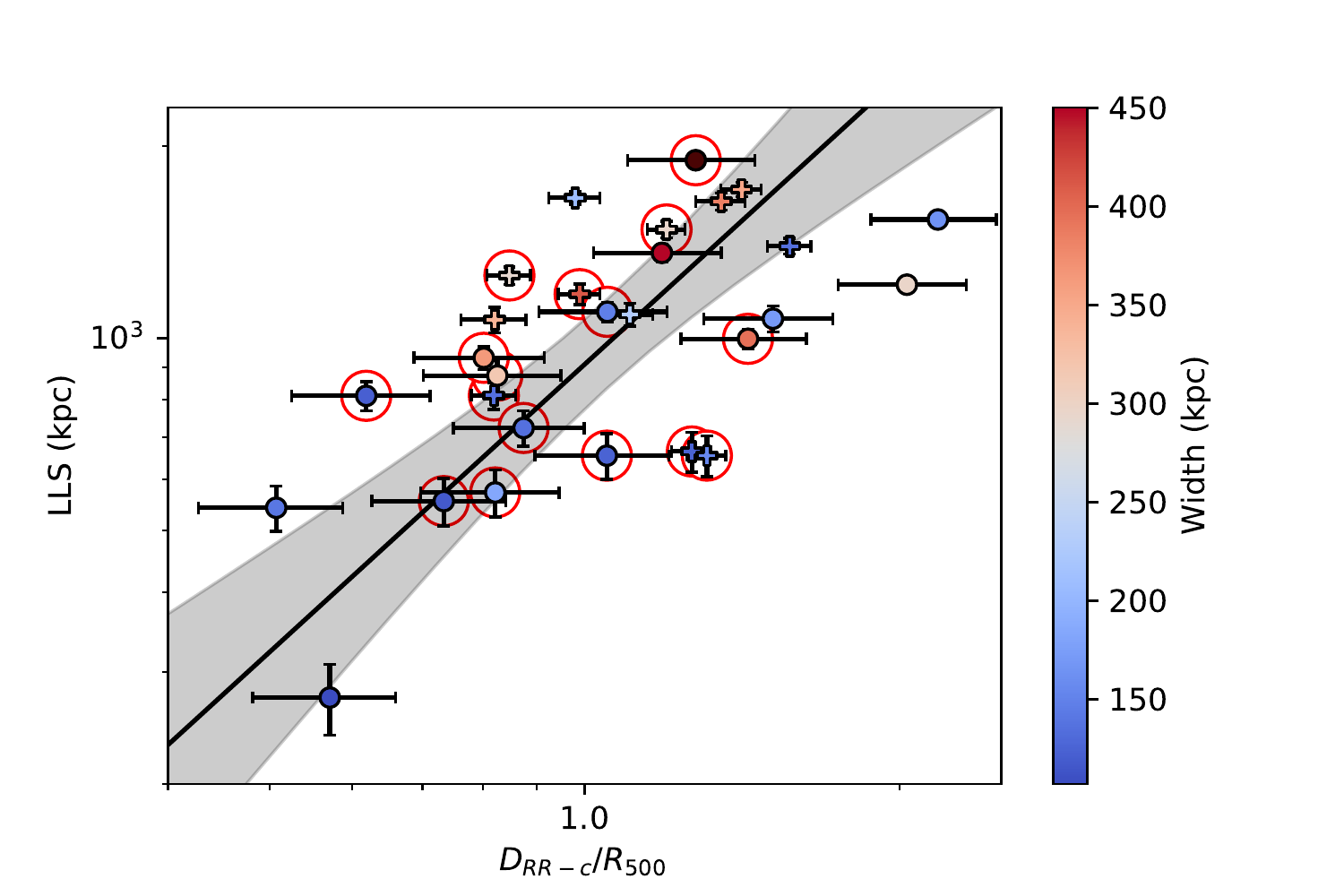}}
    \caption{Radio relic LLS as a function of its projected distance to the cluster centre. \textbf{Top:} LLS vs. cluster centre distance. The relic width is plotted on the colour bar. Plusses denote relics which are part of a double relic pair. All other relics are plotted as circles. Red circles surround relics in clusters which also host a RH. PSZ2\,G107.10+65.32 is excluded. The black line is the regression line (orthogonal fit) for our sample, with its corresponding confidence interval shaded. The dashed red line is the regression line from FdG14. We note that there are no candidates, since the cluster centre is found from X-ray observations
    \textbf{Bottom:} The same for the cluster centre distance as a fraction of the cluster $R_{500}$. There is no corresponding FdG14 correlation against which to compare.}
    \label{fig:LLS_D}
\end{figure}

 In Fig.~\ref{fig:LLS_D} we plot the LLS of our relic sample against their projected distance from the cluster centre and as a fraction of the cluster $R_{500}$. The solid black line shows the orthogonal fit to our data, with its corresponding 95\% confidence interval shown as the grey shaded region. 
 The dashed red line shows the LLS - $D_{RR-c}$ correlation of FdG14 ($\textrm{B}=1.34\pm0.38$ and $\textrm{A}=-1.04\pm1.16$) for comparison. Both the gradients and intercepts of the regression lines for each sample are within the errors of each other. There is no LLS - $D_{RR-c}/R_{500}$ correlation in FdG14 against which to compare.
 The colour bar denotes the median relic width. We see, qualitatively, that larger relics typically have larger widths. It should however be noted that the errors on the width measurements are large, since we assigned the standard deviation of the width distribution measured as the error (see Sec.~\ref{sec:sample:measurements}). For this reason, we did not perform a Spearman rank correlation coefficient test for any relic properties with the downstream width.

\subsection{Relic - cluster centre distance}
\label{sec:res:D_RR-c}

\begin{figure}
    \resizebox{\hsize}{!}{\includegraphics{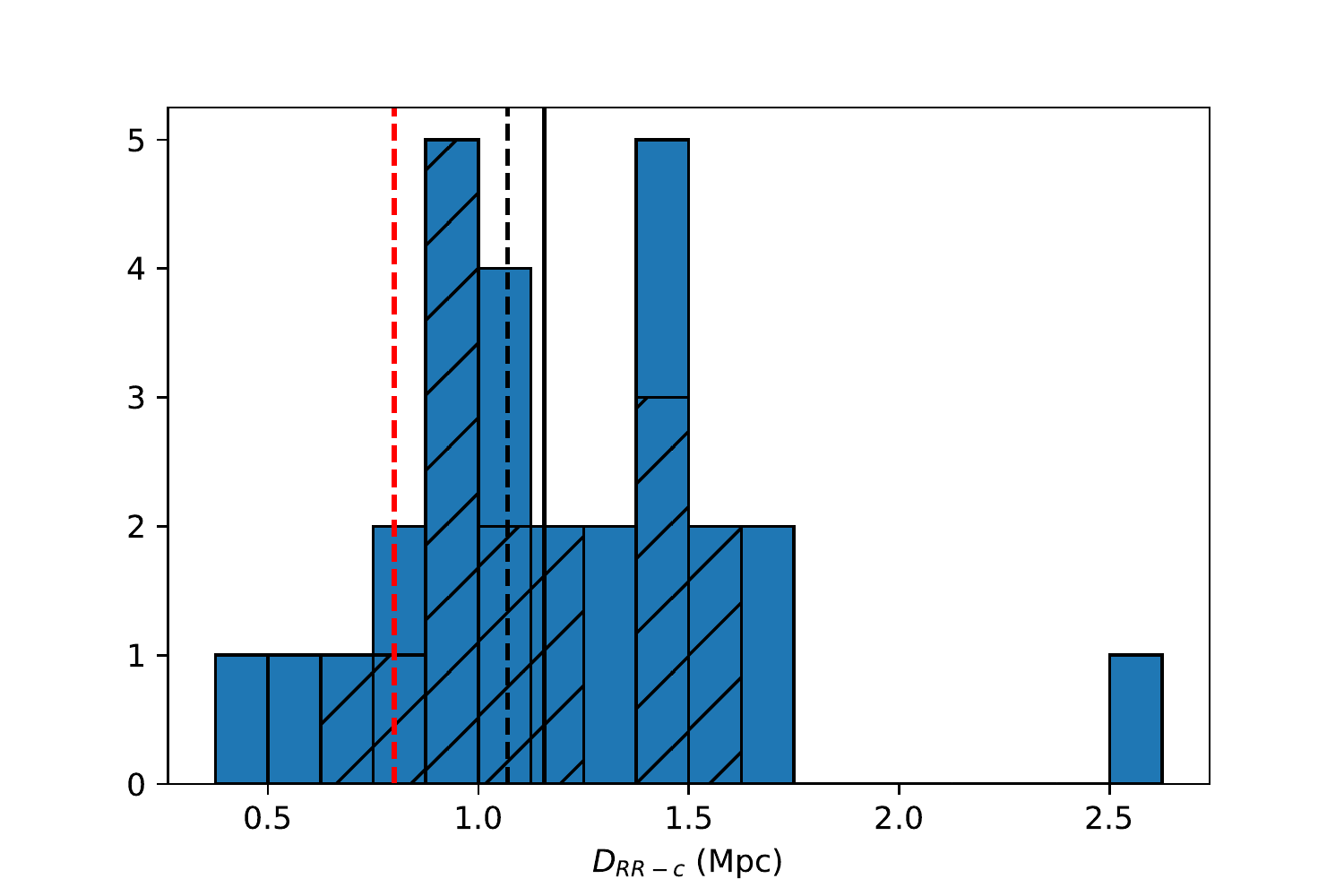}}
    \resizebox{\hsize}{!}{\includegraphics{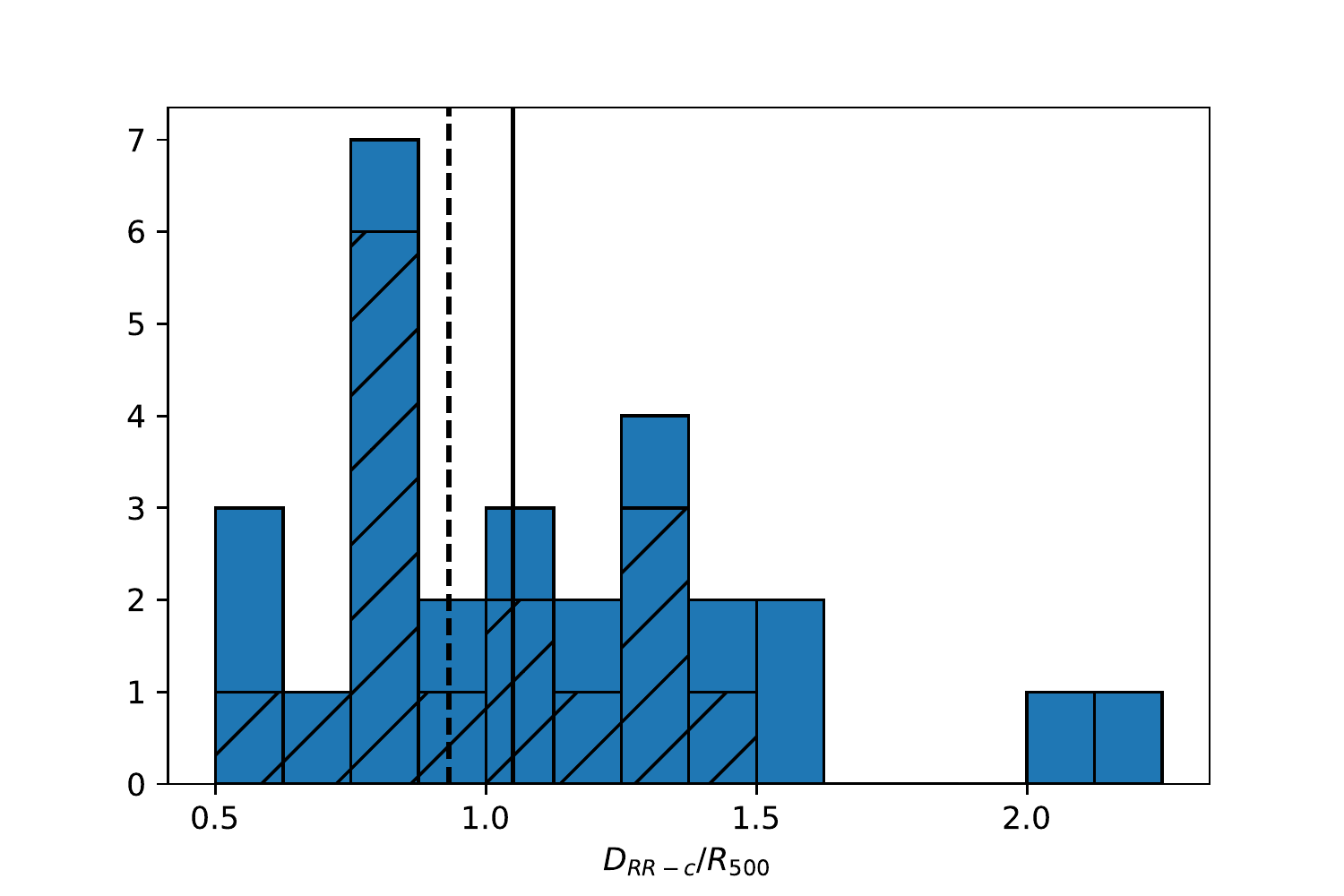}}
    \caption{Distribution of projected RR distances from their host-cluster centres. \textbf{Top}: Histogram of the RR - cluster centre distances, $D_{RR-c}$.  The red dashed line is at 800 kpc.
    \textbf{Bottom}: Histogram of the RR - cluster centre distances as a fraction of the cluster $R_{500}$.
    The hatched bars in both plots show the distribution for only relics in RH-hosting clusters.
    The solid and dashed black lines correspond to the median distances for all relics and just those which also host a RH, respectively.
    We note that PSZ2\,G107.10+65.32 is excluded for both histograms and the median calculations.}
    \label{fig:D_Hists}
\end{figure}

Fig.~\ref{fig:D_Hists} (top) shows the distribution of the projected RR-cluster centre distances, with a dashed (red) reference line at 800 kpc. The bottom panel shows the distribution as a fraction of $R_{500}$. The hatched bars show the distribution only for RH-hosting clusters. The solid and dashed black lines correspond to the median distances for all relics with $D_{RR-c}$ measurements (except PSZ2\,G107.10+65.32) and just those which also host a RH, respectively. The medians are very similar, in both cases and, in general, the distances of relics in RH-hosting clusters follow the distribution of the full sample relatively well.
For simplicity, we exclude PSZ2\,G107.10+65.32 from both plots entirely. PSZ2\,G069.39+68.05 is automatically excluded, since it is a cRR-hosting cluster, and therefore has no $D_{RR-c}$ or $R_{500}$ measurements.
Excluding PSZ2\,G107.10+65.32, $25/28=89 \pm 19 \%$ of the relics in our sample are $>800$ kpc from the cluster centre.
For the sub-sample of RR-hosting clusters which also host a RH, this is $14/15 = 93 \pm 19 \%$. One relic, in PSZ2\,G091.79-27.00, is located 2.5 Mpc from its cluster centre, which is much further than for the other relics in our sample.
$20/28 = 71 \pm 17 \%$ of relics lie within the range $0.75 \leq D_{RR-c}/R_{500} \leq 1.5$. For RH-clusters in our sample, this becomes $14/15 = 93 \pm 19 \%$. The relics  PSZ2\,G089.52+62.34 N2 and PSZ2\,G091.79-27.00 are located $\gtrsim 2R_{500}$ from their cluster centres.

\section{Discussion}
\label{sec:discussion}

\subsection{Occurrence of RRs}
\label{sec:disc:occurrence}

Radio relics are relatively uncommon phenomena. Previous studies have found that $\sim 5\%$ of galaxy clusters host a RR \citep[at 610/235 MHz,][]{Kale2015}.
To calculate the RR occurrence at 150~MHz, we considered only clusters above the \planck{} 50\% completeness line, except PSZ2\,G107.10+65.32, since we do not know the mass of the relic-hosting, S subcluster (see Sec.~\ref{sec:sample:classification}).
273 of the 309 clusters in the full \lotss{} DR2 - PSZ2 sample have adequate image quality to assess the presence of diffuse radio emission \citep[see][for further details]{Botteon2022}.
Of these, 194 lie above the \planck{} 50\% completeness line (excluding PSZ2\,G107.10+65.32). In our sample, 19/194=$10 \pm 6\%$ of galaxy clusters host at least one RR.
If we also include all cRRs, we get that 24/194=$12 \pm 7\%$ of clusters host at least one RR. We note that this is a soft lower limit at the sensitivity of \lotss{} DR2, since there may be relics which are too faint to be detected with our observations. There are 42 clusters in the \lotss{} DR2 - PSZ2 sample which contain diffuse radio emission of uncertain origin \citep[classified as U in][]{Botteon2022}, but no RR or cRR classification. 33 of these lie above the \planck{} 50\% completeness line. It is possible that some of these clusters also host RRs. However, it is unlikely that this would change our results meaningfully, as, from visual inspection, a significant fraction of these appear to have a morphology more similar to RHs.

We assessed the effect of cluster mass on RR occurrence by splitting our sample into two mass bins: high ($> 5.2 \times 10^{14} M_{\odot}$) and low ($\leq 5.2 \times 10^{14} M_{\odot}$), where $5.2 \times 10^{14} M_{\odot}$ is the median mass of our sample.
We find that 10/89=$11 \pm 7$\% of high-mass clusters host at least one RR, which becomes 13/89=$15 \pm 8$\% when cRRs are included. For low-mass clusters, these occurrences are 9/105=$9 \pm 6\%$ and 12/105=$11 \pm 7\%$. The occurrences at high and low mass are within errors, suggesting that there is no dependence of the occurrence on the cluster mass, unlike for RHs \citep[e.g.][]{Cassano2005,Cuciti2015,Cuciti2021}.

Radio relics typically have steep spectra \citep[$-1.0 \gtrsim \alpha \gtrsim -1.5$, e.g. FdG14,][]{Feretti2012}. The increase in relic occurrence at \lofar{} frequencies, as compared to higher frequencies, is therefore unsurprising.
\citet{Nuza2012} used the \textsc{marenostrum universe} cosmological simulation to estimate the expected number of RRs that \lofar{} would discover. By normalising the number of relics above a certain radio flux against the number of RRs observed at that point at 1.4~GHz, they predicted that \lofar{} would discover $\sim$ 2500 new relics, in $\sim$ 50\% of galaxy clusters.
Even including cRRs and the relics in clusters below the \planck{} 50\% completeness line, our sample contains only 35 RRs. If we extrapolate the number of relics detected in \lotss{} DR2 to the entire northern sky, that is the area that will be covered upon the completion of \lotss{}, we expect to observe $109 \pm 58$ RRs in the 835 PSZ2 detections that lie above 0\deg declination \citep[see][]{Botteon2022}.

The absolute number of RRs observed is dependent on the underlying sample of galaxy clusters. The number of RRs predicted by \citet{Nuza2012} was not calculated using PSZ2 galaxy clusters, but rather on the X-ray NORAS+REFLEX sample.
Additionally, \citet{Nuza2012} predicted that $> 50\%$ of the relics detected by \lofar{} would reside in clusters with z > 0.5. In the entire \lotss{} DR2 - PSZ2 sample, only 46 clusters lie at such high redshift, and, of those, only 2 (including PSZ2\,G069.39+68.05) host a RR.
However, whilst not a complete study of all non-PSZ2 clusters covered by the \lotss{} DR2 area, the results of \citet{Hoang2022} do not suggest that we are missing a large number of RRs because of our restriction to only PSZ2 clusters.
Also, the fraction of clusters predicted to host RRs detectable by \lofar{} is much greater than in our sample ($\sim 50\%$ vs. $10\%$). Therefore, unless the fraction of relic-hosting clusters varies significantly between the PSZ2 and NORAS+REFLEX samples, the number of relics able to be detected by \lofar{} is significantly less than predicted by simulations.

One key assumption that governs the number of RRs observable is the efficiency of CRe acceleration by the shock. If the real acceleration efficiency is much lower than that assumed in simulations, the number of observable RRs will be overestimated, since the relic power is dependent on the efficiency \citep[$\textrm{d}P(\nu)/\textrm{d}\nu \propto \xi_{\textrm{e}}$, where $\xi_{\textrm{e}}$ is the fraction of kinetic energy dissipated at the shock,][]{Hoeft2007}.
Using the same cosmological simulation as \citet{Nuza2012}, \citet{Araya-Melo2012} found that an acceleration efficiency of only $\xi_{\textrm{e}} = 0.0005$, that is to say a factor of ten lower than that assumed by \citet{Nuza2012}, was sufficient to reproduce the NRAO VLA Sky Survey (NVSS) RR luminosity function.

An alternative cause of the discrepancy between simulations and the observed number of RRs could be the assumption of shock acceleration of electrons to relativistic energies from the thermal pool (standard DSA). Numerous studies have suggested that re-acceleration of a pre-existing population of relativistic electrons is required to produce the observed brightness of RRs \citep[e.g.][]{Kang2011,Kang2012,Botteon2020Shock}.
Such a scenario relies on the existence of populations of mildly energetic electrons in the ICM available to be re-accelerated. The tails of radio galaxies could provide such a population, as has been suggested for Abell 3411-3412 \citep[][]{VanWeeren2017}.
Another potential source is fossil plasma energised by ICM motions \citep[e.g.][]{DeGasperin2017,Mandal2020}. Their ultra-steep spectra make such sources challenging to observe, so it is unclear how ubiquitous such populations are.
If re-acceleration of relativistic electrons, instead of acceleration from the thermal pool, is required in some or all cases, many fewer RRs will be observed than predicted by simulations assuming standard DSA. This is because, in this scenario, only those shocks which cross a population of relativistic electrons will produce a RR.

\subsection{Radio power of relics}
\label{sec:disc:power}

In their statistical study of RRs, FdG14 found a positive correlation between the radio power of relics at 1.4 GHz and the mass of the host cluster, similar to that found for giant RHs \citep[e.g.][]{Basu2012,Cassano2013,Cuciti2021}. The physical explanation is that RRs are driven by shock waves caused by galaxy cluster mergers, for which the energy budget is set by the total mass of the merging clusters.
The total energy released by a merger between two clusters of mass $M$ and virial radius $R_{vir} \propto M^{1/3}$ is $E \propto M^2/R_{vir}$. Assuming that the kinetic energy dissipated at the shock is a fixed fraction of the total energy, and that it scales with the RR power, $P$, the power should scale like $P \propto E/t_{cross}$, where $t_{cross}$ is the sound crossing time of the cluster and is related to the sound speed at the shock through $t_{cross} = R_{vir}/c_s$. X-ray studies show that cluster temperature, $T$, scales with $M^{2/3}$ \citep[e.g.][]{Pratt2009,Lovisari2020} and, since it also scales with $c_{s}^{2}$, we would expect the radio power of relics to be related to the cluster mass by $P \propto M^{5/3}$.

For the first time, we show that the $P$-$M$ correlation extends to frequencies below 200 MHz (see Sec.~\ref{sec:res:corrs:P-M}). However, we find that the relation is much steeper than that predicted by our simple estimate based on the total energy budget of the infalling mass, independent of the fitting method used (see Tab.~\ref{tab:P_M_fits}). This could suggest that the assumption that a constant fraction of the total energy is dissipated at the shocks does not hold. Alternatively, it could suggest a dependence of the shock magnetic field or particle acceleration efficiency on the cluster mass.
Additionally, observational bias may contribute to the measured slope of the $P$-$M$ relation. We assess this effect in more detail in Sec.~\ref{sec:disc:power:low}. 

We also find a different mass dependency than that found by FdG14, who found that $P_{1.4\rm{GHz}} \propto M^{2.8\pm 0.4}$, whereas we find $P_{150\rm{MHz}} \propto M^{5.2\pm 1.2}$ using the same fitting procedure (orthogonal least squares). Fig.~\ref{fig:P_M_fdg} (top) shows the two correlations plotted on top of each other, with all data points and the correlation scaled to 150 MHz using a constant spectral index of $\alpha = -1$. Radio relics typically have spectral indices in the range $-1.0 > \alpha > -1.5$ \citep[e.g. FdG14,][]{Feretti2012}. Even using the flattest spectrum within this range, we see that the majority of the relics in our sample lie below the correlation of FdG14. This is unsurprising, since RRs have relatively steep spectra. This, combined with the increased sensitivity of \lofar{} and low operating frequencies, makes detection of low-power RRs easier. 
For a large number of relics in the FdG14 sample, spectral information is available. We therefore took all relics for which this is the case, calculated the expected power at 150 MHz and recomputed the line of best fit on our re-scaled data (orthogonal fit). We note that this approach assumes no spectral curvature. The presence of which would cause the radio power at 150 MHz to be overestimated. The most detailed study of the integrated spectrum of a relic to-date however shows that the spectrum is straight \citep[e.g.][]{Rajpurohit2020Spectrum}. The re-scaled data, and its corresponding line of best fit, are plotted in Fig.~\ref{fig:P_M_fdg} (bottom), along with the same line of best fit for the \lotss{} DR2 - PSZ2 sample as before. We find that the lines of best fit are now remarkably similar, despite the greater number of low-power relics in our sample. These results suggest that taking the spectral index of each relic into account can entirely compensate for the discrepancy between the mass-dependence of each line of best fit.

\subsubsection{Low-power RRs}
\label{sec:disc:power:low}

\begin{figure}
    \resizebox{\hsize}{!}{\includegraphics{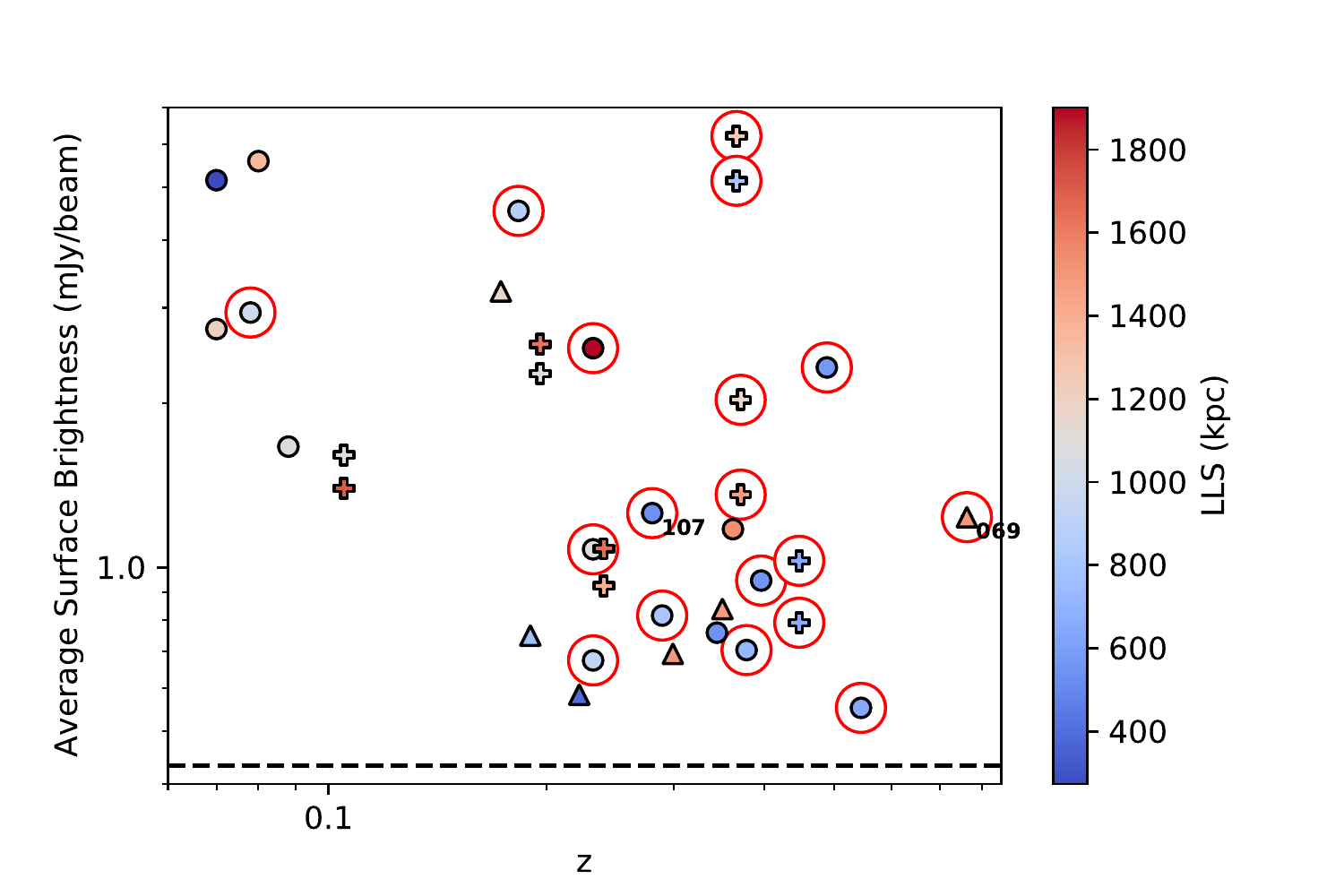}}
    \caption{Average relic surface brightness vs. redshift. The colour of the points shows the relic LLS. Triangles denote candidate relics and plusses those relics which are part of a double relic pair. All other relics are plotted as circles. Red circles surround relics in clusters which also host a RH. PSZ2\,G069.39+68.05 and PSZ2\,G107.10+65.32 are labelled, with the labels below and right of the data points. The dotted black line is at $2\sigma_{\rm{50 kpc}}$, where $\sigma_{\rm{50 kpc}}$ = 0.216 $\rm{mJybeam^{-1}}$ is the average rms noise of the 50 kpc-tapered images in our sample.}
    \label{fig:SB_z}
\end{figure}

If there exists a population of lower-power relics, especially at high redshifts, where most massive clusters are found, we may be unable to detect them. The power-mass correlation we measure would, in this case, be biased towards higher-power relics, since we only observe low-power RRs in nearby, low-mass clusters.
Unlike for RHs, we do not have a robust method for determining upper limits for RRs in clusters defined as non-RR-hosting \citep[e.g. ][]{Bonafede2017,Bruno22}. Instead, we must rely on simpler estimates from the typical noise of our observations.
Fig.~\ref{fig:SB_z} shows the surface brightness averaged across each relic as a function of cluster redshift. The black dashed line shows the approximate detection limit of our observations, given by two times the average rms noise of the 50 kpc-tapered images in our sample ($\sigma_{\rm{50 kpc}}$). Many of the higher-redshift relics lie just above the estimated detection limit.

\begin{figure}
    \resizebox{\hsize}{!}{\includegraphics{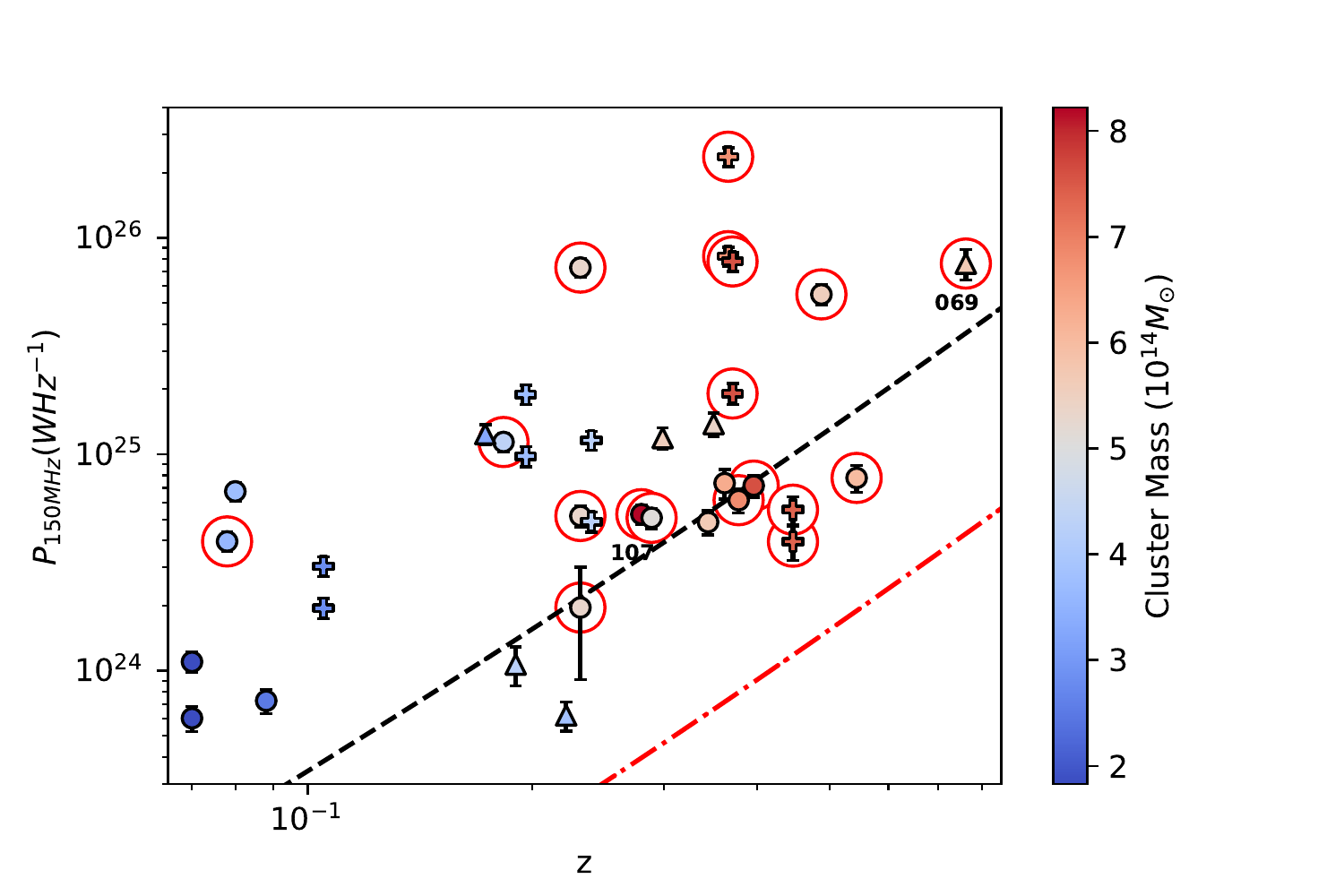}}
    \caption{Relic power vs. cluster redshift. The colour of the points shows the cluster mass. Plusses denote relics which are part of a double relic pair. All other relics are plotted as circles. Red circles surround relics in clusters which also host a RH. The black (dashed) and red (dot-dashed) lines show the estimated detection limit for the median ($\sim 1100 \textrm{kpc} \times 200 \textrm{kpc}$) and minimum ($\sim 300 \textrm{kpc} \times 100 \textrm{kpc}$) relic LLSs and widths of our sample, respectively. PSZ2\,G069.39+68.05 and PSZ2\,G107.10+65.32 are labelled, with the labels directly below the data points.}
    \label{fig:P_z}
\end{figure}

We also assess this bias by estimating the radio power required for a RR to be observable, as a function of redshift. To do this we calculated the power of a box with dimensions of the minimum ($\sim 300~\textrm{kpc} \times 100~\textrm{kpc}$) relic LLS and widths of our sample and average surface brightness equal to $2\sigma_{\rm{50 kpc}}$ (as above), for a 50 kpc beam. This is shown as a red dot-dashed line in Fig.~\ref{fig:P_z}. Since we kept the beam at a fixed physical size, independent of redshift, the relic flux is constant. The least powerful, high-z relics in our sample lie close to the $300~\textrm{kpc} \times 100~\textrm{kpc}$ line. Since the smallest relic (PSZ2\,G089.52+62.34 N1, both smallest LLS and width) is considerably smaller than the other relics in our sample (see Fig.~\ref{fig:LLS_D}), we also calculated the same sensitivity limit for the average relic. The box dimensions were, in this case, the median LLS and width values of our sample ($\sim 1100~\textrm{kpc}$ and $200~\textrm{kpc}$). This is plotted as a dashed black line in Fig.~\ref{fig:P_z}. A few relics lie between the two sensitivity lines, but most lie above the median line.
As in FdG14, we find that the faintest relics we observe are at the detection limit of our observations. This implies that the missing low-power relic population might just be a selection effect.

Additionally, the least powerful relics are found only in nearby clusters. Of the 7 relics with radio powers $< 3 \times 10^{24}~\textrm{WHz}^{-1}$ (PSZ2\,G080.16+57.65, PSZ2\,G086.58+73.11, PSZ2\,G089.52+62.34 N2, PSZ2\,G089.52+62.34 N1, PSZ2\,G099.48+55.60 S, PSZ2\,G144.99-24.64, PSZ2\,G166.62+42.13 E), all are located at $\textrm{z} < 0.25$.
This is unsurprising, since, for a given power, the flux observed decreases with distance. 
However, since in our sample more massive clusters are generally located at higher redshifts (see Fig.~\ref{fig:M_z}), this effectively places an approximate lower limit on the power detectable with \lofar{} at a given cluster mass.
If we take the approximate sensitivity limit calculated for an average relic in our sample (Fig.~\ref{fig:P_z}, black dashed line), we would expect that a relic with radio power $ 3 \times 10^{24}~\textrm{WHz}^{-1}$ would be too faint to be observable in \lotss{} DR2 if it were located $z \gtrsim 0.27$. Comparing the masses of the clusters in our sample located above and below this redshift, we find that the median mass above ($6.2 \times 10^{14}~M_{\odot}$, from 14 clusters) is much greater than below ($3.7 \times 10^{14}~M_{\odot}$, from 12 clusters).
This implies that if such a population of low-power RRs also exists at high redshift, and therefore mass, we would be unable to detect them.

Studies focussing on the properties of large samples of simulated RRs \citep[e.g.][]{Nuza2017,Bruggen2020} suggest that there is a large number of low-power RRs, especially in low-mass clusters. The low-mass systems in our sample are among the least massive clusters known to host RRs, with only the relics in Abell 168 being hosted by a lower-mass cluster \citep[$M_{500} = 1.2 \times 10^{14}~M_{\odot}$ from the X-ray mass - luminosity correlation,][]{Piffaretti2011,Dwarakanath2018} than PSZ2 G089.52+62.34 ($M_{500} = 1.8 \times 10^{14}~M_{\odot}$) in our sample. \lofar{} has allowed for the discovery of relics in low-mass clusters, such as in PSZ2 G145.92-12.53 \citep[$M_{500} = 1.9 \times 10^{14}~M_{\odot}$ from PSZ2,][]{Botteon2021}. In fact, of the 11 clusters $< 5 \times 10^{14}~M_{\odot}$ in our sample, the RRs in nine of them were discovered for the first time with \lofar{}, or in combination with another instrument. The low-mass end of our sample is nonetheless sparsely populated. 
Our results suggest that we could well be missing a number of low-power RRs, especially in low-mass clusters.
This would appear to weaken the finding of a correlation between cluster mass and the power of a RR and may suggest that, in contrast to RHs, the merging mass is not a direct driver of the relic power, as suggested by \citet{Nuza2017}. However, it should be noted that the discrepancy between observed and simulated relic counts (see Sec.~\ref{sec:disc:occurrence}) cast doubt on the existence of a large population of undetected relics and suggest that cosmological simulations are unable to fully produce the observed RR population properties.
Indeed, this would support the idea that the cluster mass sets an approximate upper limit on the power of a relic, but additional factors drive the differences in observed brightness. For example, the particle acceleration efficiency is a function of the underlying shock Mach number \citep[][]{Hoeft2007} and could therefore drive differences in RR power. Additionally, if relics are produced by re-acceleration of mildly relativistic fossil electrons, the properties of the electron population would also affect the power of the relic.

\subsubsection{On the scatter in the power-mass correlation.}
\label{sec:disc:power:scatter}
 
\citet{Cuciti2021} found that the scatter in the $P_{150 MHz} - M_{500}$ for giant RHs can be, at least in part, explained by the morphological cluster disturbance. Radio halos that lie above the correlation tend to be found in more disturbed clusters. We investigate whether this can also explain the scatter for the $P_{150 \rm{MHz}} - M_{500}$ correlation for relics.
We calculated the error in the distance from the correlation with bootstrapping methods. We assigned the cluster mass and radio power of each relic a random number drawn from a Gaussian distribution with mean of the measured value and standard deviation the corresponding errors. We then re-fitted the data with an orthogonal fit and measured the distance of each point from the new correlation. This was repeated 1000 times and we use the standard deviation of the distance measurements as the random error. The total error is from a combination of this error and the power error of each relic.

\begin{figure}
    \resizebox{\hsize}{!}{\includegraphics{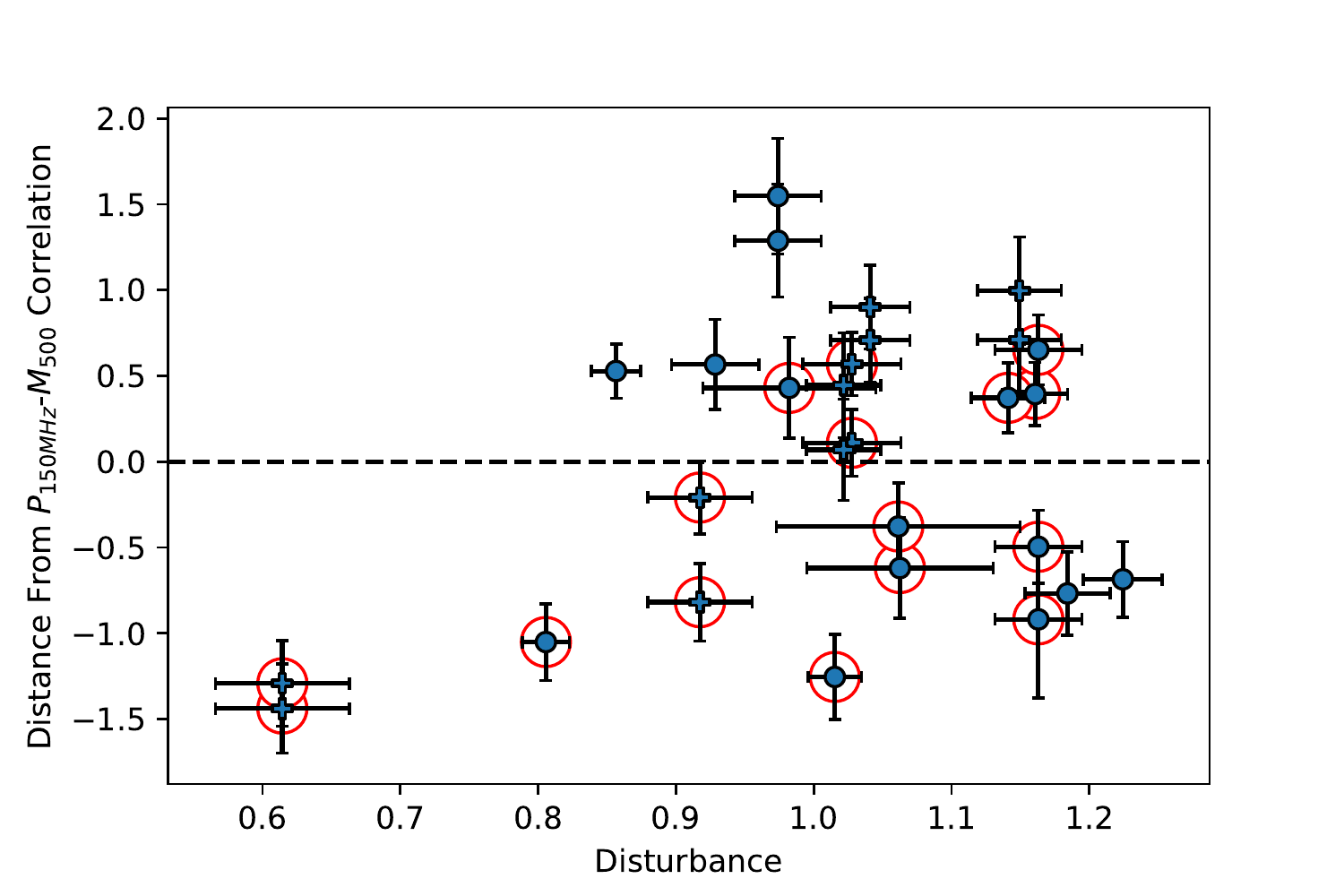}}
    \resizebox{\hsize}{!}{\includegraphics{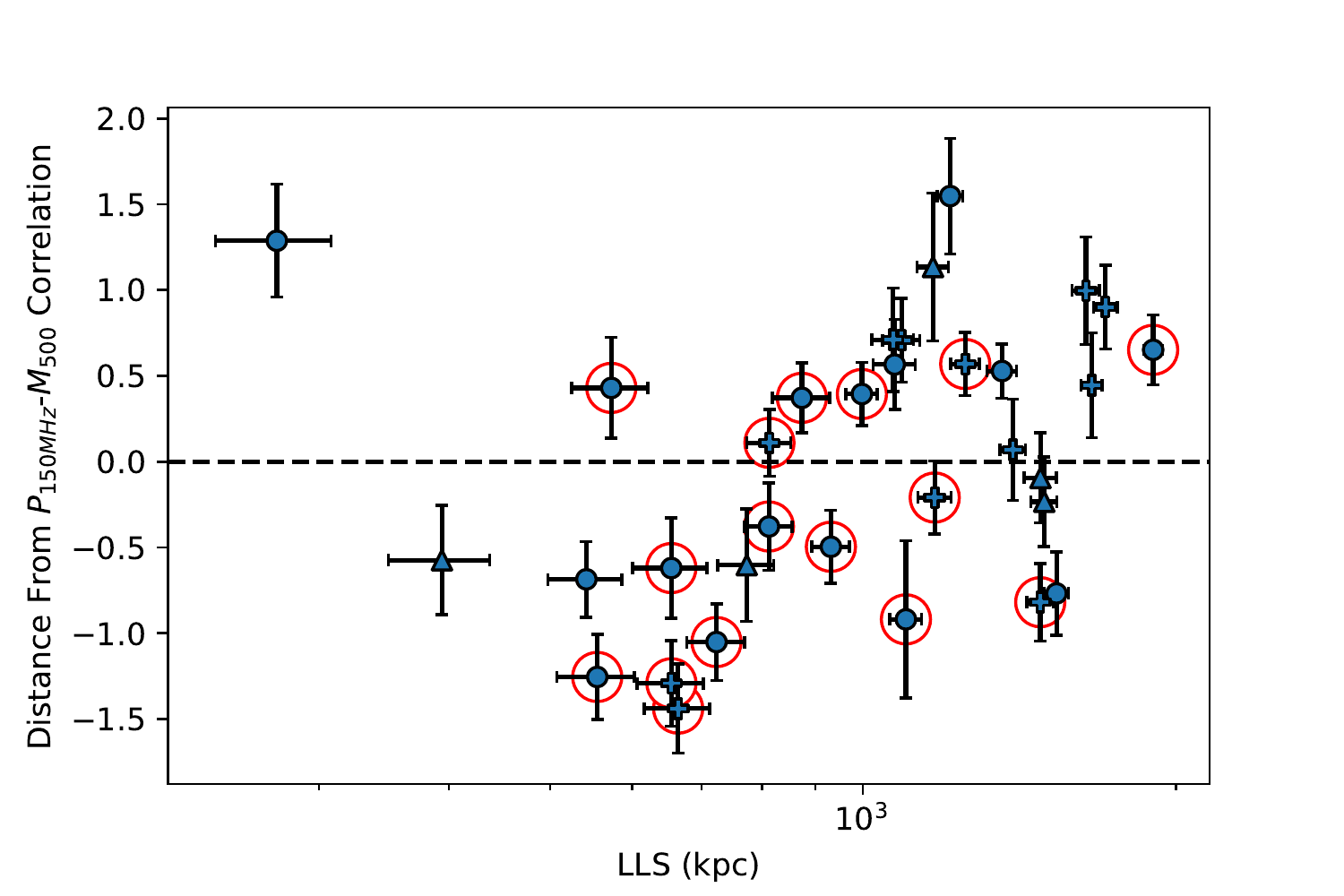}}
    \caption{Logarithmic distance from P-M correlation (along $P_{150 \rm{MHz}}$ axis, Fig.~\ref{fig:P_M_fdg}) vs. LLS (top), disturbance (bottom). The distance is calculated from the orthogonal fit on all relics (except PSZ2\,G069.39+68.05, PSZ2\,G107.10+65.32; see Tab.~\ref{tab:P_M_fits}). PSZ2\,G069.39+68.05, PSZ2\,G107.10+65.32 are excluded from the plots.}
    \label{fig:Dist_P_M}
\end{figure}

In Fig.~\ref{fig:Dist_P_M} (top) we plot the logarithmic distance, along the $P_{150 \rm{MHz}}$ axis, of a RR from the $P_{150 \rm{MHz}} - M_{500}$ correlation for all relics (orthogonal fit) against the cluster disturbance (see Sec.~\ref{sec:res:c-w}). As the lines of best fit including, and excluding, cRRs (see Tab.~\ref{tab:P_M_fits}) are so similar, we use the line of best fit including cRRs. We do not find a correlation between the disturbance and the distance from the correlation.
In Fig.~\ref{fig:Dist_P_M} (bottom) we plot the distance from the same correlation against the relic LLS. Smaller relics tend to lie below the $P_{150 \rm{MHz}} - M_{500}$ correlation and larger relics above. This finding is expected, since for a given surface brightness, a larger relic should be more powerful. The correlation found between the RR power and the LLS supports this (see Sec.~\ref{sec:res:corrs:P-LLS}), although this correlation is only found when cRRs are included.
The smallest relic (PSZ2\,G089.52+62.34 N1) is an outlier, in that it is a very small relic ($\sim$300 kpc) which lies well above the correlation. This relic is connected, at least in projection, to an AGN \citep[see][for more details]{VanWeeren2021}, which could have had an effect on the relic's brightness. Such an effect has been observed in other relics, such as PSZ2\,G096.88+24.18 \citep[][]{Jones2021}. However, PSZ2\,G089.52+62.34 is also a low-mass cluster, where we observe very few relics and therefore the correlation is poorly constrained.

\subsection{Downstream relic width}
\label{sec:disc:width}

The downstream width of a RR is set by the radiative lifetime of the CRes producing the relic. In the case of DSA, electrons at a shock front in the ICM are (re-)accelerated to relativistic energies. However, as the shock propagates through the ICM, in the absence of another acceleration mechanism, the electrons left behind the shock lose energy through inverse Compton (IC) and synchrotron emission until they become too faint to observe. 
For some relics, there is evidence that the downstream width is too large to be fully explained in this scenario. By estimating the expected width of a relic, given synchrotron and IC losses, \citet{Kang2017} found a discrepancy of a factor of $\sim 2$ in the Toothbrush relic at 610 MHz. The effect is particularly pronounced at low radio frequencies however. Performing similar estimations, \citet{DeGasperin2020} found that the width of the relic was $\sim4$ times larger than predicted at 58 MHz. A similar discrepancy was found for the Sausage relic \citep[][]{Kang2016} and in Abell 3667 \citep[][]{DeGasperin2022}.

\begin{figure*}
    \includegraphics[width=17cm]{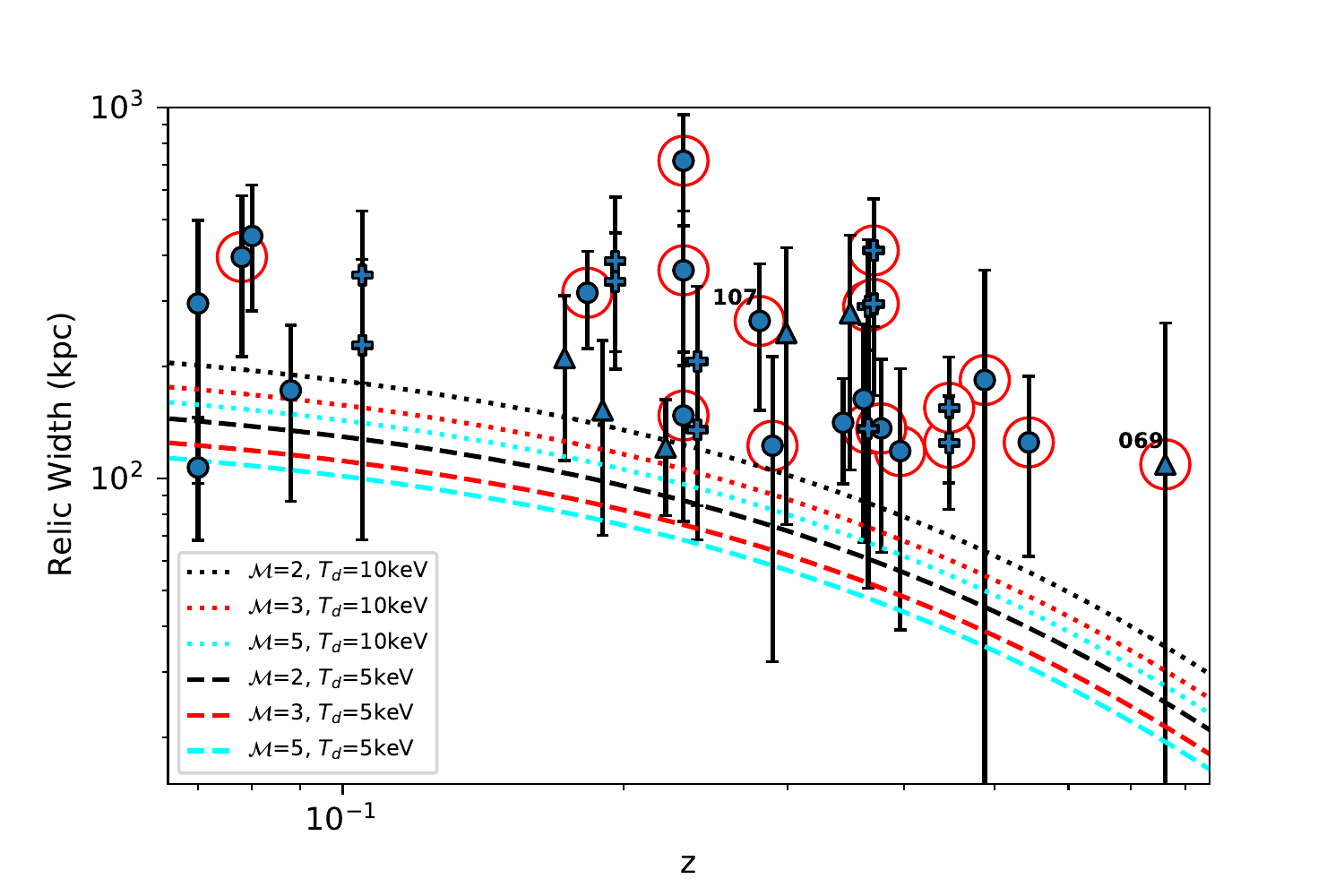}
    \caption{Median relic width vs. redshift. The error bars correspond to the standard deviation of the relic widths measured. Triangles denote candidate relics and plusses those relics which are part of a double relic pair. All other relics are plotted as circles. Red circles surround relics in clusters which also host a RH. The dashed and dotted lines show the largest expected relic width from \citet{Kang2017} for different shock properties. PSZ2\,G069.39+68.05 and PSZ2\,G107.10+65.32 are labelled, with the labels above and left of the data points.}
    \label{fig:Width_z}
\end{figure*}

For the first time, we systematically measure the downstream widths of relics, by calculating the width at many points along the relic to produce a distribution of widths (see Sec.~\ref{sec:sample:measurements}). Fig.~\ref{fig:Width_z} shows the medians of the measured relic width distributions as a function of redshift. The error bars correspond to the standard deviation of the width distributions. To compare our width measurements against theoretical expectations, we use \citet{Kang2017} (Eq. 1), who estimated the characteristic downstream width behind a spherical shock at a given frequency, due to IC and synchrotron losses. The width is then given by
\begin{equation}
    \Delta l_{\nu} \approx 120~\textrm{kpc} \left( \dfrac{u_{down}}{10^{3}~\textrm{kms}^{-1}} \right) \cdot Q \cdot \left[ \dfrac{\nu_{obs}(1 + \textrm{z})}{0.61~\textrm{GHz}} \right]^{\frac{1}{2}},
\end{equation}
where $u_{down}$ is the downstream shock speed, $\nu_{obs}$ is the observing frequency, and $Q$ depends on the downstream magnetic field, $B_{down}$, as
\begin{equation}
    Q \equiv \left[ \dfrac{\left( 5~\mu \textrm{G}\right)^{2}}{B_{down}^{2} + B_{CMB}^{2}}\right] \left( \dfrac{B_{down}}{5~\mu \textrm{G}}\right)^{\frac{1}{2}},
\end{equation}
where $B_{CMB} = 3.24(1+z)^2~\mu \rm{G}$ is the equivalent magnetic field strength of the cosmic microwave background (CMB).
Since we do not have information on the downstream magnetic field strength, we assumed the magnetic field strength which minimises radiative losses, that is $B_{down}=B_{CMB}/\sqrt{3}$. The theoretical widths we calculated therefore correspond to the \textit{maximum} expected width.
The expected width is also strongly dependent on the downstream flow speed, $u_{down}$, which depends on the shock Mach number, $\mathcal{M}$, and upstream sound speed, $c_{s,up}$. $c_{s,up}$ is then related to the downstream sound speed, $c_{s,down}$, by $c_{s,down}/c_{s,up} = \sqrt{\dfrac{(5\mathcal{M}^2-1)(\mathcal{M}^2+3)}{16\mathcal{M}^2}}$.
We estimated $u_{down}$, using $u_{down} = c_{s,up} \frac{\mathcal{M}^2+3}{4\mathcal{M}}$, and $c_{s,down} \approx 1480~\textrm{kms}^{-1}\left(\frac{T_{d}}{10^8\textrm{K}}\right)^{1/2}$.
By assuming a number of reasonable $\mathcal{M}$, and downstream temperatures, $T_d$, we could therefore estimate the maximum expected relic width. The resulting expected widths are plotted in Fig.~\ref{fig:Width_z}. The maximum downstream speed reached with these parameters is $\sim 10^3~\textrm{kms}^{-1}$. Only PSZ2\,G089.52+62.34 N1, the smallest relic in our sample, lies completely below the uppermost line. The widest relic (PSZ2\,G166.62+42.13 W, median width $\sim 700$ kpc) would require $u_{down} \sim 5000~\textrm{kms}^{-1}$ to agree with the prediction.

Clearly, the width of RRs is greater than expectations in almost all cases, for our sample. In most cases, even optimistic assumptions of the downstream temperature and Mach number are insufficient to explain the median values of our calculated width distributions. One possible solution to this is that downstream turbulence further accelerates the relic-producing CRes, extending their radiative lifetime \citep[e.g.][]{Fujita2016}. Evidence of turbulence in the downstream regions of relics has been discovered in some clusters. For example, \citet{Kale2012} and \citet{Jones2021} found patchy polarised emission in the Abell 3376 and PSZ2\,G096.88+24.18 relics, respectively, which they each attributed to turbulence. \citet{DiGennaro2021} performed an extensive analysis of the polarisation structure in the Sausage relic using RM-synthesis. They found that the observed depolarisation could be explained by a turbulent magnetic field strength of $B_{turb} \sim 5.6~\mu$G. Simulations of the Sausage and Toothbrush relics, by \citet{Kang2016} and \citet{Kang2017} respectively, found that they could be explained by shock re-acceleration of fossil electrons with post-shock turbulence. It is not clear if this can explain the systematic offset between the expected and observed relic widths, since the downstream region of only a few relics show indications of turbulence.
Another possibility is that, in general, the shock is broken and complex in shape, so multiple shocks can travel one after the other, artificially expanding the post-shock region. This is a proposed scenario to explain the filaments in Abell 3667 \citep[][]{DeGasperin2022}, and many other RRs show a complex, filamentary sub-structure \citep[e.g.][]{George2015,DiGennaro2018,Rajpurohit2022}.

\subsection{Radio halos in relic-hosting clusters}
\label{sec:disc:halos}

The connection between galaxy cluster mergers and both RRs and giant RHs is well-established. It is however unclear why, in addition to clusters which host both RR(s) and RH(s), there are numerous clusters which only host one or the other. 
In a study of dRR-hosting clusters, \citet{Bonafede2017} investigated the differences between clusters with and without RHs. In their work, they found no relation between the merger mass ratio and the presence of RHs. However, they did find a possible relation with the time since merger, though the statistics were too low to be conclusive.

In our sample, only 6 clusters host dRRs and, of these, 3 also host a giant RH. The statistics are therefore much too low to limit our analysis to only dRRs.
If the difference between RH-hosting and RR-hosting clusters is evolutionary, that is to say that they are produced at different times over the course of a merger, we might expect that there is a characteristic time since merger when RHs are produced.
MHD simulations of two clusters by \citet{Donnert2013} support this scenario. They found that the power of the RHs produced evolved throughout the merger from radio-quiet during the infall phase, to radio-loud as turbulence is driven throughout much of the cluster volume and finally decays to become radio-quiet again. 
Using the distance between a relic and the cluster centre as a proxy for the time since merger, we can investigate such an effect.
In Fig.~\ref{fig:D_Hists} we plot histograms of the distance of the relics to the cluster centre and the distance as a fraction of the cluster $R_{500}$. The black, hatched bars show those relics which are in RH-hosting clusters. There is no clear distinction in the relic distance, nor as a fraction of cluster $R_{500}$, between clusters with and without RHs. The distribution of relics in RH-hosting clusters is similar to those without, though at the most extreme distances none of the clusters also host a RH. The statistics are low, but this is similar to the results of \citet{Bonafede2017}, where clusters with a giant RH occupied the centre of the time-since-merger distribution. They suggested that, in early mergers, turbulence has not had enough time to cascade down to scales at which particles can be (re-)accelerated and, in late mergers, the CRes producing RHs may have already become too faint to observe, due to IC and synchrotron losses. Relics on the other hand, due to continued shock acceleration, would be able to stay visible for longer. Though this assumes that the shock passage lasts longer than the timescale of turbulence able to produce RHs.
In their study on the observed fraction of RHs in merging galaxy clusters, \citet{Cassano2016} found that the absence of RHs in some merging clusters could be explained by an RH-lifetime of $\sim (0.7 - 0.8) \tau_{merger}$, where $\tau_{merger}$ is the merger timescale. If RRs have longer lifetimes, this may go some way to explaining the difference between RH-hosting and RR-hosting clusters, though this would not explain why many clusters are observed with RHs but not RRs.
\citet{Cassano2016} also found evidence that RHs may instead be produced only in larger mass ratio merger events. If RR generation has a different dependence on the mass ratio, this could also contribute to the discrepancy we observe. Unfortunately, we do not have data on the merging mass ratio for our sample and are therefore unable to investigate the effect.

Another possibility is that the difference between the two populations is due to the merging state. Turbulence is injected into the ICM directly by the merger. Minor mergers may therefore be unable to inject enough turbulence into the ICM to produce visible RHs \citep[][]{Cassano2005}. If this were the case, we might expect that clusters with both relics and halos are more morphologically disturbed than those with only relics. From Fig.~\ref{fig:c_w} (bottom) and Fig.~\ref{fig:Disturb_hist} we see however that clusters with both populate the same regions as those with only relics. Furthermore, the least morphologically disturbed cluster (PSZ2\,G205.90+73.76) hosts a dRR and a RH.

In almost all plots we include red circles to denote clusters hosting both a RH and relic(s), and see no clear distinction between the two populations in any, except in the $P_{150 \rm{MHz}} - M_{500}$ plot (Fig.~\ref{fig:P_M_RH_hist}). In this plot, we see a clear split in the number of halos in relic-hosting clusters between low-mass and high-mass clusters.
The histogram in Fig.~\ref{fig:P_M_RH_hist} shows the mass distribution of clusters in our sample. The hatched bars show the distribution for only clusters which also host a RH. Of the 19 clusters which host at least one relic, 11/19=$58 \pm 15 \%$ also host a RH. Including cRRs, this becomes 11/24 = 46 $\pm 14 \%$, as no cRR-hosting cluster also hosts a cRH\footnotemark{}.
The number of RH increases sharply with cluster mass. The fraction of relic-hosting clusters below the median cluster mass, $5.2 \times 10^{14}\Msun$, which also host an RH, is 3/9= 33$\pm 12 \%$, of which none are below $3.5 \times 10^{14}\Msun$, though there are only 4 such clusters in our sample. Above $5.2 \times 10^{14}\Msun$ the fraction rises to 8/10=80$\pm 18 \%$.
If we include cRRs, these become 3/12=25 $\pm 10 \%$ and 8/12=67$ \pm 16 \%$, respectively.
\footnotetext{PSZ2\,G069.39+68.05 hosts a cRH, but we exclude it from occurrence calculations.}

It is already well-established that the occurrence of RHs drops significantly as cluster mass decreases \citep[][]{Cuciti2015,Cuciti2021}. Our results suggest that the difference between the two populations of merging galaxy clusters could be explained by the mass-dependence of RH occurrence \citep[see][]{Cassano22}. However, since the cluster redshift also plays a significant role in RH occurrence, full comparison of the occurrence in relic-hosting clusters to all clusters in the \lotss{} DR2 - PSZ2 sample requires restricting analysis to a relatively small redshift range. Due to the low numbers of RRs, and the correlation of the cluster mass with redshift (Fig.~\ref{fig:M_z}), even the most-populated redshift bin of \citet{Cassano22} does not allow for statistical analysis. Larger surveys, with greater numbers of observed relics, will be required to test if the cluster mass can explain the differences between merging cluster populations. We expect that the number of RHs and relics in PSZ2 clusters should more than double by the completion of \lotss{} \citep[see][]{Botteon2022}, which should allow greater constraints to be set on the connection between RHs and relics.

\subsection{Location and size of relics}
\label{sec:disc:dist}

Radio relics are typically found in cluster outskirts. \citet{Vazza2012} found that this could be explained by the increase in kinetic energy dissipated at the shocks with radius and the fact that relics propagating in the line of sight should be both rare and faint. In their study of simulated relics, they found that most should be located $> 800$ kpc from the cluster centre. For reference, we plot a line at 800 kpc in Fig.~\ref{fig:D_Hists} (top), which shows the distribution of our relic - cluster centre distances. In line with expectations, most (25/28=$89 \pm 19 \%$, excluding PSZ2\,G107.10+65.32) relics are located beyond this projected distance from the cluster centre, and none below 450 kpc. Additionally, only a few relics lie at very large distances from the cluster centre, in line with \citet{Vazza2012}, who showed that the kinetic flux through a shock peaks around $\sim$ 1 Mpc and subsequently decreases towards larger radii.
When plotted instead as a fraction of the cluster $R_{500}$ (Fig.~\ref{fig:D_Hists}, bottom), we see a similar picture, that is to say that all relics lie $>0.5 R_{500}$ from the cluster centre, and more than half lie $> R_{500}$ (15/28). This is in line with simulations by \citet{Zhang19}, who argued that the steep gas density profiles in cluster outskirts $\gtrsim R_{500}$ create a "habitable zone" for long-lived, runaway merger shocks.
The selection criteria used to define relics likely plays a role in the lack of relics at small radii. Since relics were classified as such by their location outside the bulk of ICM emission, our results may be biased towards larger cluster centre distances. However, relics located near the cluster centre are rare \citep[e.g.][FdG14]{Bonafede2012,Feretti2012}.

As in previous statistical studies of relics \citep[][FdG14]{VanWeeren2009,Bonafede2012}, we find that relics further from their cluster centre are typically larger, that is to say that they have a larger LLS (see Fig.~\ref{fig:LLS_D}). We also find that a larger LLS tends to be associated with a larger downstream width. This suggests that shock surfaces expand as they propagate into lower-density environments. For comparison, we also plot the LLS-$D_{RR-c}$ correlation of FdG14. We find that the slopes are very similar, despite the large difference in observing frequency. This shows that the LLS does not change much as we move to lower frequencies, which we might expect, since the line joining the most distant relic regions typically lies parallel to the shock front.

\section{Conclusions}
\label{sec:conclusions}

In this paper we have presented the first statistical sample of RRs observed at 150 MHz, systematically measuring the relic properties in a uniform manner. We used the \lotss{} DR2 - PSZ2 sample of galaxy clusters \citep[][]{Botteon2022}. Where available, archival X-ray data has been utilised to aid in source classification and cluster property measurements \citep[full anaylsis of the sample in X-rays will follow in][]{Zhang22}. 
Our main results are as follows:
\begin{itemize}
    \item We find that RRs are relatively rare phenomena, even when moving to low frequencies. In the \lotss{} DR2 - PSZ2 sample, $\sim 10\%$ of clusters host at least one relic. This is greater than at higher frequencies, however it is nonetheless much lower than predicted by simulations.
    \item We confirm the relationship between RRs and merging galaxy clusters. Radio relic-hosting clusters are among the most dynamically disturbed in the \lotss{} DR2 - PSZ2 sample.
    \item We have revisited previous correlations of RR properties. We find a positive correlation between the RR power and the cluster mass (p = 0.003). We do however find evidence that cluster mass actually sets an upper limit on the power of a relic, rather than being a direct driver. We find a correlation between relic power and its LLS, though only when cRRs are included (p = 0.029).
    We also find a correlation between relic LLS and its radial distance from the cluster (p = 0.002) and as a fraction of cluster $R_{500}$ (p < 0.001), that is to say that relics located further from the cluster centre tend to be larger.
    \item We have developed methods to measure the properties of a relic in a systematic and homogeneous way. In particular, we have introduced a statistical method of defining the relic downstream width as the median of the distribution of widths measured along the relic's extent. Using this, we have shown that, even given optimistic downstream shock properties, the width of RRs in almost all cases is too large to be explained by only synchrotron and inverse Compton losses.
    \item We have compared the properties of the relic-hosting clusters in our sample which do, and do not, also host a RH. We do not find any evidence for the two populations being at different evolutionary stages, nor differences in the merging state of the host cluster. We find that the change in halo occurrence as a function of mass and redshift could go some way to explaining the discrepancy, but the sample is too small for conclusive evidence.
    \item We find that most relics lie in the cluster outskirts. $\sim 90\%$ of relics lie >800 kpc from their cluster centre, in line with cosmological simulations. All relics lie $>0.5R_{500}$ and more than half lie above $R_{500}$.
\end{itemize}

The low occurrence of RRs means that large samples are necessary to understand their statistical properties. Extrapolating from \lotss{} DR2 to the full \lotss{} survey, the number of detected relics in \planck{} PSZ2 clusters is expected to more than double \citep[][]{Botteon2022}. This work will therefore be expanded in future, improving the constraints we can set on RR properties.

\begin{acknowledgements}
FdG and MB acknowledge support from the Deutsche Forschungsgemeinschaft under Germany's Excellence Strategy - EXC 2121 “Quantum Universe” - 390833306.
ABotteon acknowledges support from the ERC-StG DRANOEL n. 714245 and from the VIDI research programme with project number 639.042.729, which is financed by the Netherlands Organisation for Scientific Research (NWO). 
FG, MR, and RC acknowledge support from INAF mainstream project ‘Galaxy Clusters Science with LOFAR’ 1.05.01.86.05.
ABonafede acknowledges support from ERC Stg DRANOEL n. 714245 and MIUR FARE grant “SMS”.
VC and GDG acknowledge support from the Alexander von Humboldt Foundation.
AD acknowledges support by the BMBF Verbundforschung under the grant 05A20STA.
DNH acknowledges support from the ERC through the grant ERC-Stg DRANOEL n. 714245.
KR acknowledges financial support from the ERC Starting Grant ``MAGCOW“ no. 714196.
RJvW acknowledges support from the ERC Starting Grant ClusterWeb 804208.

% LOFAR
The Low Frequency Array, designed and constructed by ASTRON, has facilities in several countries, that are owned by various parties (each with their own funding sources), and that are collectively operated by the International LOFAR Telescope (ILT) foundation under a joint scientific policy.

% Julich
The Jülich LOFAR Long Term Archive and the German LOFAR network are both
coordinated and operated by the Jülich Supercomputing Centre (JSC), and
computing resources on the supercomputer JUWELS at JSC were provided by
the Gauss Centre for Supercomputing e.V. (grant CHTB00) through the John
von Neumann Institute for Computing (NIC).

% DS9
This research has made use of SAOImage DS9, developed by Smithsonian Astrophysical Observatory.
% ADS
This research has made use of NASA's Astrophysics Data System.
\end{acknowledgements}

\bibliographystyle{aa}
\bibliography{references.bib}

\begin{thebibliography}{88}
\expandafter\ifx\csname natexlab\endcsname\relax\def\natexlab#1{#1}\fi

\bibitem[{Akamatsu \& Kawahara(2013)}]{Akamatsu2013}
Akamatsu, H. \& Kawahara, H. 2013, Publications of the Astronomical Society of
  Japan, 65, 16

\bibitem[{Akamatsu {et~al.}(2017)Akamatsu, Mizuno, Ota, Zhang, van Weeren,
  Kawahara, Fukazawa, Kaastra, Kawaharada, Nakazawa, Ohashi, R{\"{o}}ttgering,
  Takizawa, Vink, \& Zandanel}]{Akamatsu2017}
Akamatsu, H., Mizuno, M., Ota, N., {et~al.} 2017, Astronomy {\&} Astrophysics,
  600, A100

\bibitem[{Akritas \& Bershady(1996)}]{Akritas1996}
Akritas, M.~G. \& Bershady, M.~A. 1996, The Astrophysical Journal, 470, 706

\bibitem[{Araya-Melo {et~al.}(2012)Araya-Melo, Arag{\'{o}}n-Calvo,
  Br{\"{u}}ggen, \& Hoeft}]{Araya-Melo2012}
Araya-Melo, P.~A., Arag{\'{o}}n-Calvo, M.~A., Br{\"{u}}ggen, M., \& Hoeft, M.
  2012, Monthly Notices of the Royal Astronomical Society, 423, 2325

\bibitem[{Basu(2012)}]{Basu2012}
Basu, K. 2012, Monthly Notices of the Royal Astronomical Society, 421, 112

\bibitem[{Blandford \& Eichler(1987)}]{Blandford1987}
Blandford, R. \& Eichler, D. 1987, Physics Reports, 154, 1

\bibitem[{Bonafede {et~al.}(2012)Bonafede, Br{\"{u}}ggen, van Weeren, Vazza,
  Giovannini, Ebeling, Edge, Hoeft, \& Klein}]{Bonafede2012}
Bonafede, A., Br{\"{u}}ggen, M., van Weeren, R., {et~al.} 2012, Monthly Notices
  of the Royal Astronomical Society, 426, 40

\bibitem[{Bonafede {et~al.}(2017)Bonafede, Cassano, Br{\"{u}}ggen, Ogrean,
  Riseley, Cuciti, de~Gasperin, Golovich, Kale, Venturi, van Weeren, Wik, \&
  Wittman}]{Bonafede2017}
Bonafede, A., Cassano, R., Br{\"{u}}ggen, M., {et~al.} 2017, Monthly Notices of
  the Royal Astronomical Society, 470, 3465

\bibitem[{Bonafede {et~al.}(2014)Bonafede, Intema, Br{\"{u}}ggen, Girardi,
  Nonino, Kantharia, van Weeren, \& R{\"{o}}ttgering}]{Bonafede2014}
Bonafede, A., Intema, H.~T., Br{\"{u}}ggen, M., {et~al.} 2014, The
  Astrophysical Journal, 785, 1

\bibitem[{Botteon {et~al.}(2020{\natexlab{a}})Botteon, Brunetti, Ryu, \&
  Roh}]{Botteon2020Shock}
Botteon, A., Brunetti, G., Ryu, D., \& Roh, S. 2020{\natexlab{a}}, Astronomy
  {\&} Astrophysics, 634, A64

\bibitem[{Botteon {et~al.}(2021)Botteon, Cassano, van Weeren, Shimwell,
  Bonafede, Br{\"{u}}ggen, Brunetti, Cuciti, Dallacasa, de~Gasperin,
  Di~Gennaro, Gastaldello, Hoang, Rossetti, \& R{\"{o}}ttgering}]{Botteon2021}
Botteon, A., Cassano, R., van Weeren, R.~J., {et~al.} 2021, The Astrophysical
  Journal Letters, 914, L29

\bibitem[{Botteon {et~al.}(2016{\natexlab{a}})Botteon, Gastaldello, Brunetti,
  \& Dallacasa}]{Botteon2016_A115}
Botteon, A., Gastaldello, F., Brunetti, G., \& Dallacasa, D.
  2016{\natexlab{a}}, Monthly Notices of the Royal Astronomical Society, 460,
  L84

\bibitem[{Botteon {et~al.}(2016{\natexlab{b}})Botteon, Gastaldello, Brunetti,
  \& Kale}]{Botteon2016}
Botteon, A., Gastaldello, F., Brunetti, G., \& Kale, R. 2016{\natexlab{b}},
  Monthly Notices of the Royal Astronomical Society, 463, 1534

\bibitem[{Botteon {et~al.}(2022)Botteon, Shimwell, Cassano, Cuciti, Zhang,
  Bruno, Camillini, Natale, Jones, Gastaldello, Simionescu, Rossetti, Akamatsu,
  van Weeren, Brunetti, Br{\"{u}}ggen, Groeneveld, Hoang, Hardcastle, Ignesti,
  Gennaro, Bonafede, Drabent, R{\"{o}}ttgering, Hoeft, \&
  de~Gasperin}]{Botteon2022}
Botteon, A., Shimwell, T.~W., Cassano, R., {et~al.} 2022, Astronomy {\&}
  Astrophysics, 660, A78

\bibitem[{Botteon {et~al.}(2020{\natexlab{b}})Botteon, van Weeren, Brunetti,
  de~Gasperin, Intema, Osinga, Di~Gennaro, Shimwell, Bonafede, Br{\"{u}}ggen,
  Cassano, Cuciti, Dallacasa, Gastaldello, Mandal, Rossetti, \&
  R{\"{o}}ttgering}]{Botteon2020A1758}
Botteon, A., van Weeren, R.~J., Brunetti, G., {et~al.} 2020{\natexlab{b}},
  Monthly Notices of the Royal Astronomical Society, 499, 11

\bibitem[{Bourdin {et~al.}(2013)Bourdin, Mazzotta, Markevitch, Giacintucci, \&
  Brunetti}]{Bourdin2013}
Bourdin, H., Mazzotta, P., Markevitch, M., Giacintucci, S., \& Brunetti, G.
  2013, The Astrophysical Journal, 486, 347

\bibitem[{Boxelaar {et~al.}(2021)Boxelaar, van Weeren, \&
  Botteon}]{Boxelaar2021}
Boxelaar, J.~M., van Weeren, R.~J., \& Botteon, A. 2021, Astronomy and
  Computing, 35, 100464

\bibitem[{Br{\"{u}}ggen \& Vazza(2020)}]{Bruggen2020}
Br{\"{u}}ggen, M. \& Vazza, F. 2020, Monthly Notices of the Royal Astronomical
  Society, 493, 2306

\bibitem[{Brunetti {et~al.}(2009)Brunetti, Cassano, Dolag, \&
  Setti}]{Brunetti2009}
Brunetti, G., Cassano, R., Dolag, K., \& Setti, G. 2009, Astronomy {\&}
  Astrophysics, 507, 661

\bibitem[{Brunetti \& Jones(2014)}]{Brunetti2014}
Brunetti, G. \& Jones, T.~W. 2014, International Journal of Modern Physics D,
  23, 1430007

\bibitem[{Bruno {et~al.}(2022)Bruno, Brunetti, Botteon, Cuciti, Dallacasa,
  Cassano, van Weeren, Shimwell, Bonafede, Br{\"{u}}ggen, Hoang,
  R{\"{o}}ttgering, \& Tasse}]{Bruno22}
Bruno, L., Brunetti, G., Botteon, A., {et~al.} 2022, Astronomy {\&}
  Astrophysics, Submitted

\bibitem[{Cassano \& Brunetti(2005)}]{Cassano2005}
Cassano, R. \& Brunetti, G. 2005, Monthly Notices of the Royal Astronomical
  Society, 357, 1313

\bibitem[{Cassano {et~al.}(2016)Cassano, Brunetti, Giocoli, \&
  Ettori}]{Cassano2016}
Cassano, R., Brunetti, G., Giocoli, C., \& Ettori, S. 2016, Astronomy {\&}
  Astrophysics, 593, A81

\bibitem[{Cassano {et~al.}(2022)Cassano, Cuciti, Brunetti, Botteon, Rossetti,
  Bruno, Simionescu, Gastaldello, van Weeren, Br{\"{u}}ggen, Dallacasa, Zhang,
  Akamatsu, Bonafede, Di~Gennaro, Shimwell, de~Gasperin, R{\"{o}}ttgering, \&
  Jones}]{Cassano22}
Cassano, R., Cuciti, V., Brunetti, G., {et~al.} 2022, Astronomy {\&}
  Astrophysics, Submitted

\bibitem[{Cassano {et~al.}(2013)Cassano, Ettori, Brunetti, Giacintucci, Pratt,
  Venturi, Kale, Dolag, \& Markevitch}]{Cassano2013}
Cassano, R., Ettori, S., Brunetti, G., {et~al.} 2013, The Astrophysical
  Journal, 777, 141

\bibitem[{Cassano {et~al.}(2010)Cassano, Ettori, Giacintucci, Brunetti,
  Markevitch, Venturi, \& Gitti}]{Cassano2010}
Cassano, R., Ettori, S., Giacintucci, S., {et~al.} 2010, The Astrophysical
  Journal Letters, 721, 82

\bibitem[{Cuciti {et~al.}(2021)Cuciti, Cassano, Brunetti, Dallacasa,
  de~Gasperin, Ettori, Giacintucci, Kale, Pratt, Van~Weeren, \&
  Venturi}]{Cuciti2021}
Cuciti, V., Cassano, R., Brunetti, G., {et~al.} 2021, Astronomy {\&}
  Astrophysics, 647, 51

\bibitem[{Cuciti {et~al.}(2015)Cuciti, Cassano, Brunetti, Dallacasa, Kale,
  Ettori, \& Venturi}]{Cuciti2015}
Cuciti, V., Cassano, R., Brunetti, G., {et~al.} 2015, Astronomy {\&}
  Astrophysics, 580, A97

\bibitem[{de~Gasperin {et~al.}(2020)de~Gasperin, Brunetti, Br{\"{u}}ggen, van
  Weeren, Williams, Botteon, Cuciti, Dijkema, Edler, Iacobelli, Kang, Offringa,
  Orru, Pizzo, Rafferty, Rottgering, \& Shimwell}]{DeGasperin2020}
de~Gasperin, F., Brunetti, G., Br{\"{u}}ggen, M., {et~al.} 2020, Astronomy {\&}
  Astrophysics, 642, A85

\bibitem[{de~Gasperin {et~al.}(2017)de~Gasperin, Intema, Shimwell, Brunetti,
  Br{\"{u}}ggen, En{\ss}lin, van Weeren, Bonafede, \&
  R{\"{o}}ttgering}]{DeGasperin2017}
de~Gasperin, F., Intema, H.~T., Shimwell, T.~W., {et~al.} 2017, Science
  Advances, 3, e1701634

\bibitem[{de~Gasperin {et~al.}(2022)de~Gasperin, Rudnick, Finoguenov, Wittor,
  Akamatsu, Br{\"{u}}ggen, Chibueze, Clarke, Cotton, Cuciti,
  Dom{\'{i}}nguez-Fern{\'{a}}ndez, Knowles, O’Sullivan, \&
  Sebokolodi}]{DeGasperin2022}
de~Gasperin, F., Rudnick, L., Finoguenov, A., {et~al.} 2022, Astronomy {\&}
  Astrophysics, 659, A146

\bibitem[{de~Gasperin {et~al.}(2014)de~Gasperin, van Weeren, Br{\"{u}}ggen,
  Vazza, Bonafede, \& Intema}]{DeGasperin2014}
de~Gasperin, F., van Weeren, R.~J., Br{\"{u}}ggen, M., {et~al.} 2014, Monthly
  Notices of the Royal Astronomical Society, 444, 3130

\bibitem[{Di~Gennaro {et~al.}(2018)Di~Gennaro, van Weeren, Hoeft, Kang, Ryu,
  Rudnick, Forman, R{\"{o}}ttgering, Br{\"{u}}ggen, Dawson, Golovich, Hoang,
  Intema, Jones, Kraft, Shimwell, \& Stroe}]{DiGennaro2018}
Di~Gennaro, G., van Weeren, R.~J., Hoeft, M., {et~al.} 2018, The Astrophysical
  Journal, 865, 24

\bibitem[{Di~Gennaro {et~al.}(2021)Di~Gennaro, van Weeren, Rudnick, Hoeft,
  Br{\"{u}}ggen, Ryu, R{\"{o}}ttgering, Forman, Stroe, Shimwell, Kraft, Jones,
  \& Hoang}]{DiGennaro2021}
Di~Gennaro, G., van Weeren, R.~J., Rudnick, L., {et~al.} 2021, The
  Astrophysical Journal, 911, 3

\bibitem[{Dom{\'{i}}nguez-Fern{\'{a}}ndez
  {et~al.}(2021)Dom{\'{i}}nguez-Fern{\'{a}}ndez, Br{\"{u}}ggen, Vazza,
  Banda-Barrag{\'{a}}n, Rajpurohit, Mignone, Mukherjee, \&
  Vaidya}]{Dominguez-Fernandez2021substructure}
Dom{\'{i}}nguez-Fern{\'{a}}ndez, P., Br{\"{u}}ggen, M., Vazza, F., {et~al.}
  2021, Monthly Notices of the Royal Astronomical Society, 500, 795

\bibitem[{Donnert {et~al.}(2013)Donnert, Dolag, Brunetti, \&
  Cassano}]{Donnert2013}
Donnert, J., Dolag, K., Brunetti, G., \& Cassano, R. 2013, Monthly Notices of
  the Royal Astronomical Society, 429, 3564

\bibitem[{Dwarakanath {et~al.}(2018)Dwarakanath, Parekh, Kale, \&
  George}]{Dwarakanath2018}
Dwarakanath, K.~S., Parekh, V., Kale, R., \& George, L.~T. 2018, Monthly
  Notices of the Royal Astronomical Society, 477, 957

\bibitem[{Eckert {et~al.}(2016)Eckert, Jauzac, Vazza, Owers, Kneib, Tchernin,
  Intema, \& Knowles}]{Eckert2016}
Eckert, D., Jauzac, M., Vazza, F., {et~al.} 2016, Monthly Notices of the Royal
  Astronomical Society, 461, 1302

\bibitem[{En{\ss}lin {et~al.}(1998)En{\ss}lin, Biermann, Klein, \&
  Kohle}]{Ensslin1998}
En{\ss}lin, T.~A., Biermann, P.~L., Klein, U., \& Kohle, S. 1998, Astronomy
  {\&} Astrophysics, 332, 395

\bibitem[{Feretti {et~al.}(2012)Feretti, Giovannini, Govoni, \&
  Murgia}]{Feretti2012}
Feretti, L., Giovannini, G., Govoni, F., \& Murgia, M. 2012, Astronomy {\&}
  Astrophysics Review, 20, 1

\bibitem[{Fermi(1949)}]{Fermi1949}
Fermi, E. 1949, Physical Review, 75, 1169

\bibitem[{Finner {et~al.}(2017)Finner, Jee, Golovich, Wittman, Dawson, Gruen,
  Koekemoer, Lemaux, \& Seitz}]{Finner2017}
Finner, K., Jee, M.~J., Golovich, N., {et~al.} 2017, The Astrophysical Journal,
  851, 46

\bibitem[{Finoguenov {et~al.}(2010)Finoguenov, Sarazin, Nakazawa, Wik, \&
  Clarke}]{Finoguenov2010}
Finoguenov, A., Sarazin, C.~L., Nakazawa, K., Wik, D.~R., \& Clarke, T.~E.
  2010, The Astrophysical Journal, 715, 1143

\bibitem[{Fujita {et~al.}(2016)Fujita, Akamatsu, \& Kimura}]{Fujita2016}
Fujita, Y., Akamatsu, H., \& Kimura, S.~S. 2016, Publications of the
  Astronomical Society of Japan, 68, 34

\bibitem[{George {et~al.}(2015)George, Dwarakanath, Johnston-Hollitt,
  Hurley-Walker, Hindson, Kapi{\'{n}}ska, Tingay, Bell, Callingham, For,
  Hancock, Lenc, McKinley, Morgan, Offringa, Procopio, Staveley-Smith, Wayth,
  Wu, Zheng, Bernardi, Bowman, Briggs, Cappallo, Corey, Deshpande, Emrich,
  Goeke, Greenhill, Hazelton, Kaplan, Kasper, Kratzenberg, Lonsdale, Lynch,
  McWhirter, Mitchell, Morales, Morgan, Oberoi, Ord, Prabu, Rogers, Roshi,
  Udaya~Shankar, Srivani, Subrahmanyan, Waterson, Webster, Whitney, Williams,
  \& Williams}]{George2015}
George, L.~T., Dwarakanath, K.~S., Johnston-Hollitt, M., {et~al.} 2015, Monthly
  Notices of the Royal Astronomical Society, 451, 4207

\bibitem[{Golovich {et~al.}(2019)Golovich, Dawson, Wittman, Jee, Benson,
  Lemaux, van Weeren, Andrade-Santos, Sobral, de~Gasperin, Br{\"{u}}ggen,
  Brada{\v{c}}, Finner, \& Peter}]{Golovich2019Spec}
Golovich, N., Dawson, W.~A., Wittman, D.~M., {et~al.} 2019, The Astrophysical
  Journal Supplement Series, 240, 39

\bibitem[{Hoang {et~al.}(2022)Hoang, Br{\"{u}}ggen, Botteon, Shimwell, Zhang,
  Bonafede, Bruno, Bonnassieux, Cassano, Cuciti, Drabent, de~Gasperin,
  Gastaldello, Di~Gennaro, Hoeft, Jones, Pignataro, R{\"{o}}ttgering,
  Simionescu, \& van Weeren}]{Hoang2022}
Hoang, D.~N., Br{\"{u}}ggen, M., Botteon, A., {et~al.} 2022, arXiv e-prints,
  arXiv2206.04666

\bibitem[{Hoeft \& Br{\"{u}}ggen(2007)}]{Hoeft2007}
Hoeft, M. \& Br{\"{u}}ggen, M. 2007, Monthly Notices of the Royal Astronomical
  Society, 375, 77

\bibitem[{Jee {et~al.}(2016)Jee, Dawson, Stroe, Wittman, Van~Weeren,
  Br{\"{u}}ggen, Brada{\v{c}}, \& R{\"{o}}ttgering}]{Jee2016}
Jee, M.~J., Dawson, W.~A., Stroe, A., {et~al.} 2016, The Astrophysical Journal,
  817, 179

\bibitem[{Jones {et~al.}(2021)Jones, de~Gasperin, Cuciti, Hoang, Botteon,
  Br{\"{u}}ggen, Brunetti, Finner, Forman, Jones, Kraft, Shimwell, \& van
  Weeren}]{Jones2021}
Jones, A., de~Gasperin, F., Cuciti, V., {et~al.} 2021, Monthly Notices of the
  Royal Astronomical Society, 505, 4762

\bibitem[{Kale {et~al.}(2012)Kale, Dwarakanath, Bagchi, \& Paul}]{Kale2012}
Kale, R., Dwarakanath, K.~S., Bagchi, J., \& Paul, S. 2012, Monthly Notices of
  the Royal Astronomical Society, 426, 1204

\bibitem[{Kale {et~al.}(2015)Kale, Venturi, Giacintucci, Dallacasa, Cassano,
  Brunetti, Cuciti, Macario, \& Athreya}]{Kale2015}
Kale, R., Venturi, T., Giacintucci, S., {et~al.} 2015, Astronomy {\&}
  Astrophysics, 579, A92

\bibitem[{Kang(2016)}]{Kang2016}
Kang, H. 2016, Journal of the Korean Astronomical Society, 49, 145

\bibitem[{Kang \& Ryu(2011)}]{Kang2011}
Kang, H. \& Ryu, D. 2011, The Astrophysical Journal, 734, 18

\bibitem[{Kang {et~al.}(2012)Kang, Ryu, \& Jones}]{Kang2012}
Kang, H., Ryu, D., \& Jones, T.~W. 2012, The Astrophysical Journal, 756, 97

\bibitem[{Kang {et~al.}(2017)Kang, Ryu, \& Jones}]{Kang2017}
Kang, H., Ryu, D., \& Jones, T.~W. 2017, The Astrophysical Journal, 840, 42

\bibitem[{Liang {et~al.}(2000)Liang, Hunstead, Birkinshaw, \&
  Andreani}]{Liang2000}
Liang, H., Hunstead, R.~W., Birkinshaw, M., \& Andreani, P. 2000, The
  Astrophysical Journal, 544, 686

\bibitem[{Lovisari {et~al.}(2020)Lovisari, Schellenberger, Sereno, Ettori,
  Pratt, Forman, Jones, Andrade-Santos, Randall, \& Kraft}]{Lovisari2020}
Lovisari, L., Schellenberger, G., Sereno, M., {et~al.} 2020, The Astrophysical
  Journal, 892, 102

\bibitem[{Mandal {et~al.}(2020)Mandal, Intema, van Weeren, Shimwell, Botteon,
  Brunetti, de~Gasperin, Br{\"{u}}ggen, Di~Gennaro, Kraft, R{\"{o}}ttgering,
  Hardcastle, \& Tasse}]{Mandal2020}
Mandal, S., Intema, H.~T., van Weeren, R.~J., {et~al.} 2020, Astronomy {\&}
  Astrophysics, 634, A4

\bibitem[{Markevitch {et~al.}(2002)Markevitch, Gonzalez, David, Vikhlinin,
  Murray, Forman, Jones, \& Tucker}]{Markevitch2002}
Markevitch, M., Gonzalez, A.~H., David, L., {et~al.} 2002, The Astrophysical
  Journal, 567, 27

\bibitem[{Markevitch {et~al.}(2005)Markevitch, Govoni, Brunetti, \&
  Jerius}]{Markevitch2005}
Markevitch, M., Govoni, F., Brunetti, G., \& Jerius, D. 2005, The Astrophysical
  Journal, 627, 733

\bibitem[{Mohr {et~al.}(1993)Mohr, Fabricant, \& Geller}]{Mohr1993}
Mohr, J.~J., Fabricant, D.~G., \& Geller, M.~J. 1993, The Astrophysical
  Journal, 413, 492

\bibitem[{Nuza {et~al.}(2017)Nuza, Gelszinnis, Hoeft, \& Yepes}]{Nuza2017}
Nuza, S.~E., Gelszinnis, J., Hoeft, M., \& Yepes, G. 2017, Monthly Notices of
  the Royal Astronomical Society, 470, 240

\bibitem[{Nuza {et~al.}(2012)Nuza, Hoeft, van Weeren, Gottl{\"{o}}ber, \&
  Yepes}]{Nuza2012}
Nuza, S.~E., Hoeft, M., van Weeren, R.~J., Gottl{\"{o}}ber, S., \& Yepes, G.
  2012, Monthly Notices of the Royal Astronomical Society, 420, 2006

\bibitem[{Ogrean \& Br{\"{u}}ggen(2013)}]{Ogrean2013Coma}
Ogrean, G.~A. \& Br{\"{u}}ggen, M. 2013, Monthly Notices of the Royal
  Astronomical Society, 433, 1701

\bibitem[{Ogrean {et~al.}(2013)Ogrean, Br{\"{u}}ggen, R{\"{o}}ttgering,
  Simionescu, Croston, van Weeren, \& Hoeft}]{Ogrean2013XMM}
Ogrean, G.~A., Br{\"{u}}ggen, M., R{\"{o}}ttgering, H., {et~al.} 2013, Monthly
  Notices of the Royal Astronomical Society, 429, 2617

\bibitem[{Piffaretti {et~al.}(2011)Piffaretti, Arnaud, Pratt, Pointecouteau, \&
  Melin}]{Piffaretti2011}
Piffaretti, R., Arnaud, M., Pratt, G.~W., Pointecouteau, E., \& Melin, J.-B.
  2011, Astronomy {\&} Astrophysics, 534, A109

\bibitem[{Pinzke {et~al.}(2013)Pinzke, Oh, \& Pfrommer}]{Pinzke2013}
Pinzke, A., Oh, S.~P., \& Pfrommer, C. 2013, Monthly Notices of the Royal
  Astronomical Society, 435, 1061

\bibitem[{{Planck Collaboration} {et~al.}(2014){Planck Collaboration}, Ade,
  Aghanim, Armitage-Caplan, Arnaud, Ashdown, Atrio-Barandela, Aumont, Aussel,
  Baccigalupi, Banday, Barreiro, Barrena, Bartelmann, Bartlett, Battaner,
  Benabed, Beno{\^{i}}t, Benoit-L{\'{e}}vy, Bernard, Bersanelli, Bielewicz,
  Bikmaev, Bobin, Bock, B{\"{o}}hringer, Bonaldi, Bond, Borrill, Bouchet,
  Bridges, Bucher, Burenin, Burigana, Butler, Cardoso, Carvalho, Catalano,
  Challinor, Chamballu, Chary, Chen, Chiang, Chiang, Chon, Christensen,
  Churazov, Church, Clements, Colombi, Colombo, Comis, Couchot, Coulais, Crill,
  Curto, Cuttaia, Da~Silva, Dahle, Danese, Davies, Davis, De~Bernardis,
  De~Rosa, De~Zotti, Delabrouille, Delouis, D{\'{e}}mocl{\`{e}}s, D{\'{e}}sert,
  Dickinson, Diego, Dolag, Dole, Donzelli, Dor{\'{e}}, Douspis, Dupac,
  Efstathiou, Eisenhardt, En{\ss}lin, Eriksen, Feroz, Finelli, Flores-Cacho,
  Forni, Frailis, Franceschi, Fromenteau, Galeotta, Ganga, G{\'{e}}nova-Santos,
  Giard, Giardino, Gilfanov, Giraud-H{\'{e}}raud, Gonz{\'{a}}lez-Nuevo,
  G{\'{o}}rski, Grainge, Gratton, Gregorio, Groeneboom, Gruppuso, Hansen,
  Hanson, Harrison, Hempel, Henrot-Versill{\'{e}}, Hern{\'{a}}ndez-Monteagudo,
  Herranz, Hildebrandt, Hivon, Hobson, Holmes, Hornstrup, Hovest, Huffenberger,
  Hurier, Hurley-Walker, Jaffe, Jaffe, Jones, Juvela, Keih{\"{a}}nen,
  Keskitalo, Khamitov, Kisner, Kneissl, Knoche, Knox, Kunz, Kurki-Suonio,
  Lagache, L{\"{a}}hteenm{\"{a}}ki, Lamarre, Lasenby, Laureijs, Lawrence,
  Leahy, Leonardi, Le{\'{o}}n-Tavares, Lesgourgues, Li, Liddle, Liguori, Lilje,
  Linden-V{\o}rnle, L{\'{o}}pez-Caniego, Lubin, Maci{\'{a}}s-P{\'{e}}rez,
  Mactavish, Maffei, Maino, Mandolesi, Maris, Marshall, Martin,
  Mart{\'{i}}nez-Gonz{\'{a}}lez, Masi, Massardi, Matarrese, Matthai, Mazzotta,
  Mei, Meinhold, Melchiorri, Melin, Mendes, Mennella, Migliaccio, Mikkelsen,
  Mitra, Miville-Desch{\^{e}}nes, Moneti, Montier, Morgante, Mortlock, Munshi,
  Murphy, Naselsky, Nati, Natoli, Nesvadba, Netterfield, N{\o}rgaard-Nielsen,
  Noviello, Novikov, Novikov, O'dwyer, Olamaie, Osborne, Oxborrow, Paci,
  Pagano, Pajot, Paoletti, Pasian, Patanchon, Pearson, Perdereau, Perotto,
  Perrott, Perrotta, Piacentini, Piat, Pierpaoli, Pietrobon, Plaszczynski,
  Pointecouteau, Polenta, Ponthieu, Popa, Poutanen, Pratt, Pr{\'{e}}zeau,
  Prunet, Puget, Rachen, Reach, Rebolo, Reinecke, Remazeilles, Renault,
  Ricciardi, Riller, Ristorcelli, Rocha, Rosset, Roudier, Rowan-Robinson,
  Rubinõ-Mart{\'{i}}n, Rumsey, Rusholme, Sandri, Santos, Saunders, Savini,
  Schammel, Scott, Seiffert, Shellard, Shimwell, Spencer, Stanford, Starck,
  Stolyarov, Stompor, Sudiwala, Sunyaev, Sureau, Sutton, Suur-Uski, Sygnet,
  Tauber, Tavagnacco, Terenzi, Toffolatti, Tomasi, Tristram, Tucci, Tuovinen,
  T{\"{u}}rler, Umana, Valenziano, Valiviita, Van~Tent, Vibert, Vielva, Villa,
  Vittorio, Wade, Wandelt, White, White, Yvon, Zacchei, \& Zonca}]{Planck2014}
{Planck Collaboration}, Ade, P.~A., Aghanim, N., {et~al.} 2014, Astronomy {\&}
  Astrophysics, 571, A29

\bibitem[{{Planck Collaboration} {et~al.}(2016){Planck Collaboration}, Ade,
  Aghanim, Arnaud, Ashdown, Aumont, Baccigalupi, Banday, Barreiro, Barrena,
  Bartlett, Bartolo, Battaner, Battye, Benabed, Beno{\^{i}}t,
  Benoit-L{\'{e}}vy, Bernard, Bersanelli, Bielewicz, Bikmaev, B{\"{o}}hringer,
  Bonaldi, Bonavera, Bond, Borrill, Bouchet, Bucher, Burenin, Burigana, Butler,
  Calabrese, Cardoso, Carvalho, Catalano, Challinor, Chamballu, Chary, Chiang,
  Chon, Christensen, Clements, Colombi, Colombo, Combet, Comis, Couchot,
  Coulais, Crill, Curto, Cuttaia, Dahle, Danese, Davies, Davis, De~Bernardis,
  De~Rosa, De~Zotti, Delabrouille, D{\'{e}}sert, Dickinson, Diego, Dolag, Dole,
  Donzelli, Dor{\'{e}}, Douspis, Ducout, Dupac, Efstathiou, Eisenhardt, Elsner,
  En{\ss}lin, Eriksen, Falgarone, Fergusson, Feroz, Ferragamo, Finelli, Forni,
  Frailis, Fraisse, Franceschi, Frejsel, Galeotta, Galli, Ganga,
  G{\'{e}}nova-Santos, Giard, Giraud-H{\'{e}}raud, Gjerl{\o}w,
  Gonz{\'{a}}lez-Nuevo, G{\'{o}}rski, Grainge, Gratton, Gregorio, Gruppuso,
  Gudmundsson, Hansen, Hanson, Harrison, Hempel, Henrot-Versill{\'{e}},
  Hern{\'{a}}ndez-Monteagudo, Herranz, Hildebrandt, Hivon, Hobson, Holmes,
  Hornstrup, Hovest, Huffenberger, Hurier, Jaffe, Jaffe, Jin, Jones, Juvela,
  Keih{\"{a}}nen, Keskitalo, Khamitov, Kisner, Kneissl, Knoche, Kunz,
  Kurki-Suonio, Lagache, Lamarre, Lasenby, Lattanzi, Lawrence, Leonardi,
  Lesgourgues, Levrier, Liguori, Lilje, Linden-V{\o}rnle, L{\'{o}}pez-Caniego,
  Lubin, Maci{\'{a}}s-P{\'{e}}rez, Maggio, Maino, Mak, Mandolesi, Mangilli,
  Martin, Mart{\'{i}}nez-Gonz{\'{a}}lez, Masi, Matarrese, Mazzotta, Mcgehee,
  Mei, Melchiorri, Melin, Mendes, Mennella, Migliaccio, Mitra,
  Miville-Desch{\^{e}}nes, Moneti, Montier, Morgante, Mortlock, Moss, Munshi,
  Murphy, Naselsky, Nastasi, Nati, Natoli, Netterfield, N{\o}rgaard-Nielsen,
  Noviello, Novikov, Novikov, Olamaie, Oxborrow, Paci, Pagano, Pajot, Paoletti,
  Pasian, Patanchon, Pearson, Perdereau, Perotto, Perrott, Perrotta, Pettorino,
  Piacentini, Piat, Pierpaoli, Pietrobon, Plaszczynski, Pointecouteau, Polenta,
  Pratt, Pr{\'{e}}zeau, Prunet, Puget, Rachen, Reach, Rebolo, Reinecke,
  Remazeilles, Renault, Renzi, Ristorcelli, Rocha, Rosset, Rossetti, Roudier,
  Rozo, Rubinõ-Mart{\'{i}}n, Rumsey, Rusholme, Rykoff, Sandri, Santos,
  Saunders, Savelainen, Savini, Schammel, Scott, Seiffert, Shellard, Shimwell,
  Spencer, Stanford, Stern, Stolyarov, Stompor, Streblyanska, Sudiwala,
  Sunyaev, Sutton, Suur-Uski, Sygnet, Tauber, Terenzi, Toffolatti, Tomasi,
  Tramonte, Tristram, Tucci, Tuovinen, Umana, Valenziano, Valiviita, Van~Tent,
  Vielva, Villa, Wade, Wandelt, Wehus, White, Wright, Yvon, Zacchei, \&
  Zonca}]{Planck2016}
{Planck Collaboration}, Ade, P.~A., Aghanim, N., {et~al.} 2016, Astronomy {\&}
  Astrophysics, 594, A27

\bibitem[{Poole {et~al.}(2006)Poole, Fardal, Babul, McCarthy, Quinn, \&
  Wadsley}]{Poole2006}
Poole, G.~B., Fardal, M.~A., Babul, A., {et~al.} 2006, Monthly Notices of the
  Royal Astronomical Society, 373, 881

\bibitem[{Pratt {et~al.}(2009)Pratt, Croston, Arnaud, \&
  B{\"{o}}hringer}]{Pratt2009}
Pratt, G.~W., Croston, J.~H., Arnaud, M., \& B{\"{o}}hringer, H. 2009,
  Astronomy {\&} Astrophysics, 498, 361

\bibitem[{Rajpurohit {et~al.}(2022)Rajpurohit, van Weeren, Hoeft, Vazza,
  Brienza, Forman, Wittor, Dom{\'{i}}nguez-Fern{\'{a}}ndez, Rajpurohit,
  Riseley, Botteon, Osinga, Brunetti, Bonnassieux, Bonafede, Rajpurohit,
  Stuardi, Drabent, Br{\"{u}}ggen, Dallacasa, Shimwell, R{\"{o}}ttgering,
  de~Gasperin, Miley, \& Rossetti}]{Rajpurohit2022}
Rajpurohit, K., van Weeren, R.~J., Hoeft, M., {et~al.} 2022, The Astrophysical
  Journal, 927, 80

\bibitem[{Rajpurohit {et~al.}(2020)Rajpurohit, Vazza, Hoeft, Loi, Beck, Vacca,
  Kierdorf, van Weeren, Wittor, Govoni, Murgia, Riseley, Locatelli, Drabent, \&
  Bonnassieux}]{Rajpurohit2020Spectrum}
Rajpurohit, K., Vazza, F., Hoeft, M., {et~al.} 2020, Astronomy {\&}
  Astrophysics, 642, L13

\bibitem[{Santos {et~al.}(2008)Santos, Rosati, Tozzi, B{\"{o}}hringer, Ettori,
  \& Bignamini}]{Santos2008}
Santos, J.~S., Rosati, P., Tozzi, P., {et~al.} 2008, Astronomy {\&}
  Astrophysics, 483, 35

\bibitem[{Shimwell {et~al.}(2022)Shimwell, Hardcastle, Tasse, Best, Rottgering,
  Williams, Botteon, Drabent, Mechev, Shulevski, van Weeren, Bester,
  Br{\"{u}}ggen, Brunetti, Callingham, Chyzy, Conway, Dijkema, Duncan,
  de~Gasperin, Hale, Haverkorn, Hugo, Jackson, Mevius, Miley, Morabito,
  Morganti, Offringa, Oonk, Rafferty, Sabater, Smith, Schwarz, Smirnov,
  Oasullivan, Vedantham, White, Albert, Alegre, Asabere, Bacon, Bonafede,
  Bonnassieux, Brienza, Bilicki, Bonato, Calistro~Rivera, Cassano, Cochrane,
  Croston, Cuciti, Dallacasa, Danezi, Dettmar, Di~Gennaro, Edler, Enalin, Emig,
  Franzen, Garcia-Vergara, Grange, Gurkan, Hajduk, Heald, Heesen, Hoang, Hoeft,
  Horellou, Iacobelli, Jamrozy, Jelia, Kondapally, Kukreti,
  Kunert-Bajraszewska, Magliocchetti, Mahatma, Maaek, Mandal, Massaro,
  Meyer-Zhao, Mingo, Mostert, Nair, Nakoneczny, Nikiel-Wroczyaski, Orru,
  Pajdosz-Amierciak, Pasini, Prandoni, van Piggelen, Rajpurohit,
  Retana-Montenegro, Riseley, Rowlinson, Saxena, Schrijvers, Sweijen, Siewert,
  Timmerman, Vaccari, Vink, West, Woaowska, Zhang, \& Zheng}]{Shimwell2022}
Shimwell, T.~W., Hardcastle, M.~J., Tasse, C., {et~al.} 2022, Astronomy {\&}
  Astrophysics, 659, A1

\bibitem[{Shimwell {et~al.}(2015)Shimwell, Markevitch, Brown, Feretti,
  Gaensler, Johnston-Hollitt, Lage, \& Srinivasan}]{Shimwell2015}
Shimwell, T.~W., Markevitch, M., Brown, S., {et~al.} 2015, Monthly Notices of
  the Royal Astronomical Society, 449, 1486

\bibitem[{Tasse {et~al.}(2021)Tasse, Shimwell, Hardcastle, O'Sullivan, van
  Weeren, Best, Bester, Hugo, Smirnov, Sabater, Calistro-Rivera, de~Gasperin,
  Morabito, R{\"{o}}ttgering, Williams, Bonato, Bondi, Botteon, Br{\"{u}}ggen,
  Brunetti, Chyay, Garrett, G{\"{u}}rkan, Jarvis, Kondapally, Mandal, Prandoni,
  Repetti, Retana-Montenegro, Schwarz, Shulevski, \& Wiaux}]{Tasse2021}
Tasse, C., Shimwell, T., Hardcastle, M.~J., {et~al.} 2021, Astronomy {\&}
  Astrophysics, 648, A1

\bibitem[{Urdampilleta {et~al.}(2018)Urdampilleta, Akamatsu, Mernier, Kaastra,
  De~Plaa, Ohashi, Ishisaki, \& Kawahara}]{Urdampilleta2018}
Urdampilleta, I., Akamatsu, H., Mernier, F., {et~al.} 2018, Astronomy {\&}
  Astrophysics, 618, A74

\bibitem[{van Weeren {et~al.}(2017)van Weeren, Andrade-Santos, Dawson,
  Golovich, Lal, Kang, Ryu, Br{\`{i}}ggen, Ogrean, Forman, Jones, Placco,
  Santucci, Wittman, Jee, Kraft, Sobral, Stroe, \& Fogarty}]{VanWeeren2017}
van Weeren, R.~J., Andrade-Santos, F., Dawson, W.~A., {et~al.} 2017, Nature
  Astronomy, 1, 0005

\bibitem[{van Weeren {et~al.}(2011)van Weeren, Br{\"{u}}ggen, R{\"{o}}ttgering,
  \& Hoeft}]{VanWeeren2011}
van Weeren, R.~J., Br{\"{u}}ggen, M., R{\"{o}}ttgering, H.~J., \& Hoeft, M.
  2011, Monthly Notices of the Royal Astronomical Society, 418, 230

\bibitem[{van Weeren {et~al.}(2019)van Weeren, de~Gasperin, Akamatsu,
  Br{\"{u}}ggen, Feretti, Kang, Stroe, \& Zandanel}]{VanWeeren2019}
van Weeren, R.~J., de~Gasperin, F., Akamatsu, H., {et~al.} 2019, Space Science
  Reviews, 215, 16

\bibitem[{van Weeren {et~al.}(2009)van Weeren, R{\"{o}}ttgering, Br{\"{u}}ggen,
  \& Cohen}]{VanWeeren2009}
van Weeren, R.~J., R{\"{o}}ttgering, H. J.~A., Br{\"{u}}ggen, M., \& Cohen, A.
  2009, Astronomy {\&} Astrophysics, 508, 75

\bibitem[{van Weeren {et~al.}(2021)van Weeren, Shimwell, Botteon, Brunetti,
  Br{\"{u}}ggen, Boxelaar, Cassano, Di~Gennaro, Andrade-Santos, Bonnassieux,
  Bonafede, Cuciti, Dallacasa, De~Gasperin, Gastaldello, Hardcastle, Hoeft,
  Kraft, Mandal, Rossetti, R{\"{o}}ttgering, Tasse, \& Wilber}]{VanWeeren2021}
van Weeren, R.~J., Shimwell, T.~W., Botteon, A., {et~al.} 2021, Astronomy {\&}
  Astrophysics, 651, A115

\bibitem[{Vazza {et~al.}(2012)Vazza, Br{\"{u}}ggen, van Weeren, Bonafede,
  Dolag, \& Brunetti}]{Vazza2012}
Vazza, F., Br{\"{u}}ggen, M., van Weeren, R., {et~al.} 2012, Monthly Notices of
  the Royal Astronomical Society, 421, 1868

\bibitem[{Wittor {et~al.}(2021)Wittor, Ettori, Vazza, Rajpurohit, Hoeft, \&
  Dom{\'{i}}nguez-Fern{\'{a}}ndez}]{Wittor2021}
Wittor, D., Ettori, S., Vazza, F., {et~al.} 2021, Monthly Notices of the Royal
  Astronomical Society, 506, 396

\bibitem[{Zhang {et~al.}(2019)Zhang, Churazov, Forman, \& Lyskova}]{Zhang19}
Zhang, C., Churazov, E., Forman, W.~R., \& Lyskova, N. 2019, Monthly Notices of
  the Royal Astronomical Society, 488, 5259

\bibitem[{Zhang {et~al.}(2022)Zhang, Simionescu, Gastaldello, Eckert,
  Camillini, Natale, Rossetti, Brunetti, Akamatsu, Botteon, Cassano, Cuciti,
  Bruno, Shimwell, Jones, Kaastra, Ettori, Br{\"{u}}ggen, de~Gasperin, Drabent,
  van Weeren, \& R{\"{o}}ttgering}]{Zhang22}
Zhang, X., Simionescu, A., Gastaldello, F., {et~al.} 2022, Astronomy {\&}
  Astrophysics, Submitted

\end{thebibliography}

\begin{appendix}
\section{Radio relic image gallery}
\label{app:collages}
Fig.~\ref{fig:radio_collage} shows \lofar{} 50kpc-taper, discrete source-subtracted images of all RRs in the \lotss{} DR2 - PSZ2 galaxy cluster sample. The images are taken from \citet{Botteon2022} and are also available on the project website\footnote{\url{https://lofar-surveys.org/planck_dr2.html}}. All images are centred on the PSZ2 cluster coordinates, except PSZ2\,G165.46+66.15 which is centred on the X-ray centroid.
The RRs in PSZ2\,G089.52+62.34, PSZ2\,G091.79-27.00, and PSZ2\,G165.46+66.15 were not fully covered by the images, so we have re-imaged these clusters.
The black circle denotes the cluster $R_{500}$, centred on the PSZ2 coordinates. Each image includes a 1 Mpc scalebar and information on the cluster mass, redshift, and image r.m.s. The blue ellipse shows the image beam size.
White crosses denote the coordinates used as the relic positions (see Sec.~\ref{sec:sample:measurements}).

Fig.~\ref{fig:xray_collage} shows the same radio images as white contours, overlaid on \chandra{}/\xmm{} X-ray images. The instrument used is shown in the image title. The radio contours are spaced by factor of 2, starting at 2$\sigma_{rms}$. PSZ2\,G089.52+62.34, PSZ2\,G091.79-27.00, and PSZ2\,G165.46+66.15 are again re-imaged. The $R_{500}$, centred on the PSZ2 coordinates, is shown as a white circle. The position of the X-ray centroids, used as the cluster centres in our analysis, are denoted by black crosses. We note that, for PSZ2\,G107.10+65.32, only the X-ray centroid of the S subcluster is plotted. All other information is as in Fig.~\ref{fig:radio_collage}.

\begin{figure*}
  \centering
  \includegraphics[width=17cm, page=1,trim={0cm 1cm 0cm 0cm},clip]{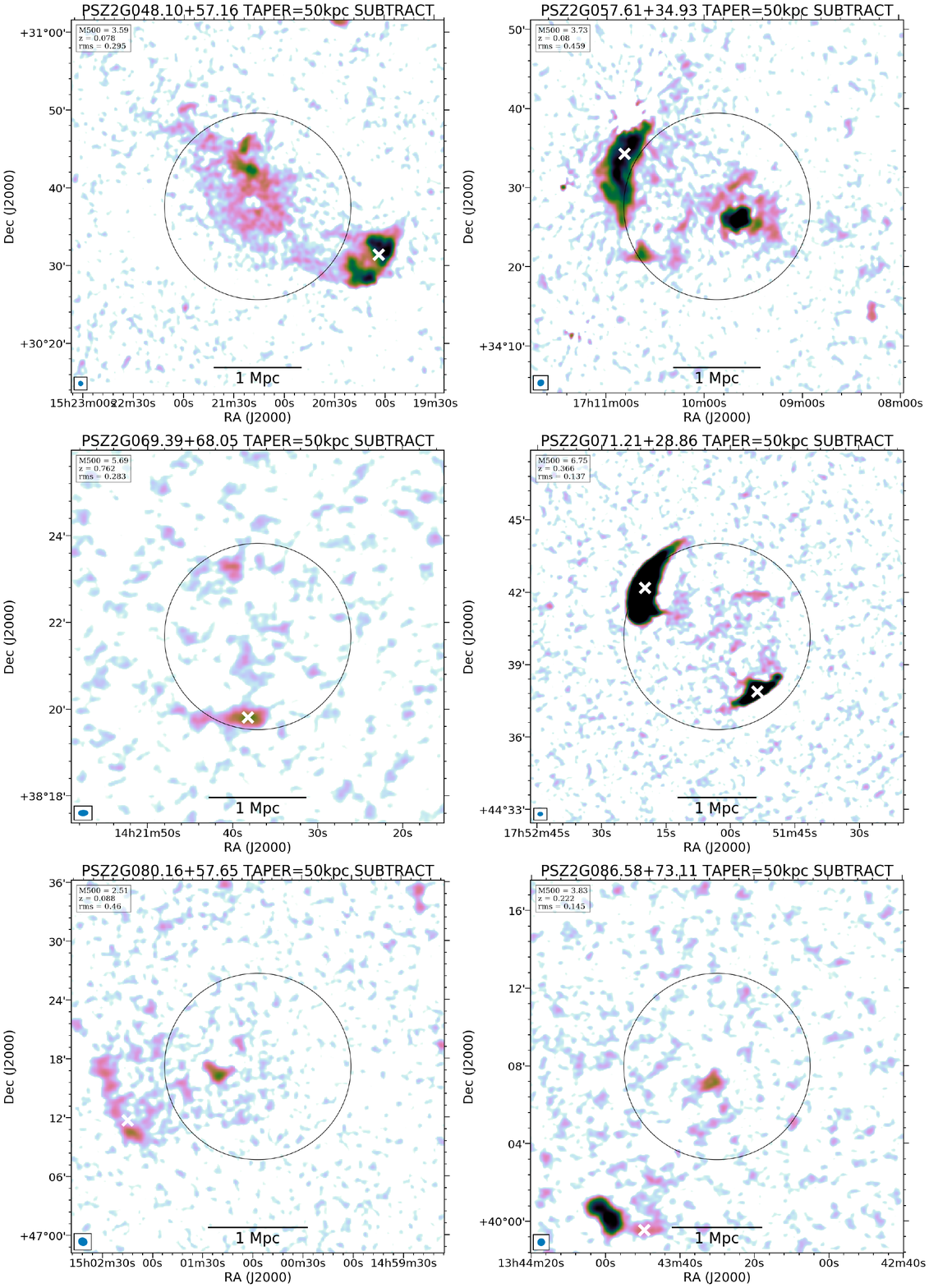}
  \caption{\lofar{} 50kpc-taper, discrete source-subtracted images of the RRs in the \lotss{} DR2 - PSZ2 sample, from \citet{Botteon2022}. The images and $R_{500}$ (black circles) are centred on the PSZ2 coordinate. The image of PSZ2\,G165.46+66.15 is centred on the X-ray centroid. PSZ2\,G089.52+62.34, PSZ2\,G091.79-27.00, and PSZ2\,G165.46+66.15 have been re-imaged. A scalebar, denoting 1 Mpc, is shown in black. The beam is shown in the bottom-left corner, and the mass ($M_{500}$, in units of $10^{14} M_{\odot}$), redshift (z), and image noise (rms, in units of \mjybeam{}) are reported in the top-left corner. White crosses mark the location of the relics, as used throughout the paper (see Sec.~\ref{sec:sample:measurements} for a description of their calculation).}
  \label{fig:radio_collage}
  \vspace{-23.06346pt}
\end{figure*}

\begin{figure*}
  \addtocounter{figure}{-1}
  \includegraphics[width=17cm, page=2]{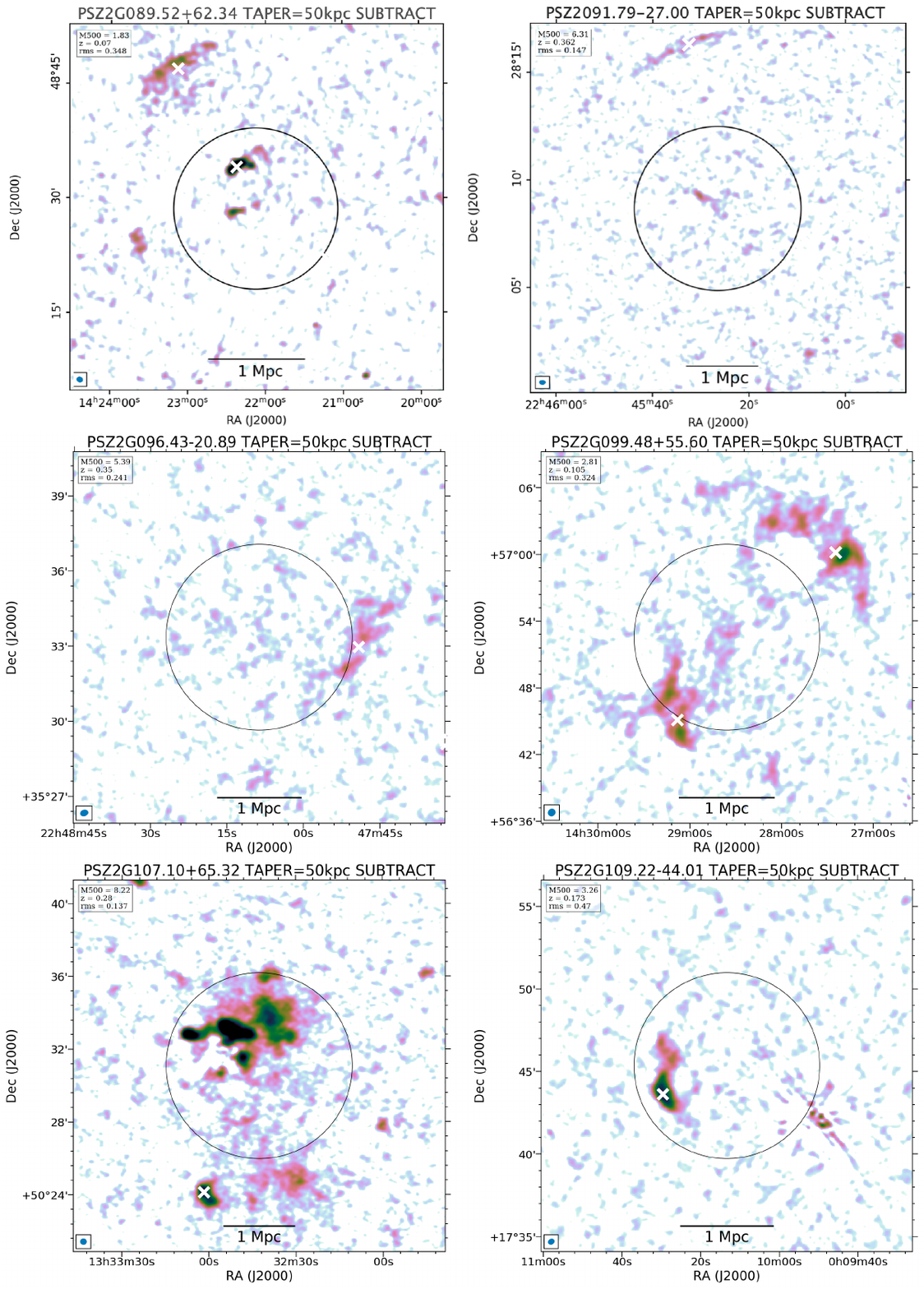}
  \caption{continued.}
\end{figure*}

\begin{figure*}
  \addtocounter{figure}{-1}
  \includegraphics[width=17cm, page=3]{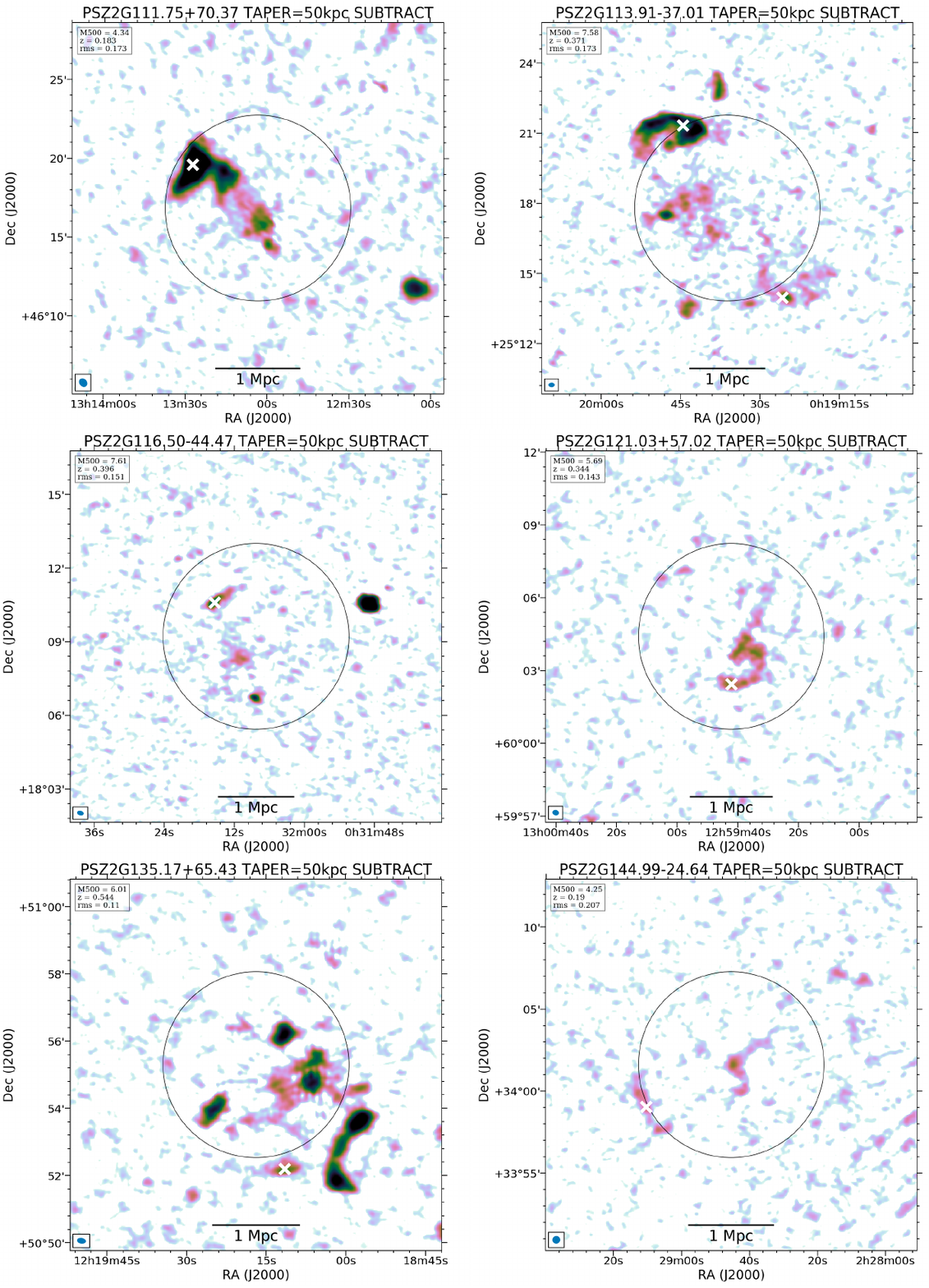}
  \caption{continued.}
\end{figure*}

\begin{figure*}
  \addtocounter{figure}{-1}
  \includegraphics[width=17cm, page=4]{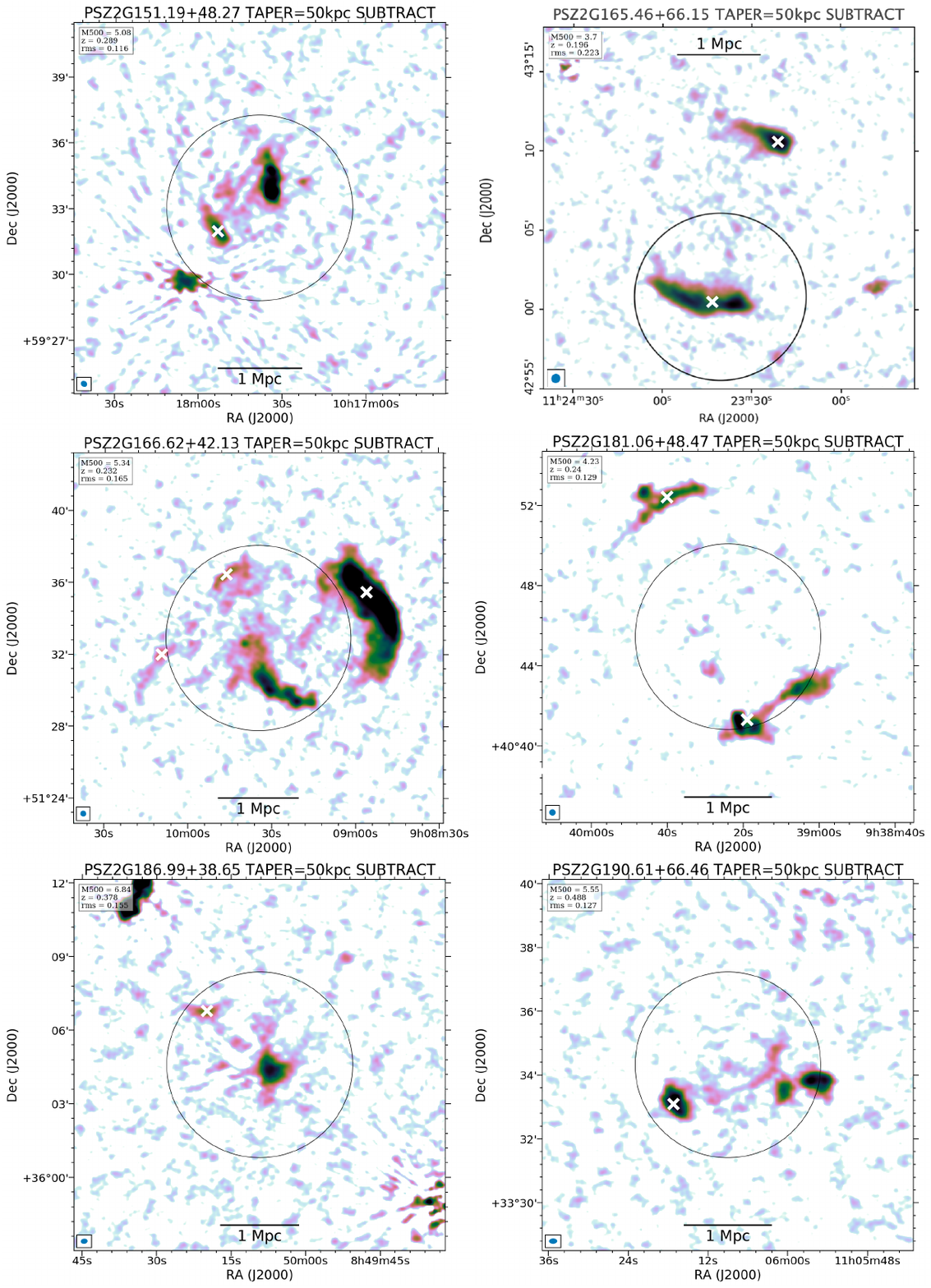}
  \caption{continued.}
\end{figure*}

\begin{figure*}
  \addtocounter{figure}{-1}
  \includegraphics[width=17cm, page=5,trim={0cm 19cm 0cm 0cm},clip]{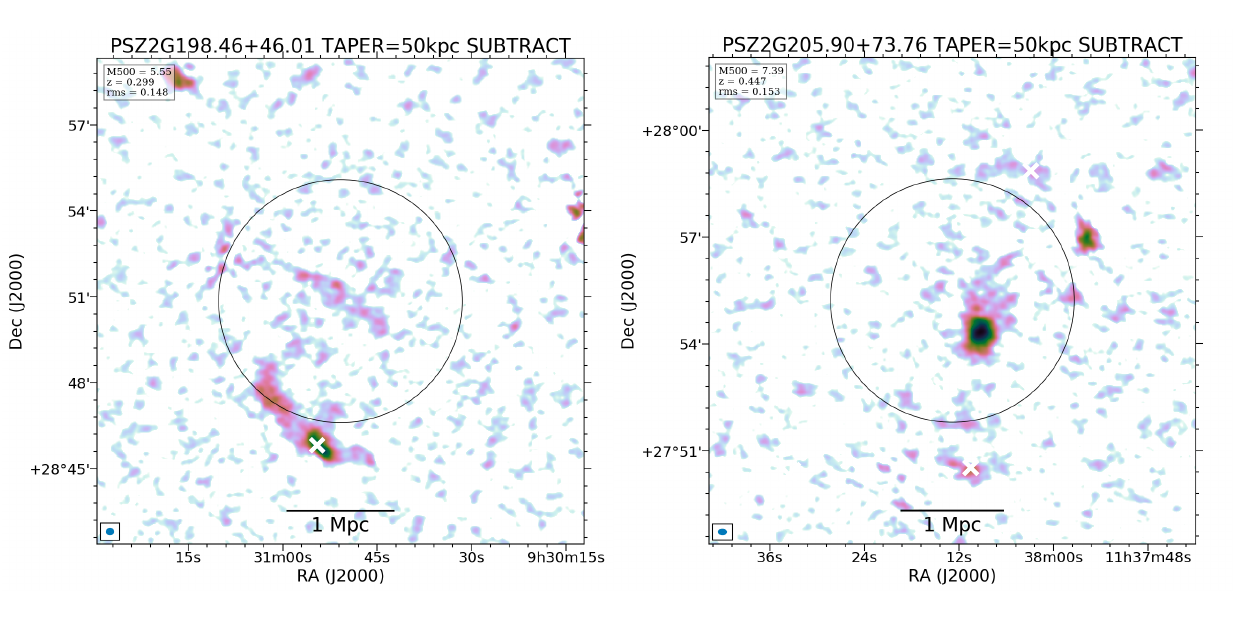}
  \caption{continued.}
\end{figure*}

\begin{figure*}
  \includegraphics[width=17cm, page=1]{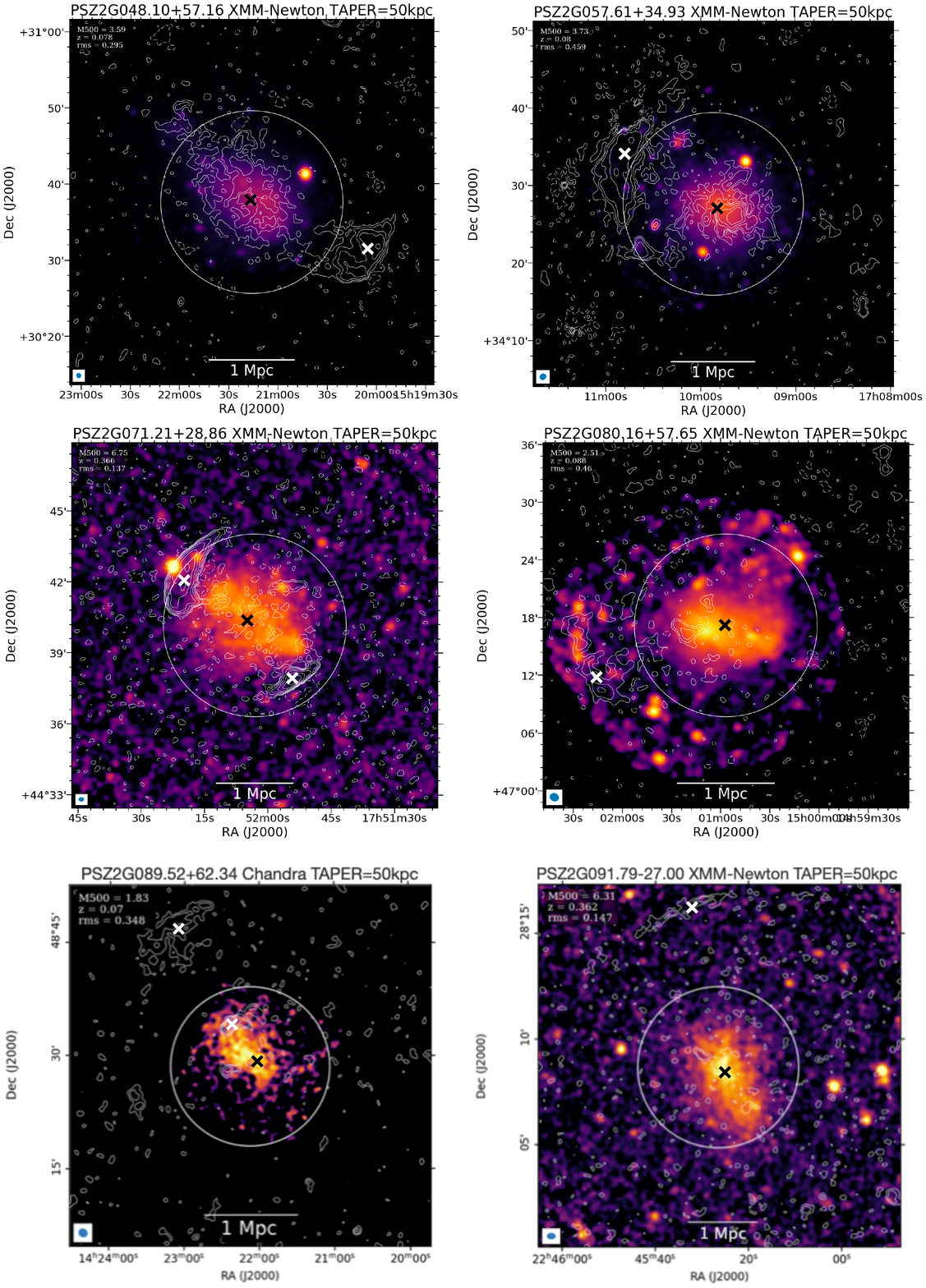}
  \caption{\chandra{}/\xmm{} images of the relic-hosting clusters in our sample. The white contours show the corresponding radio images (Fig.~\ref{fig:radio_collage}) spaced by a factor of 2, starting at 2$\sigma_{rms}$. Black crosses mark the position of the X-ray centroid. We note that, for PSZ2\,G107.10+65.32, only the X-ray centroid of the S subcluster is plotted. All other information is as in Fig.~\ref{fig:radio_collage}.}
  \label{fig:xray_collage}
  \vspace{-16.44496pt}
\end{figure*}

\begin{figure*}
  \addtocounter{figure}{-1}
  \includegraphics[width=17cm, page=2]{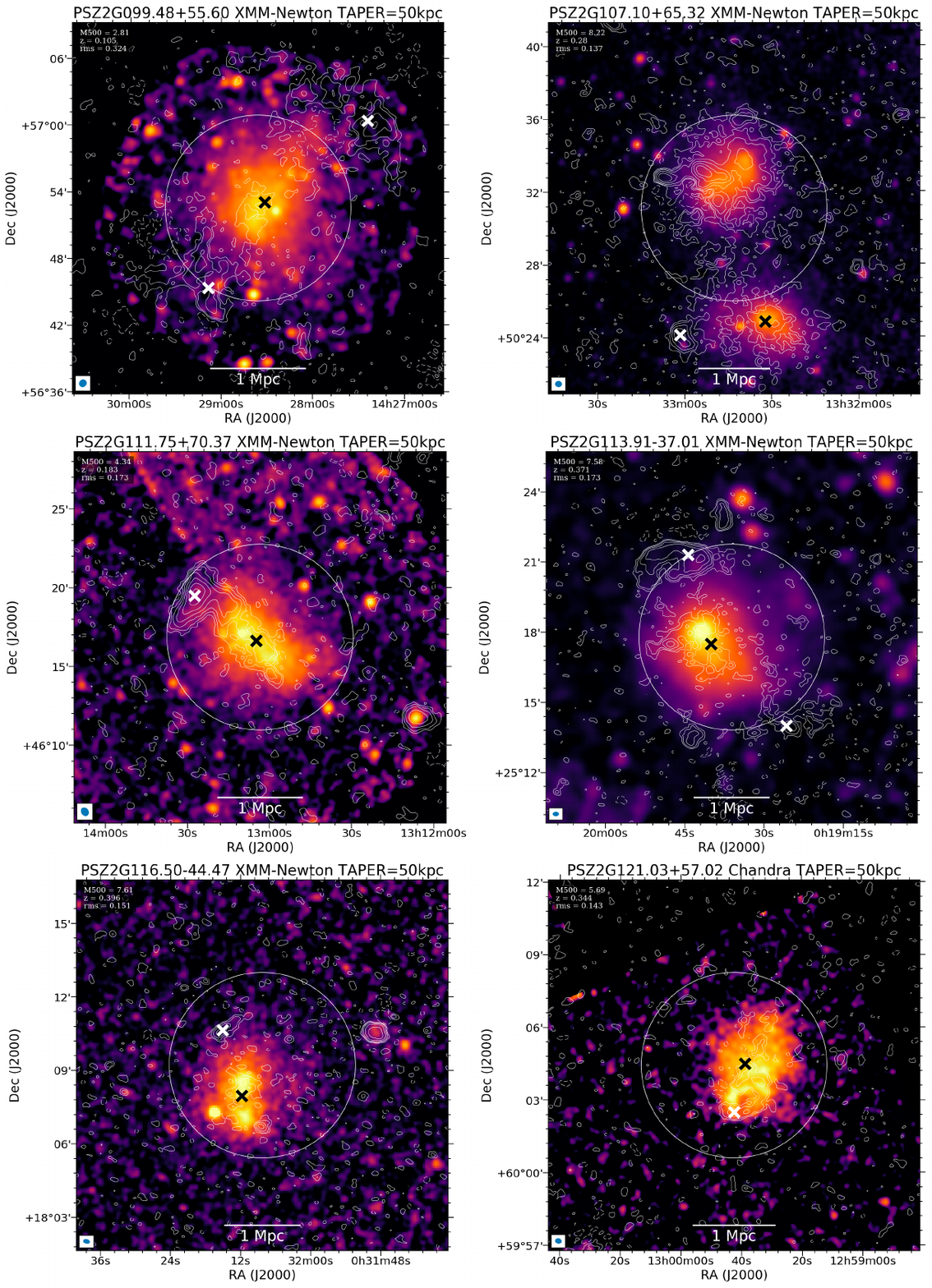}
  \caption{continued.}
\end{figure*}

\begin{figure*}
  \addtocounter{figure}{-1}
  \includegraphics[width=17cm, page=3]{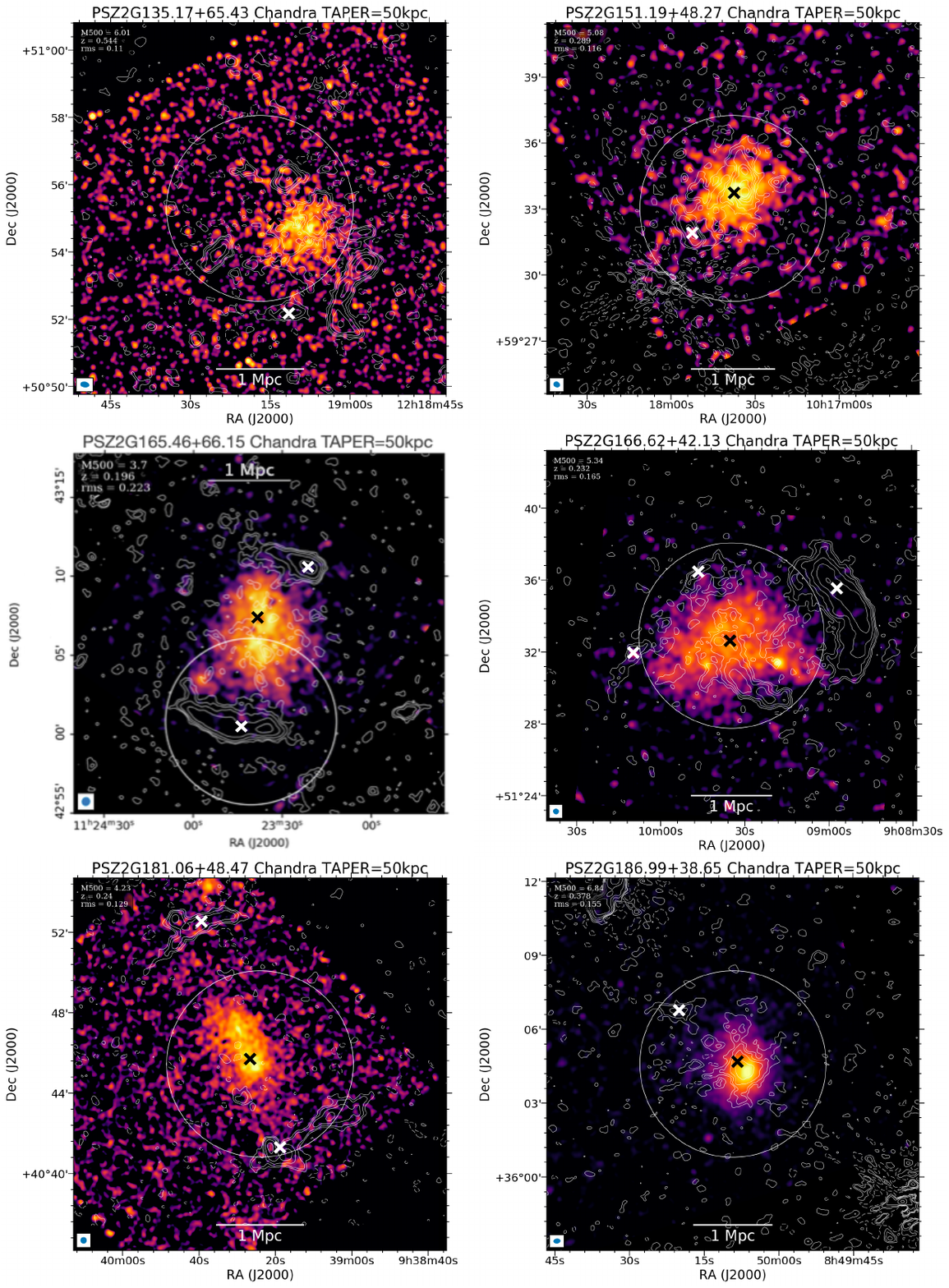}
  \caption{continued.}
\end{figure*}

\begin{figure*}
  \addtocounter{figure}{-1}
  \includegraphics[width=17cm, page=4,trim={0cm 19cm 0cm 0cm},clip]{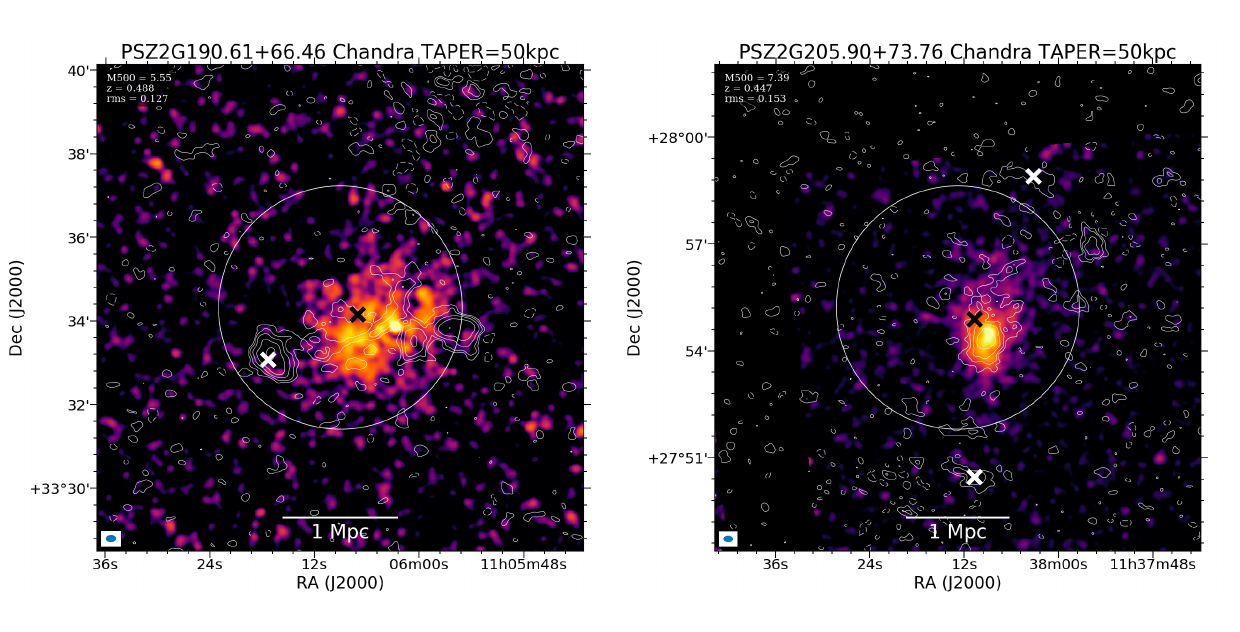}
  \caption{continued.}
\end{figure*}
\FloatBarrier

\section{Comparison with FdG14 powers}
\label{app:fdg+14 P comparison}
\begin{figure}
    \centering
    \resizebox{\hsize}{!}{\includegraphics{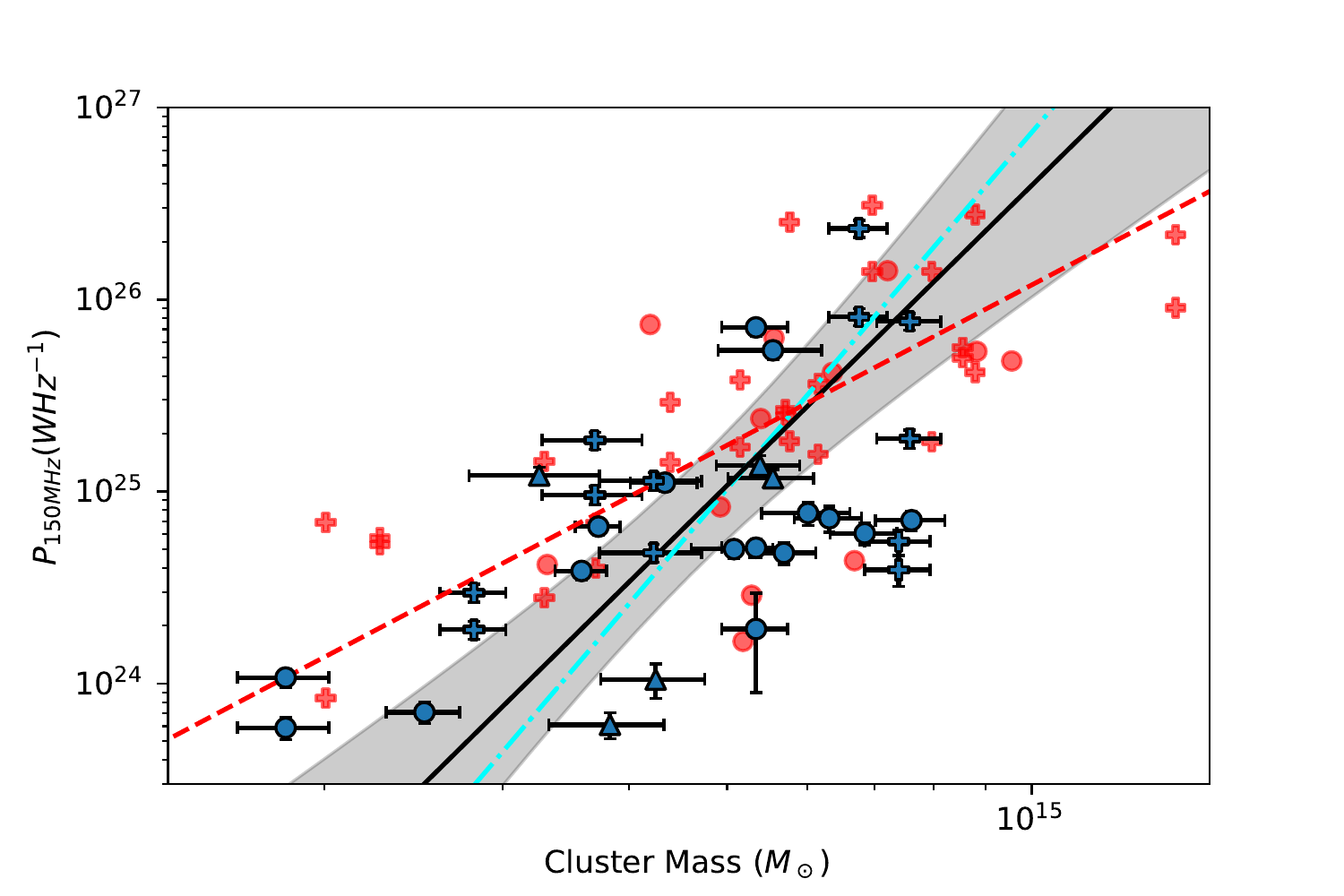}}
    \caption{Same as Fig.~\ref{fig:P_M_fdg} (top) but the relic powers from the DR2 sample have been recomputed using the cosmology used by FdG14.}
    \label{fig:P_M_fdg_cosmo}
\end{figure}

In Fig.~\ref{fig:P_M_fdg} we compare the power vs. mass scaling relations of the relics in this sample (150 MHz) and those used by FdG14 (1.4 GHz). To make a completely fair comparison, the cosmology used to calculate the relic powers should be the same. In Fig.~\ref{fig:P_M_fdg_cosmo} we reproduce Fig.~\ref{fig:P_M_fdg} (top) with the relic powers from our sample recomputed with the same cosmology as FdG14 (\lcdm\ cosmology with $\omegal = 0.73$, $\omegam = 0.27$, and $\hzero = 71$ \kmsmpc), since the scaling relation in FdG14 was calculated with this cosmology. The slope and intercept (orthogonal fit) of the line of best fit for the recomputed powers are $\textrm{B}=5.84\pm1.31$ and $\textrm{A}=-60.74\pm19.27$.

\FloatBarrier

\section{Fitting methods}
\label{app:fit methods}
\begin{figure}
    \resizebox{\hsize}{!}{\includegraphics{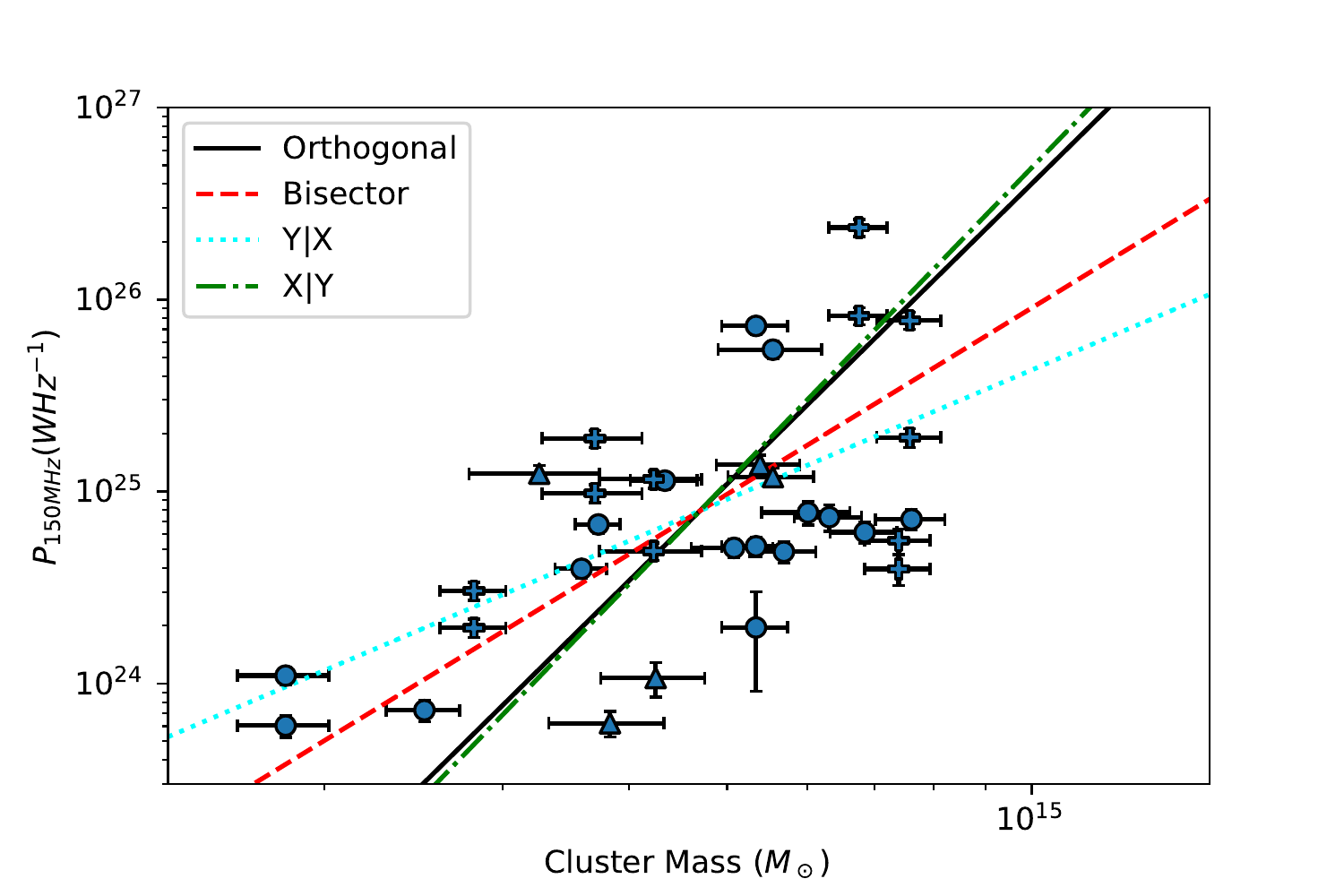}}
    \caption{Relic power vs. cluster mass for confirmed relics only. The regression lines are shown for all four fitting methods.}
    \label{fig:P_M_fits}
\end{figure}

The slope and intercept obtained when fitting a regression line depends on the fitting method (see Tab.~\ref{tab:P_M_fits}. In this paper we report the parameters for four different methods: orthogonal; Y$\vert$X; X$\vert$Y and the bisector of Y$\vert$X and X$\vert$Y. In Fig.~\ref{fig:P_M_fits} we show the regression line (confirmed relics only) for each fitting method.  

\end{appendix}
\end{document}